\documentclass[aps,prd,reprint,superscriptaddress]{revtex4-2} 

\usepackage{color}
\usepackage{graphicx}
\usepackage{physics}
\usepackage{lineno}
\usepackage{amsmath}
\begin{document}


\title{Measurements of the charge ratio and polarization of cosmic-ray muons \\ with the Super-Kamiokande detector}




\newcommand{\AFFicrr}{\affiliation{Kamioka Observatory, Institute for Cosmic Ray Research, University of Tokyo, Kamioka, Gifu 506-1205, Japan}}
\newcommand{\AFFkashiwa}{\affiliation{Research Center for Cosmic Neutrinos, Institute for Cosmic Ray Research, University of Tokyo, Kashiwa, Chiba 277-8582, Japan}}
\newcommand{\AFFicrronly}{\affiliation{Institute for Cosmic Ray Research, University of Tokyo, Kashiwa, Chiba 277-8582, Japan}}
\newcommand{\AFFipmu}{\affiliation{Kavli Institute for the Physics and
Mathematics of the Universe (WPI), The University of Tokyo Institutes for Advanced Study,
University of Tokyo, Kashiwa, Chiba 277-8583, Japan }}
\newcommand{\AFFmad}{\affiliation{Department of Theoretical Physics, University Autonoma Madrid, 28049 Madrid, Spain}}
\newcommand{\AFFubc}{\affiliation{Department of Physics and Astronomy, University of British Columbia, Vancouver, BC, V6T1Z4, Canada}}
\newcommand{\AFFbu}{\affiliation{Department of Physics, Boston University, Boston, MA 02215, USA}}
\newcommand{\AFFuci}{\affiliation{Department of Physics and Astronomy, University of California, Irvine, Irvine, CA 92697-4575, USA }}
\newcommand{\AFFcsu}{\affiliation{Department of Physics, California State University, Dominguez Hills, Carson, CA 90747, USA}}
\newcommand{\AFFcnm}{\affiliation{Institute for Universe and Elementary Particles, Chonnam National University, Gwangju 61186, Korea}}
\newcommand{\AFFduke}{\affiliation{Department of Physics, Duke University, Durham NC 27708, USA}}
\newcommand{\AFFfukuoka}{\affiliation{Junior College, Fukuoka Institute of Technology, Fukuoka, Fukuoka 811-0295, Japan}}
\newcommand{\AFFgifu}{\affiliation{Department of Physics, Gifu University, Gifu, Gifu 501-1193, Japan}}
\newcommand{\AFFgist}{\affiliation{GIST College, Gwangju Institute of Science and Technology, Gwangju 500-712, Korea}}
\newcommand{\AFFuh}{\affiliation{Department of Physics and Astronomy, University of Hawaii, Honolulu, HI 96822, USA}}
\newcommand{\AFFicl}{\affiliation{Department of Physics, Imperial College London , London, SW7 2AZ, United Kingdom }}
\newcommand{\AFFkek}{\affiliation{High Energy Accelerator Research Organization (KEK), Tsukuba, Ibaraki 305-0801, Japan }}
\newcommand{\AFFkobe}{\affiliation{Department of Physics, Kobe University, Kobe, Hyogo 657-8501, Japan}}
\newcommand{\AFFkyoto}{\affiliation{Department of Physics, Kyoto University, Kyoto, Kyoto 606-8502, Japan}}
\newcommand{\AFFliv}{\affiliation{Department of Physics, University of Liverpool, Liverpool, L69 7ZE, United Kingdom}}
\newcommand{\AFFmiyagi}{\affiliation{Department of Physics, Miyagi University of Education, Sendai, Miyagi 980-0845, Japan}}
\newcommand{\AFFnagoya}{\affiliation{Institute for Space-Earth Environmental Research, Nagoya University, Nagoya, Aichi 464-8602, Japan}}
\newcommand{\AFFkmi}{\affiliation{Kobayashi-Maskawa Institute for the Origin of Particles and the Universe, Nagoya University, Nagoya, Aichi 464-8602, Japan}}
\newcommand{\AFFpol}{\affiliation{National Centre For Nuclear Research, 02-093 Warsaw, Poland}}
\newcommand{\AFFsuny}{\affiliation{Department of Physics and Astronomy, State University of New York at Stony Brook, NY 11794-3800, USA}}
\newcommand{\AFFokayama}{\affiliation{Department of Physics, Okayama University, Okayama, Okayama 700-8530, Japan }}
\newcommand{\AFFosaka}{\affiliation{Department of Physics, Osaka University, Toyonaka, Osaka 560-0043, Japan}}
\newcommand{\AFFox}{\affiliation{Department of Physics, Oxford University, Oxford, OX1 3PU, United Kingdom}}
\newcommand{\AFFqmul}{\affiliation{School of Physics and Astronomy, Queen Mary University of London, London, E1 4NS, United Kingdom}}
\newcommand{\AFFregina}{\affiliation{Department of Physics, University of Regina, 3737 Wascana Parkway, Regina, SK, S4SOA2, Canada}}
\newcommand{\AFFseoul}{\affiliation{Department of Physics, Seoul National University, Seoul 151-742, Korea}}
\newcommand{\AFFsheff}{\affiliation{Department of Physics and Astronomy, University of Sheffield, S3 7RH, Sheffield, United Kingdom}}
\newcommand{\AFFshizuokasc}{\affiliation{Department of Informatics in
Social Welfare, Shizuoka University of Welfare, Yaizu, Shizuoka, 425-8611, Japan}}
\newcommand{\AFFstfc}{\affiliation{STFC, Rutherford Appleton Laboratory, Harwell Oxford, and Daresbury Laboratory, Warrington, OX11 0QX, United Kingdom}}
\newcommand{\AFFskk}{\affiliation{Department of Physics, Sungkyunkwan University, Suwon 440-746, Korea}}
\newcommand{\AFFtodai}{\affiliation{Department of Physics, University of Tokyo, Bunkyo, Tokyo 113-0033, Japan }}
\newcommand{\AFFtit}{\affiliation{Department of Physics,Tokyo Institute of Technology, Meguro, Tokyo 152-8551, Japan }}
\newcommand{\AFFtus}{\affiliation{Department of Physics, Faculty of Science and Technology, Tokyo University of Science, Noda, Chiba 278-8510, Japan }}
\newcommand{\AFFtoronto}{\affiliation{Department of Physics, University of Toronto, ON, M5S 1A7, Canada }}
\newcommand{\AFFtriumf}{\affiliation{TRIUMF, 4004 Wesbrook Mall, Vancouver, BC, V6T2A3, Canada }}
\newcommand{\AFFtokai}{\affiliation{Department of Physics, Tokai University, Hiratsuka, Kanagawa 259-1292, Japan}}
\newcommand{\AFFtsinghua}{\affiliation{Department of Engineering Physics, Tsinghua University, Beijing, 100084, China}}
\newcommand{\AFFynu}{\affiliation{Department of Physics, Yokohama National University, Yokohama, Kanagawa, 240-8501, Japan}}
\newcommand{\AFFllr}{\affiliation{Ecole Polytechnique, IN2P3-CNRS, Laboratoire Leprince-Ringuet, F-91120 Palaiseau, France }}
\newcommand{\AFFbari}{\affiliation{ Dipartimento Interuniversitario di Fisica, INFN Sezione di Bari and Universit\`a e Politecnico di Bari, I-70125, Bari, Italy}}
\newcommand{\AFFnapoli}{\affiliation{Dipartimento di Fisica, INFN Sezione di Napoli and Universit\`a di Napoli, I-80126, Napoli, Italy}}
\newcommand{\AFFroma}{\affiliation{INFN Sezione di Roma and Universit\`a di Roma ``La Sapienza'', I-00185, Roma, Italy}}
\newcommand{\AFFpadova}{\affiliation{Dipartimento di Fisica, INFN Sezione di Padova and Universit\`a di Padova, I-35131, Padova, Italy}}
\newcommand{\AFFkeio}{\affiliation{Department of Physics, Keio University, Yokohama, Kanagawa, 223-8522, Japan}}
\newcommand{\AFFwinnipeg}{\affiliation{Department of Physics, University of Winnipeg, MB R3J 3L8, Canada }}
\newcommand{\AFFkcl}{\affiliation{Department of Physics, King's College London, London, WC2R 2LS, UK }}
\newcommand{\AFFwarwick}{\affiliation{Department of Physics, University of Warwick, Coventry, CV4 7AL, UK }}
\newcommand{\AFFral}{\affiliation{Rutherford Appleton Laboratory, Harwell, Oxford, OX11 0QX, UK }}
\newcommand{\AFFwu}{\affiliation{Faculty of Physics, University of Warsaw, Warsaw, 02-093, Poland }}
\newcommand{\AFFbcit}{\affiliation{Department of Physics, British Columbia Institute of Technology, Burnaby, BC, V5G 3H2, Canada }}
\newcommand{\AFFtohoku}{\affiliation{Department of Physics, Faculty of Science, Tohoku University, Sendai, Miyagi, 980-8578, Japan }}
\newcommand{\AFFicise}{\affiliation{Institute For Interdisciplinary Research in Science and Education, ICISE, Quy Nhon, 55121, Vietnam }}
\newcommand{\AFFilance}{\affiliation{ILANCE, CNRS - University of Tokyo International Research Laboratory, Kashiwa, Chiba 277-8582, Japan}}
\newcommand{\AFFibs}{\affiliation{Institute for Basic Science (IBS), Daejeon, 34126, Korea}}
\newcommand{\AFFoecu}{\affiliation{Media Communication Center, Osaka Electro-Communication University, Neyagawa, Osaka, 572-8530, Japan}}
\newcommand{\AFFglasgow}{\affiliation{School of Physics and Astronomy, University of Glasgow, Glasgow, Scotland, G12 8QQ, United Kingdom}}
\newcommand{\AFFminn}{\affiliation{School of Physics and Astronomy, University of Minnesota, Minneapolis, MN  55455, USA}}
\newcommand{\AFFsilesia}{\affiliation{August Che\l{}kowski Institute of Physics, University of Silesia in Katowice, 75 Pu\l{}ku Piechoty 1, 41-500 Chorz\'{o}w, Poland}}

\AFFicrr
\AFFkashiwa
\AFFicrronly
\AFFmad
\AFFbu
\AFFbcit
\AFFuci
\AFFcsu
\AFFcnm
\AFFduke
\AFFllr
\AFFgifu
\AFFgist
\AFFglasgow
\AFFuh
\AFFibs
\AFFicise
\AFFicl
\AFFbari
\AFFnapoli
\AFFpadova
\AFFroma
\AFFilance
\AFFkeio
\AFFkek
\AFFkcl
\AFFkobe
\AFFkyoto
\AFFliv
\AFFminn
\AFFmiyagi
\AFFnagoya
\AFFkmi
\AFFpol
\AFFsuny
\AFFokayama
\AFFoecu
\AFFox
\AFFral
\AFFseoul
\AFFsheff
\AFFshizuokasc
\AFFsilesia
\AFFstfc
\AFFskk
\AFFtohoku
\AFFtokai
\AFFtodai
\AFFipmu
\AFFtit
\AFFtus
\AFFtoronto
\AFFtriumf
\AFFtsinghua
\AFFwu
\AFFwarwick
\AFFwinnipeg
\AFFynu

\author{H.~Kitagawa}
\author{T.~Tada}
\AFFokayama

\author{K.~Abe}
\AFFicrr
\AFFipmu
\author{C.~Bronner}
\AFFicrr
\author{Y.~Hayato}
\author{K.~Hiraide}
\AFFicrr
\AFFipmu
\author{K.~Hosokawa}
\AFFicrr
\author{K.~Ieki}
\author{M.~Ikeda}
\author{J.~Kameda}
\AFFicrr
\AFFipmu
% ...


\collaboration{The Super-Kamiokande Collaboration}
\noaffiliation

\date{\today}

\begin{abstract}

We present the results of the charge ratio~($R$) and polarization~($P^{\mu}_{0}$) measurements using decay electron events collected between September 2008 and June 2022 with the Super-Kamiokande detector. Because of its underground location and long operation, we are able to perform high-precision measurements by accumulating cosmic-ray muons. We measured the muon charge ratio to be $R=1.32 \pm 0.02~(\mathrm{stat.}{+}\mathrm{syst.})$ at $E_{\mu}\cos \theta_{\mathrm{Zenith}}=0.7^{+0.3}_{-0.2}~\mathrm{TeV}$, where $E_{\mu}$ is the muon energy and $\theta_{\mathrm{Zenith}}$ is the zenith angle of incoming cosmic-ray muons. This result is consistent with the Honda flux model while indicating a tension with the $\pi K$ model of $1.9\sigma$. We also measured the muon polarization at the production location to be $P^{\mu}_{0}=0.52 \pm 0.02~(\mathrm{stat.}{+}\mathrm{syst.})$ at the muon momentum of $0.9^{+0.6}_{-0.1}~\mathrm{TeV}/c$ at the surface of the mountain; this also suggests a tension with the Honda flux model of $1.5\sigma$. This is the most precise measurement ever to experimentally determine the cosmic-ray muon polarization near $1~\mathrm{TeV}/c$. These measurement results are useful to improve atmospheric neutrino simulations. 
\end{abstract}


\maketitle


\section{Introduction} \label{sec:intro}

Three-flavor neutrino mixing, described by the Pontecorvo-Maki-Nakagawa-Sakata~(PMNS) matrix~\cite{Maki:1962mu, Pontecorvo:1967fh}, is generally parameterized by three mixing angles, two neutrino mass-squared differences, and one $CP$-violating phase. Neutrino oscillation was first confirmed in atmospheric neutrino data observed with the Super-Kamiokande~(SK) detector~\cite{Super-Kamiokande:1998kpq}. Since then neutrino oscillation has been observed not only in atmospheric neutrinos~\cite{IceCube:2017lak, Super-Kamiokande:2017yvm, ANTARES:2018rtf} but also with other neutrino sources such as solar neutrinos~\cite{Fukuda:2001nj, Ahmad:2001an, Ahmad:2002jz}, accelerator neutrinos~\cite{K2K:2004iot, MINOS:2011amj, NOvA:2019cyt, T2K:2019bcf}, and reactor neutrinos~\cite{KamLAND:2002uet, DoubleChooz:2011ymz, DayaBay:2012fng}. However, several oscillation parameters remain unknown; the mass hierarchy of $\Delta m^{2}_{23}$, the octant of $\theta_{23}$, and the value of $CP$-violating phase.

In the SK detector, the approach to measuring these oscillation parameters is by comparing the energy, angle, and flavor distributions of atmospheric neutrino interaction products observed in data with that predicted by atmospheric neutrino simulations~\cite{Gaisser:1988ar, Honda:1995hz}. For atmospheric neutrino measurements, a precise knowledge of the ratio of neutrinos to anti-neutrinos~($\nu/\bar{\nu}$) is of particular importance in the extraction of accurate oscillation parameters. However, the accuracy of the atmospheric neutrino predictions in the absolute flux and the flavor proportion is limited by uncertainties in both the primary cosmic-ray flux and the hadronic production in the air shower~\cite{Barr:2006it, Honda:2006qj, Fedynitch:2012fs}. In particular, air shower simulations show a significant cosmic-ray muon deficit in experimental measurement. This deficit is the so-called Muon Puzzle~\cite{Albrecht:2021cxw} and such discrepancies must be reduced for precise neutrino measurement.

Cosmic-ray muons predominantly come from the decay of mesons produced in hadronic showers that follow from the interaction between primary cosmic-rays and nuclei in the atmosphere. Accordingly, these muons carry essential information on pion and kaon production in the hadronic interaction. 

\begin{figure}[!h]
    \begin{center}
        \includegraphics[width=\linewidth]{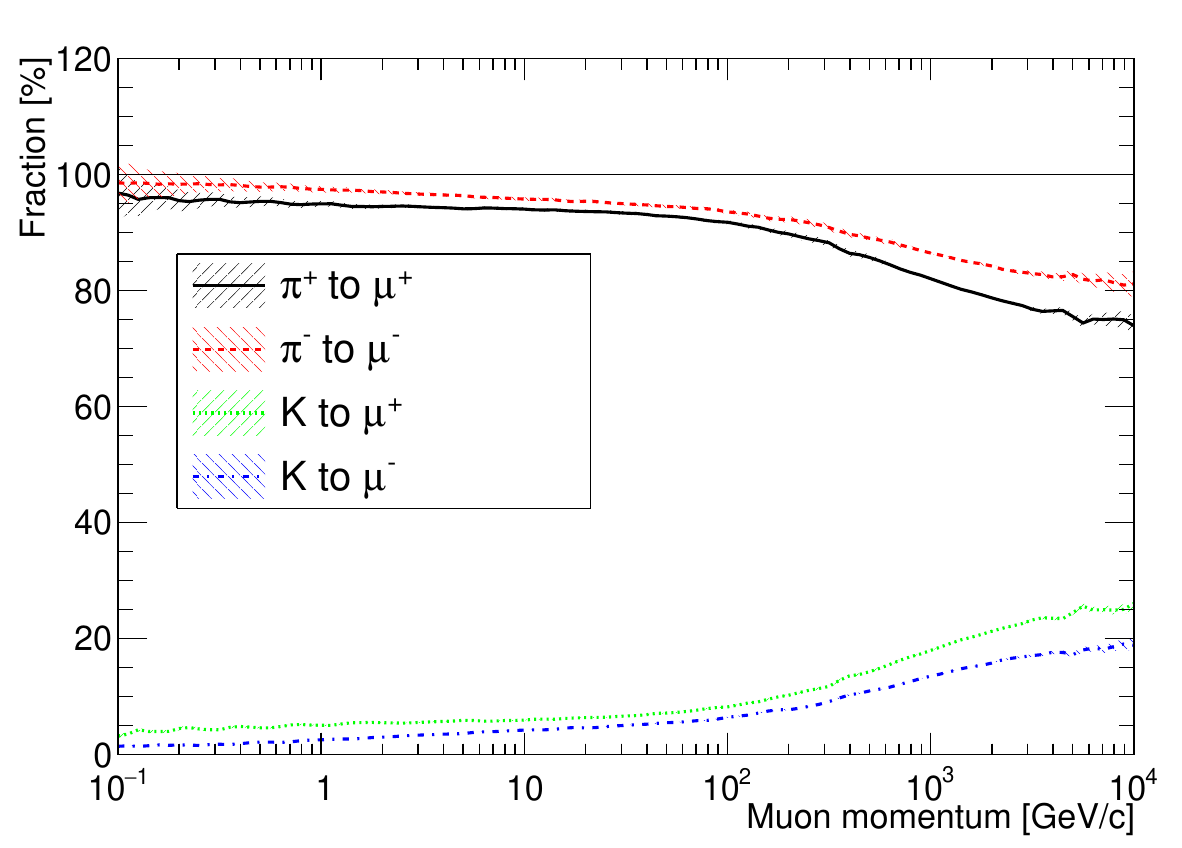}
    \end{center}
\caption{The fraction of parent particles that produce cosmic-ray muons at the Kamioka site, based on the Honda flux simulation~\cite{Honda:2015fha}. The fraction is calculated by integrating over the zenith angle. The solid black, dashed red, dotted green, and blue dashed-dotted lines show the fraction of positive pion to positive muon, negative pion to negative muon, kaon to positive muon, and kaon to negative muon, respectively. \label{fig:mu_contribution}}
\end{figure}

Figure~\ref{fig:mu_contribution} shows the expected fraction of parent particles that produce cosmic-ray muons at the Kamioka site by the Honda flux model~\cite{Honda:2015fha}. The main source of cosmic-ray muons is pion decays at low energy. On the other hand, the contribution of pion decays is suppressed at high energy because the interactions with atmospheric nuclei before their decays increase due to their relatively long lifetime. Thus, the contribution of kaon decays relatively increases at high energy. This phenomenon is parameterized in Ref.~\cite{Gaisser:2011klf} by using critical energy of $\varepsilon_{\pi} = 115$~GeV~($\varepsilon_{K} = 850$~GeV), which is the energy of pion~(kaon) where the probabilities of causing interaction and decay are equal. For the muon energy of $E_{\mu}>\varepsilon_{\pi}/\cos\theta_{\mathrm{Zenith}}$, the contribution of pion decays is suppressed while the contribution of kaon decays increases, where the dependence of zenith angle~($\theta_{\mathrm{Zenith}}$) due to the air density profile change is taken into account. This tendency continues until $E_{\mu}\gg \varepsilon_{K}/\cos \theta_{\mathrm{Zenith}}$. 

\begin{figure}[!h]
    \begin{center}
        \includegraphics[width=\linewidth]{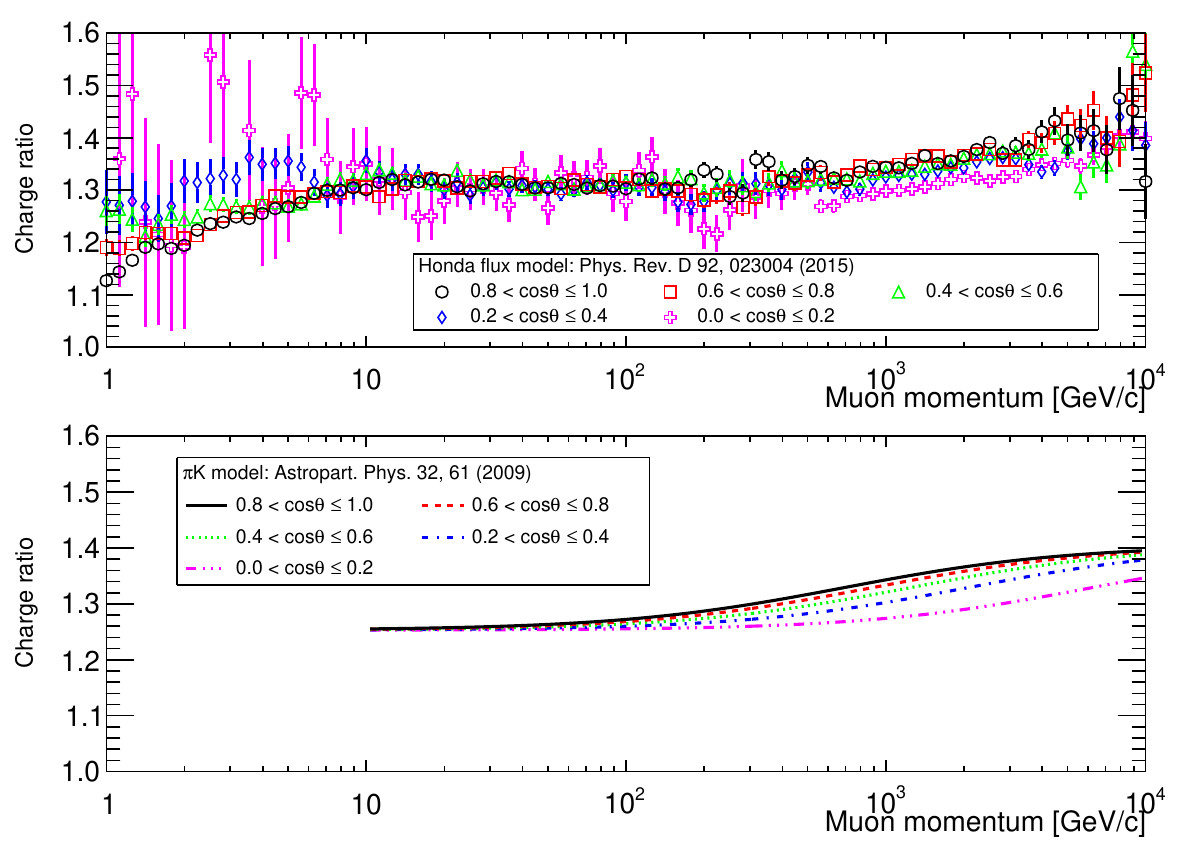}
    \end{center}
\caption{The expected muon charge ratio from two simulation models. The top panel shows the expectation at the Kamioka site based on the Honda flux simulation~\cite{Honda:2015fha}. The vertical uncertainties are the statistical uncertainties of the MC simulation. The bottom panel shows the expectation from the $\pi K$ model~\cite{Schreiner:2009pf}. In both models, the charge ratio depends on the zenith angle in the muon momentum above $100~\mathrm{GeV}/c$. The $\pi K$ model considers the energy range from $10$~GeV to PeV~\cite{Gaisser:2011klf}. \label{fig:cr_honda}}
\end{figure}

The charge ratio, which is defined as the ratio of the number of positive particles to negative particles, is an important observable to constrain the flavor ratio of atmospheric neutrinos. The kaon charge ratio is larger than that of pions because positive kaons production is associated with lambda particles in air showers, and this results in a higher production rate than that of negative kaons. The contribution of kaon decays to muon production increases at high $E_{\mu} \cos \theta_{\mathrm{Zenith}}$ as explained above, and this results in the rise of muon charge ratio~\cite{Schreiner:2009pf, Gaisser:2011klf}. Figure~\ref{fig:cr_honda} shows the muon charge ratio expected by two theoretical models; the Honda flux model~\cite{Honda:2015fha} and the $\pi K$ model~\cite{Schreiner:2009pf}. The Honda flux model is a full simulation of the atmospheric neutrino and muon fluxes. Since it simulates the interaction between primary cosmic-rays and the nucleus in the atmosphere considering the modern spectrum and composition of primary cosmic-rays, this model can predict not only the muon charge ratio but also the polarization of cosmic-ray muons in the energy range from MeV to PeV. The simulation of muon polarization is detailed in Appendix~\ref{sec:mc-pol}. On the other hand, the $\pi K$ model predicts only the muon charge ratio in the energy range from $10$~GeV to PeV because this model has been empirically constructed from past experimental data above $10$~GeV. The $\pi K$ model parameterizes the charge ratio considering the critical energy of pion and kaon decays and the zenith angle of cosmic ray muons. Details of the parameterization are described in Sect.~\ref{sec:charge_result}.

Since the interaction length of the parent meson depends on $E_{\mu} \cos \theta_{\mathrm{Zenith}}$, the charge ratio for vertically down-going muons is higher than that for horizontally going muons, as shown in Fig.~\ref{fig:cr_honda}.

\begin{figure}[!h]
    \begin{center}
        \includegraphics[width=\linewidth]{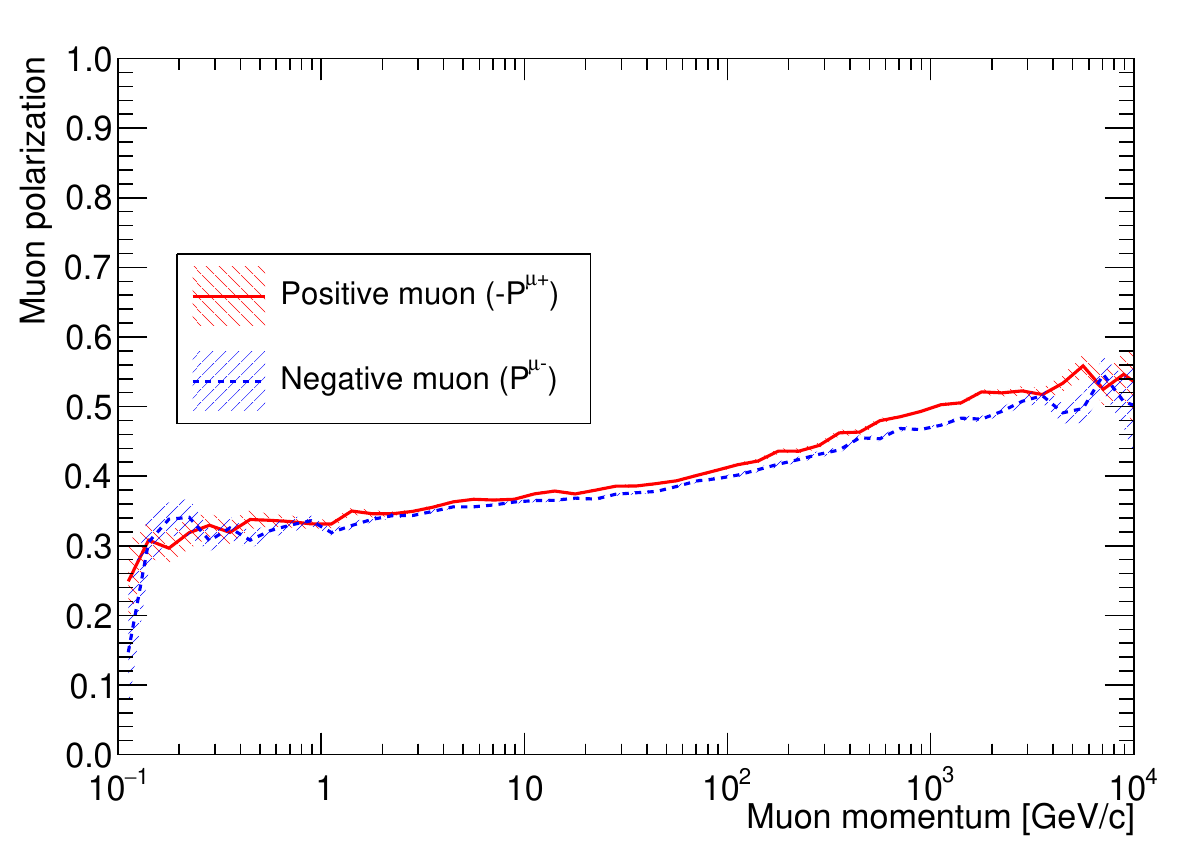}
    \end{center}
\caption{The expected polarization as a function of muon momentum based on the Honda flux simulation~\cite{Honda:2015fha}. The red solid and blue dashed lines show the polarization of positive and negative muons, respectively. For positive muons the sign of the polarization is inverted. Because of the differing contributions from charged kaons in the atmosphere as shown in Fig.~\ref{fig:mu_contribution}, the absolute values of polarization are slightly larger for positive muons than for negative muons above $1$~GeV/$c$. \label{fig:mu_pol_expect}}
\end{figure}

In addition to the muon charge ratio, the polarization of cosmic-ray muons is also an important observable to constrain the contribution from kaon decays. The muon produced in the two-body decay of a meson is completely polarized in the direction of motion of the muon in the rest frame of the meson~\cite{Barr:1989ru, Lee:1989bw, Lipari:1993hd}. The polarization of muons from kaon decays in the laboratory frame is therefore much larger than those from pion decays because the polarization reflects the rest mass of the parent meson~\cite{Hayakawa:1957}. Hence, a measurement of the magnitude of muon polarization constrains the relative contribution from kaons and pions to the muon flux~\cite{clark:1957, Osborne:1964, Turner:1971}. Figure~\ref{fig:mu_pol_expect} shows the momentum dependence of the muon polarization simulated with the Honda flux model~\cite{Honda:2015fha}. 

The SK detector is a water Cherenkov detector located $1000$~m beneath the top of Mt Ikenoyama in Japan~\cite{Fukuda:2002uc}. Cosmic-ray muons must penetrate the mountain to reach the SK detector, selecting a typical momentum of $0.9~\mathrm{TeV}$ at sea level. Although the detector cannot distinguish the charge of the penetrating cosmic-ray muons, the muon charge ratio can be statistically determined by measuring the decay time of stopping muons, since negative muons tend to have a shorter decay time due to the formation of muonic atoms with oxygen in the water. Furthermore, the angular distribution between the parent muon and the decay electron gives the magnitude of the muon polarization. To this end, we analyzed stopping muons observed in the SK detector to provide new information for the simulation of atmospheric neutrinos~\cite{Sanuki:2006yd, Honda:2006qj, Honda:2011nf, Honda:2015fha, Honda:2019ymh, Fedynitch:2018cbl}.

This paper is organized as follows. In Sec.~\ref{sec:sk} we briefly describe the SK detector and its reconstruction performance. In Sec.~\ref{sec:simulation} we describe the development of the MC simulation for muon decays with polarization. In Sec.~\ref{sec:opt}, we describe the data analysis methods to identify pairs of parent muons and decay electrons, the definition of the $\chi^{2}$ method used to determine the charge ratio and the polarization of cosmic-ray muons, and the systematic uncertainties. In Sec.~\ref{sec:result} we present the analysis results, including the incoming muon directional dependence and a search for periodicity using yearly data, and make comparisons to results from other experiments. In the final section, we conclude this study and give prospects.

\section{Super-Kamiokande detector} \label{sec:sk}
\subsection{Detector} \label{sec:detector}
Super-Kamiokande is a water Cherenkov detector in the Gifu prefecture of central Japan. The detector was constructed $1000$~meters~(m) underground, which corresponds to $2700$~m water equivalent~(m.w.e.). It is a cylindrical stainless steel tank structure and contains $50$~kilotons~(ktons) of ultra-pure water. The detector is divided into two regions by an inner structure that optically separates the two with Tyvek sheets. One region is the inner detector~(ID) and the other is the outer detector~(OD). The ID serves as the target volume for neutrino interactions and the OD is used to veto external cosmic-ray muons as well as $\gamma$-rays from the surrounding rock. In the ID, the diameter~(height) of the cylindrical tank is $33.8$~m~($36.2$~m). It contains $32$~ktons of water and holds about eleven thousand inward-facing $20$-inch photomultiplier tubes~(PMTs) to observe the Cherenkov light produced by relativistic particles. The diameter~(height) of the OD is $39.3$~m~($41.4$~m). Further details of the detector can be found elsewhere~\cite{Fukuda:2002uc, Abe:2013gga}.

\subsection{Data acquisition and data set} \label{sec:dataset}

The SK data set is separated into seven distinct periods, from SK-I to SK-VII. The SK-VI and SK-VII phases, starting in July 2020 and June 2022 respectively, are the first and second phases in which gadolinium sulfate was dissolved into the detector. For the prior SK-I to SK-V phases, spanning April 1996 to July 2020, the detector operated with ultra-pure water~\cite{Beacom:2003nk, Super-Kamiokande:2021the}.

From SK-I to SK-III~(until 2008 August), Analog Timing Modules~(ATMs), based on the TKO~(Tristan KEK Online) standards, were used as front-end electronics~\cite{Tanimori:1988qi, Ikeda:1990}. However, some charge could be leaked during charge integration after triggering a cosmic-ray muon, such that the number of hit PMTs and hit times were not always correct for decay electrons, resulting in an inaccurate reconstruction of their energy.

From SK-IV, starting in September 2008, new front-end electronics denoted QBEEs~\cite{Nishino:2009zu} were installed. These are capable of very high-speed signal processing, enabling the integration and recording of charge and time for every PMT signal. Since all PMT signals are digitized and recorded, there is no dead time for the detector. Furthermore, a new online data acquisition~(DAQ) system was implemented that generates multiple software triggers depending on the number of hit PMTs within $200$~ns~\cite{Super-Kamiokande:2010kjr}. For every trigger all PMT signals in a [$-5$, $+35$]~$\mu$s window around the trigger time are recorded. This window is long enough to capture the vast majority of electrons from muon decay; the corresponding decay electron identification procedure is detailed in Sec.~\ref{sec:tag_e}. The analysis presented here used data collected from SK-IV to SK-VI, acquired with the QBEE electronics and associated DAQ system. Table~\ref{tb:phase} summarizes the period of operation, the livetime used for this analysis, and the water status.

\begin{table}[h]
    \begin{center}
    \caption{The summary of data sets used in this analysis. The livetime is the total duration of data samples used for this analysis after removing calibration and bad condition runs. The gadolinium~(Gd) concentration in SK-VI is $0.011\%$~\cite{Super-Kamiokande:2021the}.}
        \label{tb:phase}
            \begin{tabular}{cccc}
                \hline
                \hline
                SK phase & SK-IV & SK-V & SK-VI  \\ \hline
                Period & Sep. 2008  & Jan. 2019  & Aug. 2020 \\
                       & -- May 2018 & -- Jul. 2020 & -- Jun. 2022 \\
                Livetime~[days] & $2970.1$ & $379.2$ & $560.6$  \\
                Water &  \multicolumn{2}{c}{Ultra-pure water} & Gd loaded water \\
            \hline
            \hline
        \end{tabular}
    \end{center}
\end{table}

\subsection{Reconstruction of the muon track} \label{sec:recon_mu}

Muon track reconstruction is performed by a dedicated muon fitter called MUBOY~(detailed in Ref.~\cite{Zoa, Desai}). MUBOY classifies reconstructed muons into four groups depending on the number of muon tracks and the event topology; (I)~Single through-going muons, (II)~Single stopping muons, (III)~Multiple muons, and (IV)~Corner-clipping muons. Here, the reconstruction procedure for stopping muons is briefly described.  

The fitter uses information from ID PMT hits to reconstruct the muon entry point and exit point. MUBOY begins with an initial entry point determined by selecting the earliest hit PMT which has at least three nearest neighbor hits within a $10$~ns window. It determines an initial exit point by selecting the center of the nine PMTs~(one tube and eight surrounding neighbors) that have the maximum total charge. If the muon penetrates the corner of the water tank or stops inside the ID without an exit point, the trial entry and exit points are located close to each other. In such cases, the charge-weighted center of mass of all remaining PMTs is used as the trial exit point. At this stage, the trial direction of the muon track between the entry point and exit point is determined. Here, if the entry point is close to the exit point, the event is recognized as a corner-clipping muon. To finalize the entry point and the direction of the muon track, MUBOY maximizes a likelihood function that depends on the expected Cherenkov light pattern from the muon track, by changing the direction and entry time within the PMT timing resolution. 

To classify the reconstructed muon as a stopping muon, MUBOY evaluates the amount of charge generated by the tail of the muon track. The number of photo-electrons~(p.e.) observed from ID PMTs within $2$~m from the projected exit point is defined as
\begin{equation}
    Q_{\mathrm{ID}}^{\mathrm{Exit}} = \sum_{i}^{N_{\mathrm{ID}}^{\mathrm{2m}}} Q_{i},
\end{equation}
\noindent where $N_{\mathrm{ID}}^{\mathrm{2m}}$ is the number of selected PMTs and $Q_{i}$ is the observed p.e. in each of those PMTs. In the same way, the total number of p.e. observed on OD PMTs within $4$~m from the projected exit point~(the number of selected OD PMTs) is defined as $Q_{\mathrm{OD}}^{\mathrm{Exit}}$~($N^{4\mathrm{m}}_{\mathrm{OD}}$). Table~\ref{tb:stop} summarizes the requirements for tagging an event as a stopping muon. Note that MUBOY may issue a stopping muon flag even when the stopping point is located in the OD region.

\begin{table}[h]
    \begin{center}
    \caption{The summary of the requirement of the total charge of PMTs near the projected exit point to select the stopping muons by MUBOY. The stopping muon is selected by the first criterion. The second and third criteria can select stopping muon near the wall depending on the number of OD hit PMTs.}
        \label{tb:stop}
            \begin{tabular}{ccc}
                \hline
                \hline
                $Q_{\mathrm{ID}}^{\mathrm{Exit}}$ & $Q_{\mathrm{OD}}^{\mathrm{Exit}}$  & $N^{4\mathrm{m}}_{\mathrm{OD}}$\\ \hline
                $Q_{\mathrm{ID}}^{\mathrm{Exit}}<200$~p.e. & -- & --  \\
                $200 \le Q_{\mathrm{ID}}^{\mathrm{Exit}}<300$~p.e. & -- & $N^{4\mathrm{m}}_{\mathrm{OD}}=0$ \\
                $200 \le Q_{\mathrm{ID}}^{\mathrm{Exit}}<400$~p.e. & $<30$~p.e. & $N^{4\mathrm{m}}_{\mathrm{OD}}>0$ \\
                \hline
                \hline
        \end{tabular}
    \end{center}
\end{table}

\begin{figure}[!h]
    \begin{center}
        \includegraphics[width=\linewidth]{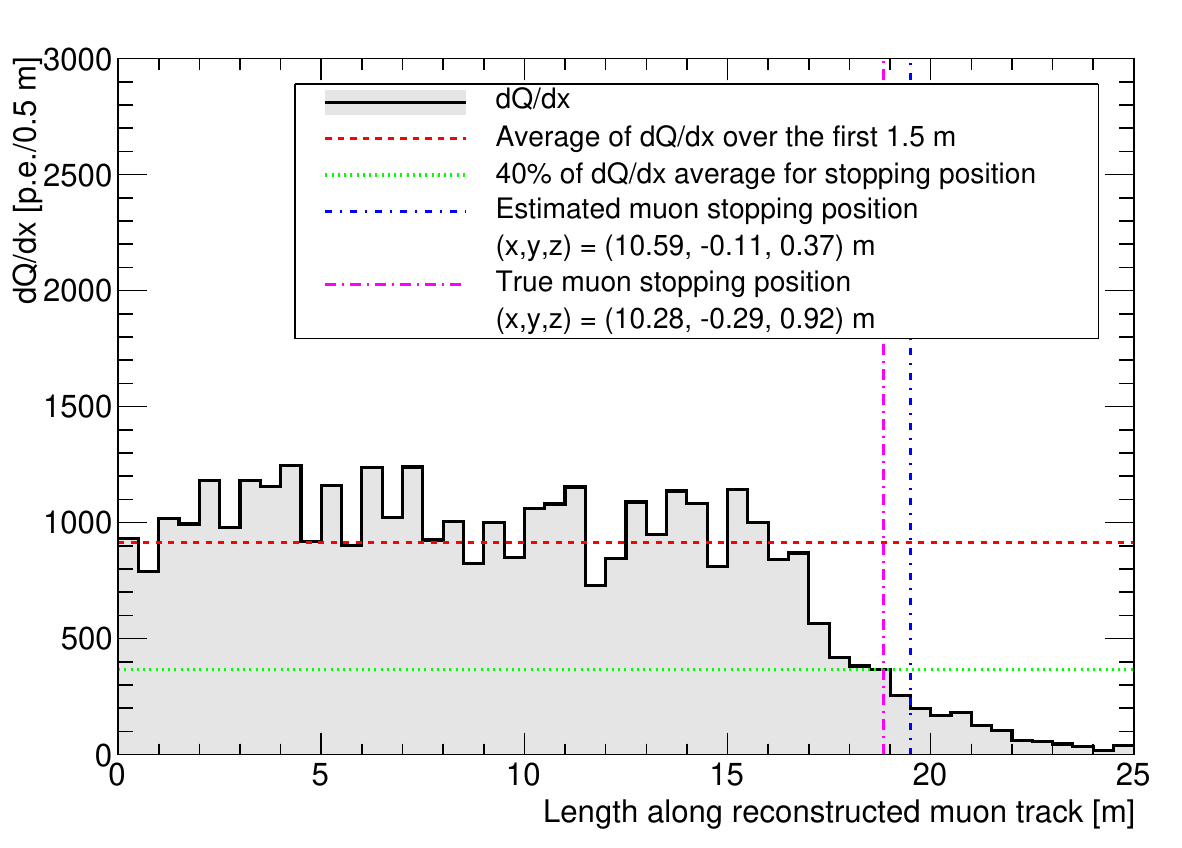}
    \end{center}
\caption{Example of $dQ/dx$ distribution for a stopping muon generated from MC simulation. The gray filled histogram shows the distribution of $dQ/dx$ in units of p.e./$0.5$~m. The horizontal red dashed line shows the average $dQ/dx$ in the first $1.5$~m of the muon track, while the green dotted line shows $40\%$ of the average over this region. The true stopping position in the MC simulation of $(x,y,z)=(10.28,-0.29,0.92)$~m is indicated by the vertical purple line, while the estimated stopping position of $(x,y,z)=(10.59,-0.11,0.37)$~m is indicated by the vertical blue line. \label{fig:dedx}}
\end{figure}

When flagging a stopping muon MUBOY also estimates the stopping point. For that purpose it calculates the track length based on $dQ/dx$, the observed energy loss per unit track length, with a resolution of $0.5$~m. When the muon stops inside the tank, the amount of charge detected from each unit track length drops off at the end of the muon track. Figure~\ref{fig:dedx} shows a typical distribution of $dQ/dx$ and the true vs estimated muon stopping point.

The track length is determined to be within $0.5$~m by finding the point at which $dQ/dx$ falls below $40\%$ of the average value over the first $1.5$~m of the reconstructed track. Then, the stopping position is estimated based on the reconstructed entry position, the reconstructed muon direction, and the track length. Based on the MC simulation described in Sec.~\ref{sec:simulation}, the efficiency for correctly identifying stopping muons in the fiducial volume is $(95.06 \pm 0.01)\%$ and its resolution of estimating the stopping position is $(0.49 \pm 0.04)$~m. Table~\ref{tb:muboy_stop} summarizes the efficiency of stopping muon reconstruction and the resolution of the reconstructed stopping position. 

\begin{table}[h]
    \begin{center}
    \caption{Summary of the reconstruction efficiency for stopping muons in SK-IV using MUBOY. The resolution of the stopping muon position is estimated by calculating the difference between the true stopping position and the estimated stopping position based on the MC simulation.}
        \label{tb:muboy_stop}
            \begin{tabular}{ccc}
            \hline
            \hline
            Event category & Efficiency & Resolution \\
             & [$\%$] & [m]\\ \hline
            Full volume~(OD+ID) & $62.55\pm0.02$ & --\\
            ID only~($32.5$~ktons) & $77.11\pm0.02$ & $0.49 \pm 0.05$ \\
            Fiducial volume~($22.5$~ktons) & $95.06 \pm 0.01$ & $0.49\pm0.04$ \\
            \hline
            \hline
        \end{tabular}
    \end{center}
\end{table}

\subsection{Tagging the decay electron} \label{sec:tag_e}

After an event is identified as a stopping muon with MUBOY, a decay electron signal is searched in a window of $+35$~$\mu$s after the muon trigger time, where the threshold of PMT hits within $200$~ns is $20$. If a decay electron event is found, its position, direction, and energy are reconstructed using the BONSAI fitter~\cite{Smy:2007maa}. This fitter reconstructs the particle's position based on the time residual of PMT hits after subtracting the time of flight, and their direction by maximizing a likelihood function that considers the angle between the direction of the decay electron and the direction of the observed photon, taken from the vertex position, with a correction for PMT acceptance. The energy is determined based on the number of hit PMTs with factors to account for delayed hits due to reflection and scattering in water, dark noise on un-hit PMTs, photo-cathode coverage, PMT gain, and water transparency. Further details on the performance of BONSAI may be found in Ref.~\cite{Hosaka:2005um, Cravens:2008aa, Abe:2010hy, Abe:2016nxk}.

\begin{figure}[!h]
    \begin{tabular}{cc}
        \begin{minipage}{0.5\textwidth}
        \centering
		\includegraphics[width=0.95\textwidth]{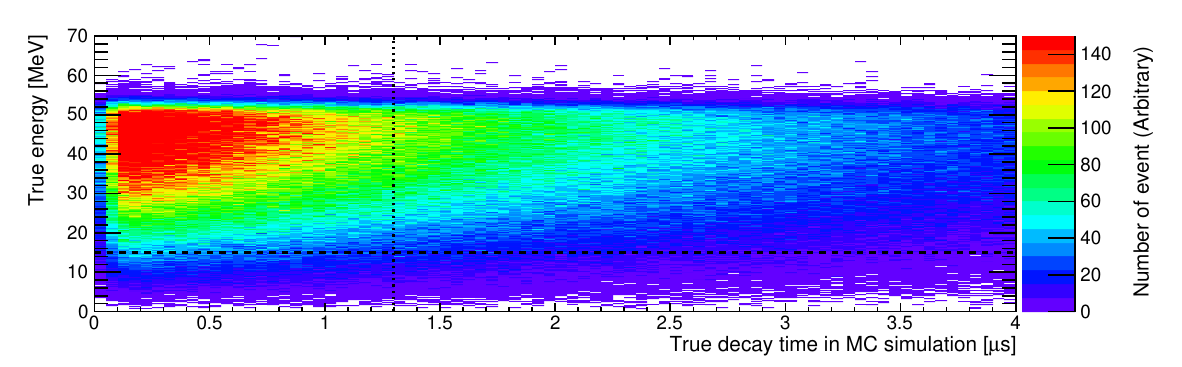}
	\end{minipage} \\
	\begin{minipage}{0.5\textwidth}
		\centering
		\includegraphics[width=0.95\textwidth]{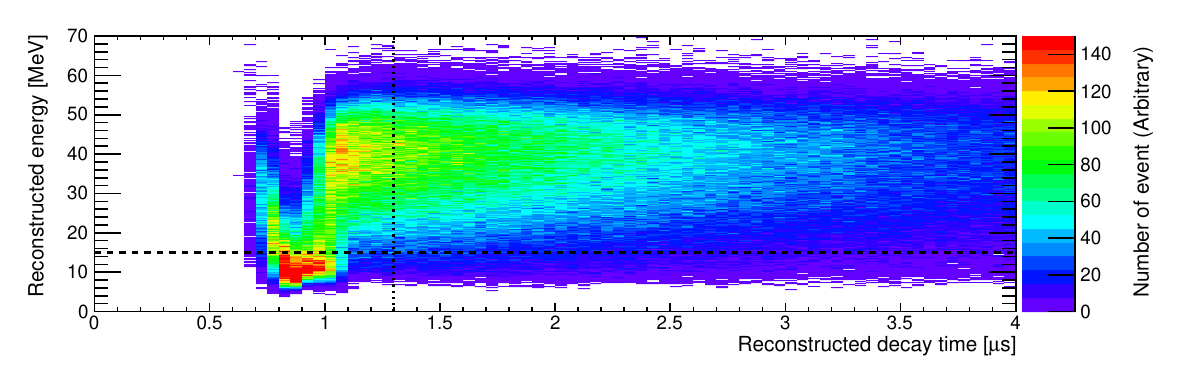}
	\end{minipage}
    \end{tabular}
    \caption{The relationship between decay time and total energy of decay electrons in MC simulation. The top~(bottom) panel shows the true~(reconstructed) energy of the decay electron as a function of the true~(reconstructed) time interval between the parent muon and the decay electron. The vertical~(horizontal) dashed line shows the cut criteria for the timing difference between the parent muon and the decay electron~(the energy cut to reject $\gamma$-rays from excited nitrogen) used in the selection cut described in Sec.~\ref{sec:reduction}. Hits from decay electrons with a short time interval overlap with those from the parent muon, preventing accurate energy reconstruction. A cut of time interval $>1.3~\mu\mathrm{s}$~(vertical dashed line) is used to remove these events from the analysis sample.\label{fig:time-energy}}
\end{figure}

Even though the QBEE front-end electronics can digitize hit timing and charge information with high speed, some decay electrons will nonetheless be reconstructed incorrectly due to the overlap of PMT hits from the cosmic-ray muon with those from the decay electron. Scattering and reflection of photons originating from the parent muon produce a long tail of hits that sometimes exceeds the threshold for tagging a decay electron event~($20$~hits within $200$~ns). This results in a false delayed event before the true decay electron. Such mis-identified events have a shorter time difference between muon and electron (typically less than $1.3~\mu$sec) and tend to under-estimate the decay electron energy. Figure~\ref{fig:time-energy} top~(bottom) shows the relationship between the true~(reconstructed) energy and true~(reconstructed) decay time obtained from MC simulation. To minimize contamination from such mis-identified decay electrons in the analysis sample a cut is applied to remove events whose time difference is shorter than $1.3~\mu$s. The efficiency of stopping muon event selection, including this cut, is described in Sec.~\ref{sec:opt}.

\section{Development of MC simulation} \label{sec:simulation}

To understand the detector response to cosmic-ray muons and their associated decay electrons, a detector simulation based on the \textsc{Geant3} toolkit~\cite{Brun:1994aa} was used for this study \footnote{The Super-Kamiokande detector started its operation on April 1996. Around that time, only \textsc{Geant3} was available. After that, we continued to use \textsc{Geant3}, which has been well optimized by many calibration inputs for more than $25$~years. To keep the consistency with old measurement results, we used \textsc{Geant3} for this study. Note that we have started to develop a new simulation based on \textsc{Geant4} for further updates of the MC simulation used for the SK analysis.}. This simulation has been tuned by comparing calibration data with outputs from the MC simulation. The simulation models particle interactions in the water and electronic systems response.

\subsection{Intensity of incident muons}
The cosmic muon intensity at the underground detector depends on the thickness and density of the surrounding rock~\cite{Bugaev:1998bi, Mei:2005gm}. This implies that the muon energy threshold varies with direction due to the variation in overburden~\cite{Super-Kamiokande:2005plj}. To obtain the directional dependence of the muon intensity at the detector site we used the MUSIC~(MUon SImulation Code) package to simulate muon propagation through the rock surrounding the SK detector~\cite{Antonioli:1997qw, Kudryavtsev:1999zu, Tang:2006uu, Kudryavtsev:2008qh}.

\begin{figure}[!h]
    \begin{center}
        \includegraphics[width=\linewidth]{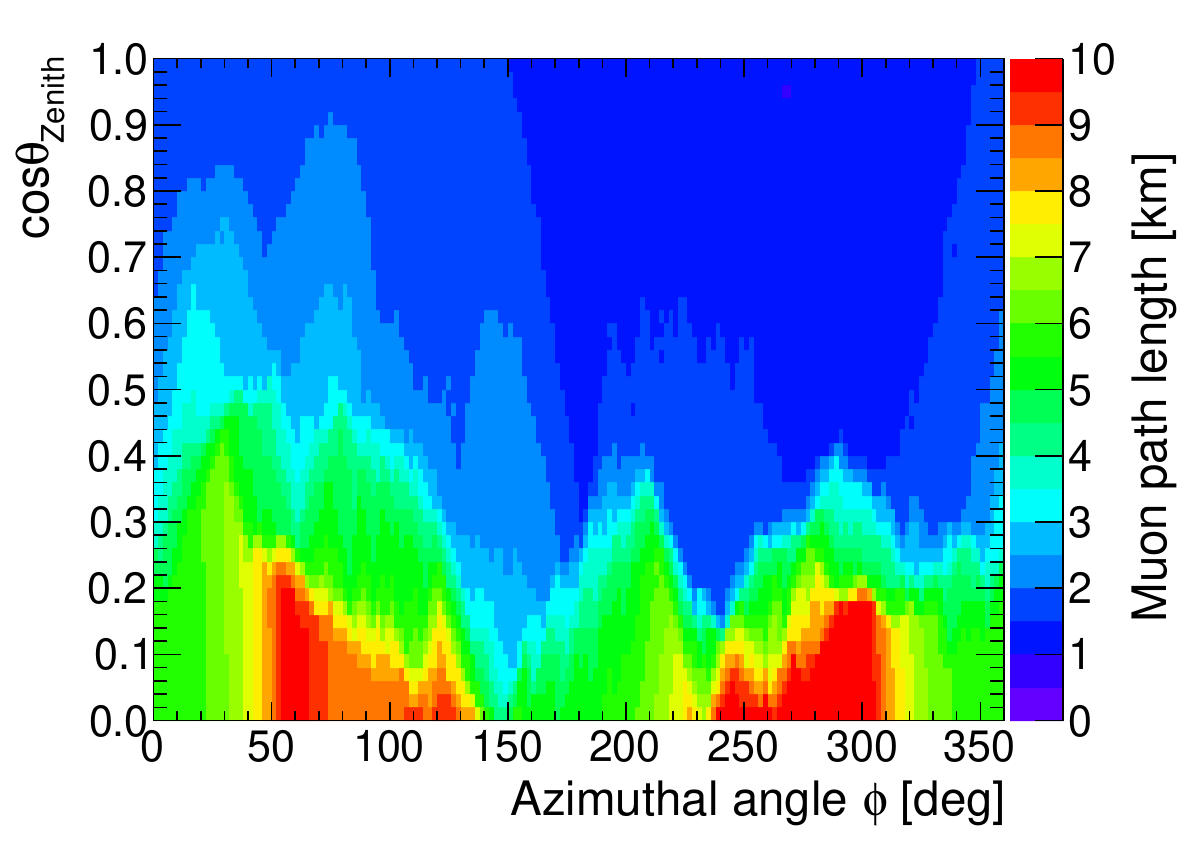}
    \end{center}
\caption{Directional dependence of the muon path length from the surface of the mountain to the SK detector based on the Fundamental Geographic Data published by Geospatial Information Authority of Japan. \label{fig:e-threshold}}
\end{figure}

Figure~\ref{fig:e-threshold} shows the directional dependence of the muon path length from the surface of the mountain to the SK detector site. The minimum path length is $0.85$~km from the direction of $(\phi, \cos \theta_{\mathrm{Zenith}})=(265.0^{\circ}, 0.81)$ near the top of the mountain while the maximum is $11.36$~km from the horizontal direction of $(\phi, \cos \theta_{\mathrm{Zenith}})=(299.0^{\circ}, 0.02)$, where $\phi$ is defined as the azimuthal angle.

\begin{figure}[!h]
    \begin{center}
        \includegraphics[width=\linewidth]{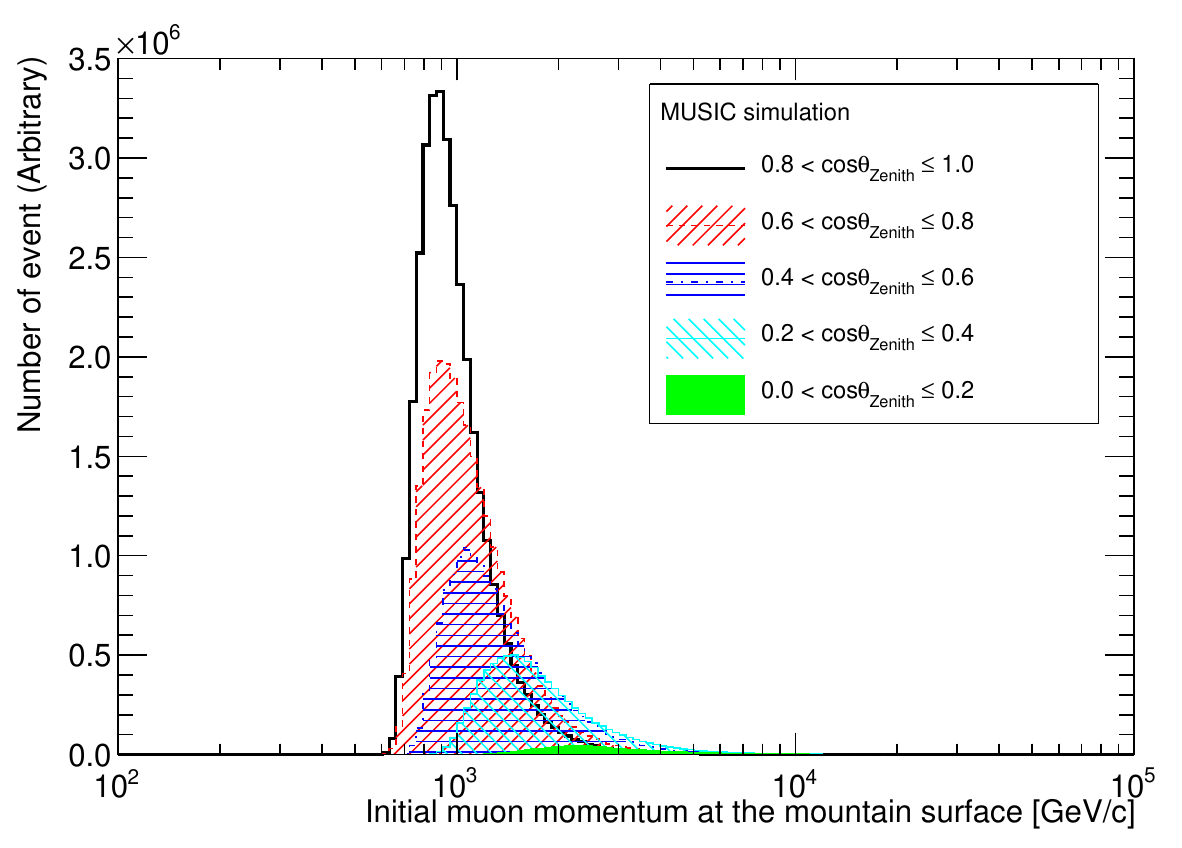}
    \end{center}
\caption{Distributions of cosmic-ray muon initial momenta at the surface of the mountain, obtained from MUSIC simulation~\cite{Antonioli:1997qw, Kudryavtsev:2008qh}, where we selected muons stopping at the SK detector. We show five sample directions depending on the zenith angle as summarized in Table~\ref{tb:ranges}. The different propagation lengths for each trajectory~(shown in Fig.~\ref{fig:e-threshold}) affect the shape of the energy distribution as well as the number of events at the SK detector. \label{fig:mu-initial}}
\end{figure}

Figure~\ref{fig:mu-initial} shows the distribution of initial momenta at the surface of the mountain for cosmic-ray muons that stop at the SK detector, based on the MUSIC simulations~\cite{Antonioli:1997qw, Tang:2006uu}. In the calculation of the muon flux at the surface of the mountain, we used the modified Gaisser parameterization defined in Ref.~\cite{Tang:2006uu} with a muon spectral index of $\gamma=-2.7$. For muons that enter the SK detector horizontally, the momentum is greater than for other directions, and the intensity is low due to the longer muon path length and increased attenuation, as expected.

Table~\ref{tb:ranges} summarizes the ranges of muon momentum and $E_{\mu}\cos \theta_{\mathrm{Zenith}}$ for stopping muons at the surface of the mountain, coming from different incident directions, as simulated by MUSIC. By sampling muons with different $E_{\mu}\cos \theta_{\mathrm{Zenith}}$, we test the directional dependence of charge ratio and polarization. However, the number of muons from a low incident angle is limited due to their long propagation length in the mountain, resulting in large uncertainties on their range of muon momenta and $E_{\mu} \cos \theta_{\mathrm{Zenith}}$, as indicated in Table~\ref{tb:ranges}. In the analysis, we removed muons whose $\cos \theta_{\mathrm{Zenith}}$ is less than $0.2$.

\begin{table}[!h]
    \begin{center}
    \caption{The summary of ranges of muon momentum and $E_{\mu}\cos \theta_{\mathrm{Zenith}}$ of stopping muons at the surface of the mountain in different directions around the SK detector simulated by MUSIC. Due to the long propagation length in the rock of the mountain, there are few muons and large statistical uncertainties in the region of $\cos \theta_{\mathrm{Zenith}}<0.2$. The symbol $\dagger$ indicates that the sample does not include muons from the horizontal direction of $\cos \theta_{\mathrm{Zenith}} \le 0.2$.}
        \label{tb:ranges}
            \begin{tabular}{lcc}
                \hline \hline
                Direction & Momentum & $E_{\mu}\cos \theta_{\mathrm{Zenith}}$ \\
                & [$\mathrm{TeV}/c$] & [$\mathrm{TeV}$] \\ \hline
                $0.0<\cos \theta_{\mathrm{Zenith}} \le 0.2$ & $2.1^{+1.5}_{-0.4}$ & $0.3^{+0.2}_{-0.1}$ \\
                $0.2<\cos \theta_{\mathrm{Zenith}} \le 0.4$ & $1.4^{+0.8}_{-0.3}$ & $0.4^{+0.2}_{-0.1}$ \\
                $0.4<\cos \theta_{\mathrm{Zenith}} \le 0.6$ & $1.0^{+0.6}_{-0.1}$ & $0.5^{+0.3}_{-0.1}$ \\
                $0.6<\cos \theta_{\mathrm{Zenith}} \le 0.8$ & $0.8^{+0.4}_{-0.1}$ & $0.6^{+0.3}_{-0.1}$ \\
                $0.8<\cos \theta_{\mathrm{Zenith}} \le 1.0$ & $0.8^{+0.3}_{-0.1}$ & $0.8^{+0.2}_{-0.1}$ \\ \hline 
                North$^{\dagger}$~($45^{\circ}<\phi \le 135^{\circ}$) & $1.0^{+0.7}_{-0.1}$ & $0.9^{+0.3}_{-0.1}$ \\
                West~$^{\dagger}$($135^{\circ}<\phi \le 225^{\circ}$)  & $0.8^{+0.5}_{-0.1}$ & $0.7^{+0.2}_{-0.1}$ \\
                South$^{\dagger}$~($225^{\circ}<\phi \le 315^{\circ}$)  & $0.8^{+0.4}_{-0.1}$ & $0.6^{+0.2}_{-0.1}$ \\
                East$^{\dagger}$~($\phi \le 45^{\circ}$ or $\phi>315^{\circ}$)  & $0.9^{+0.5}_{-0.1}$ & $0.8^{+0.3}_{-0.1}$ \\ \hline
                All direction$^{\dagger}$ & $0.9^{+0.7}_{-0.1}$ & $0.7^{+0.4}_{-0.2}$ \\
            \hline
            \hline
        \end{tabular}
    \end{center}
\end{table}

\begin{figure}[!h]
    \begin{center}
        \includegraphics[width=\linewidth]{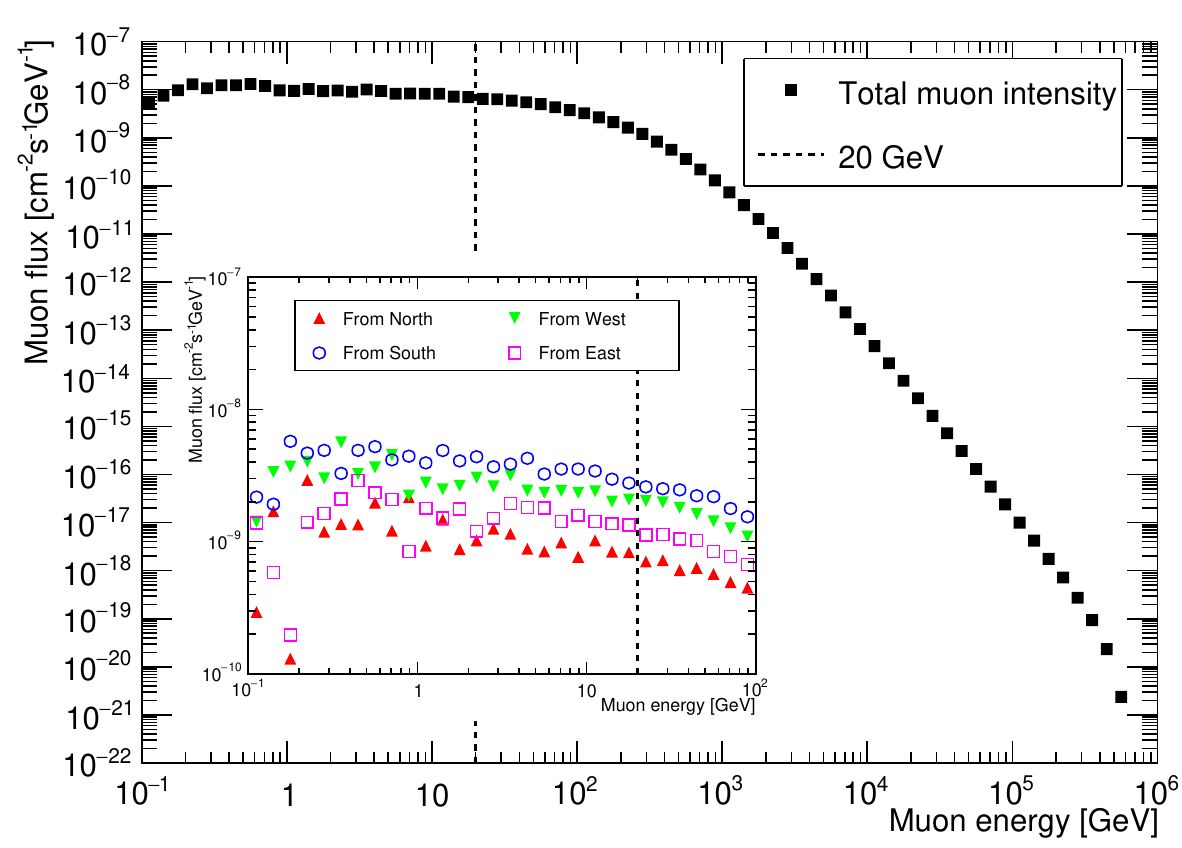}
    \end{center}
\caption{The total muon flux as a function of muon energy at the SK detector, from simulations with the MUSIC package~\cite{Antonioli:1997qw, Kudryavtsev:2008qh}. Black squares show the predicted muon flux; the vertical dashed line indicates an energy of $20$~GeV, a threshold that encapsulates the majority of stopping muons used in this analysis.
The inset panel shows the directional dependence of the muon flux in the energy range of $0.1$~GeV to $100$~GeV. The different colored points illustrate the muon flux from various directions. \label{fig:mu-energy}}
\end{figure}

\begin{figure}[!h]
    \begin{center}
        \includegraphics[width=\linewidth]{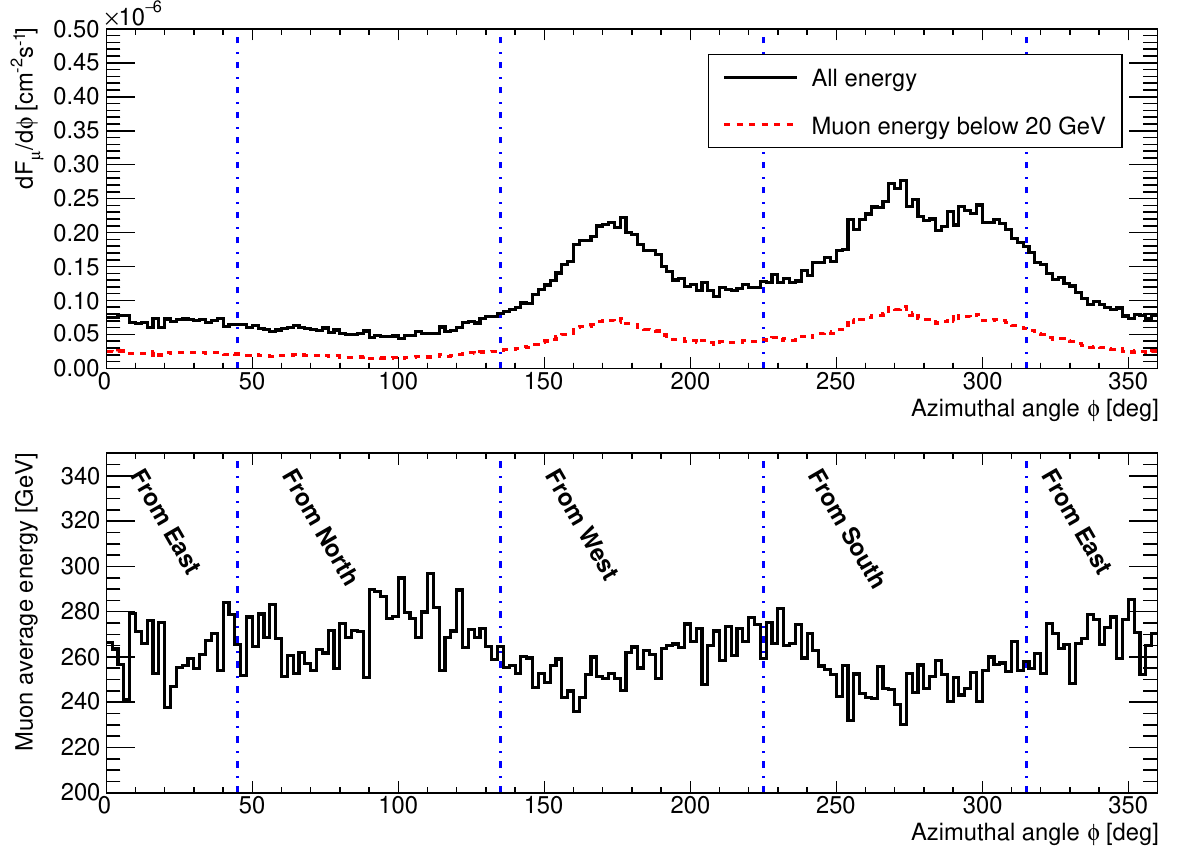}
    \end{center}
\caption{Muon flux and average muon energy as a function of azimuthal angle, where the azimuthal angle is the angle about the central axis of the SK detector in cylindrical coordinates.
The black solid~(red dotted) histogram in the top panel shows the directional dependence of the muon flux~(flux below $20$~GeV); the histogram in the bottom panel shows the directional dependence of average muon energy at the detector site. This azimuthal dependence arises from the difference in the overburden of the mountain. The average energy of cosmic-ray muons that reach the SK detector is approximately $271$~GeV. \label{fig:music-dist}}
\end{figure}

Figure~\ref{fig:mu-energy} shows the energy dependence of total muon flux at the detector site. The integrated muon flux is estimated to be $1.48\times 10^{-7}~\mathrm{cm^{-2}s^{-1}}$, corresponding to an event rate of $2.4$~Hz. Figure~\ref{fig:music-dist} shows the azimuthal angular dependence of the muon flux and the mean muon energy at the SK site, as simulated by MUSIC. The average energy of cosmic-ray muons that reach the detector is about $271$~GeV, of which the majority fully penetrate the SK detector.

For the simulation of muon decay events inside the SK detector we used the initial azimuthal angular distribution shown in Fig.~\ref{fig:music-dist}, but with a truncated energy distribution spanning $0.1$~MeV to $20$~GeV to avoid generating many muons that penetrate through the SK detector without producing a decay electron event.

\subsection{Muon decay with polarization} \label{sec:dcy_po}

Cosmic-ray muons are mainly produced via the two-body decay of charged pions and kaons. The kinematics of these decays result in muons that are polarized in the rest frame of the parent meson. The direction of the muon spin is either parallel~(for negative muons) or anti-parallel~(for positive muons) to its direction of propagation. The different degrees of contribution from charged kaons to positive and negative muon production, together with the greater polarization of muons from parent kaons, means that the level of polarization of positive muons is expected to be higher than that of negative muons~\cite{Lipari:1993hd}, as shown in Fig.~\ref{fig:mu_pol_expect}. The implementation of this muon polarization in the simulation is detailed in Appendix~\ref{sec:mc-pol}.

\begin{figure}[!h]
    \begin{center}
        \includegraphics[width=\linewidth]{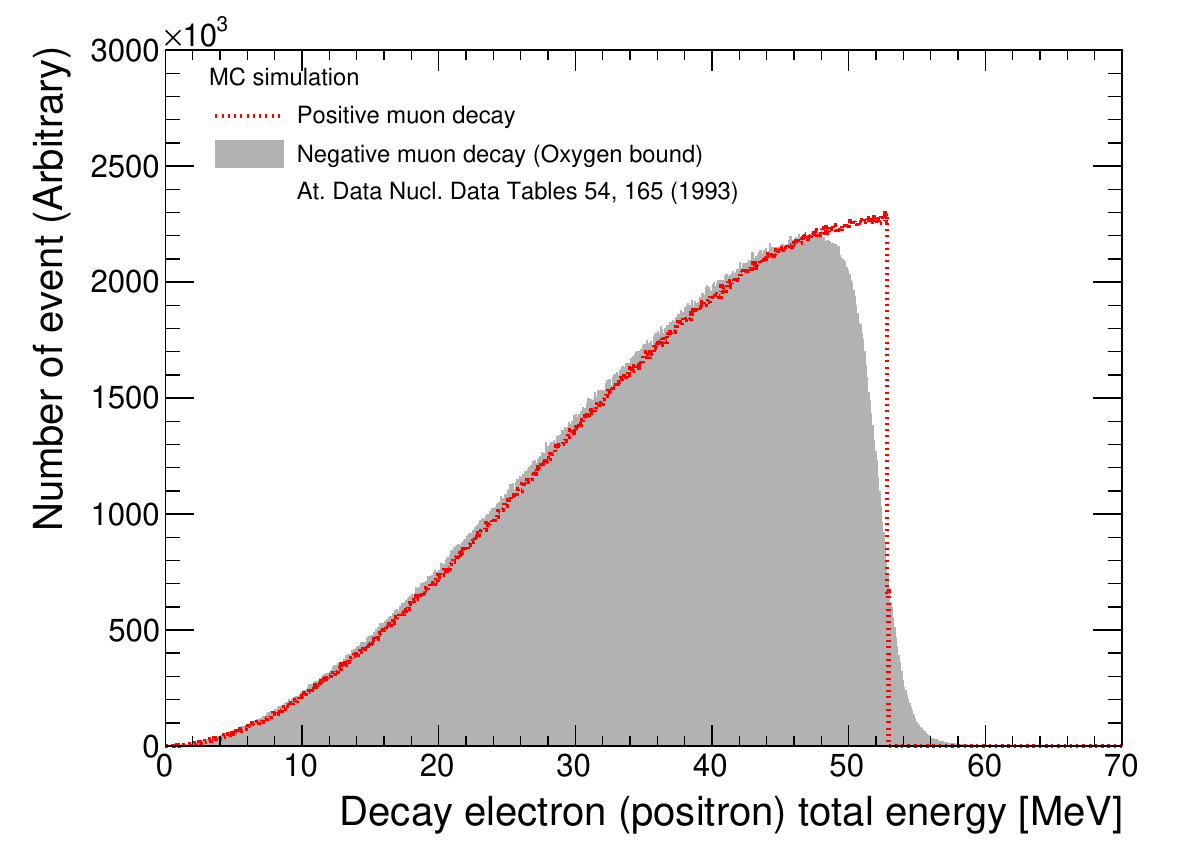}
    \end{center}
\caption{The energy distribution of decay electrons, used as input for the MC simulation. The red dashed line shows the energy distribution for free decay of muons, based on the Michel parameters~\cite{Michel:1949qe}.
The gray filled histogram shows the energy distribution of electrons emitted in the decay of negative muons atomically bound with oxygen, according to Ref.~\cite{Watanabe:1993emp}. \label{fig:decaye-energy-dist}}
\end{figure}

Muon decay is a purely leptonic process mediated by the charged current weak interaction and is generally characterized by the Michel parameters~\cite{Michel:1949qe, Bouchiat:1957zz, Kinoshita:1957zz, Kinoshita:1957zza}. In the case of free muon decay the direction of the emitted electron is highly correlated with the spin of the parent muon, due to the maximally violating nature of parity in the weak interaction.
The expected decay rate~($\Gamma$) is described as

\begin{equation}
    \frac{d^{2} \Gamma}{dx \, d \cos \theta} \sim  N(x)[1 + P^{\mu} \beta(x) \cos \theta], \label{eq:free-rate} 
\end{equation}

\noindent where $x=2 E_{e}/m_{\mu}$ is the reduced energy of the emitted electron~($E_{e}$ is the total energy of the electron and $m_{\mu}$ is the mass of muon), $\theta$ is the angle between the spin-direction of the parent muon and that of the emitted electron momentum, $N(x)$ is the expected energy spectrum, $P^{\mu}$ is the polarization of the parent muon, and $\beta(x)$ is the degree of correlation between the electron momentum and the muon spin direction. In the case of free muon decay the parameters $N(x)$ and $\beta(x)$ are simply described as $N(x)=x^{2}(3-2x)$ and $\beta(x) = - (1-2x)/(3-2x)$ for positive muons, with the sign inverted in the case of negative muons. Figure~\ref{fig:decaye-energy-dist} shows the expected energy spectrum of emitted electrons, obtained by integrating Eq.~(\ref{eq:free-rate}) over $\cos \theta$.

Negative muons traversing matter may be captured by the Coulomb potential of atoms in the material, subsequently decaying in orbit. When a muonic hydrogen atom is formed it freely diffuses in water because its charge is strongly shielded by the compact muon orbit~\cite{Fermi:1947uv, Frank:1947}. When the muonic hydrogen atom approaches an oxygen it transfers its muon to form a new muonic oxygen atom due to the stronger binding energy. This transfer process occurs on a much shorter timescale than that of muon decay, so all muon decays in orbit occur within the orbit of an oxygen atom. In the decay in orbit, both the nuclear charge distribution~(Coulomb potential) and the finite size of the nucleus affect the direction and energy of the emitted electrons~\cite{Gilinsky1960}. Furthermore, nuclear recoil alters the kinematics such that electrons from muon decays in orbit have a higher energy than those from muon decay in vacuum~\cite{Haenggi:1974hp, Czarnecki:2011mx}, as shown in Fig.~\ref{fig:decaye-energy-dist}. To include the effects of oxygen capture in the MC simulation the parameters $N(x)$ and $\beta(x)$ are modified based on the studies in Refs.~\cite{Watanabe:1987} and \cite{Watanabe:1993emp}. 

\begin{figure}[!h]
    \begin{center}
        \includegraphics[width=\linewidth]{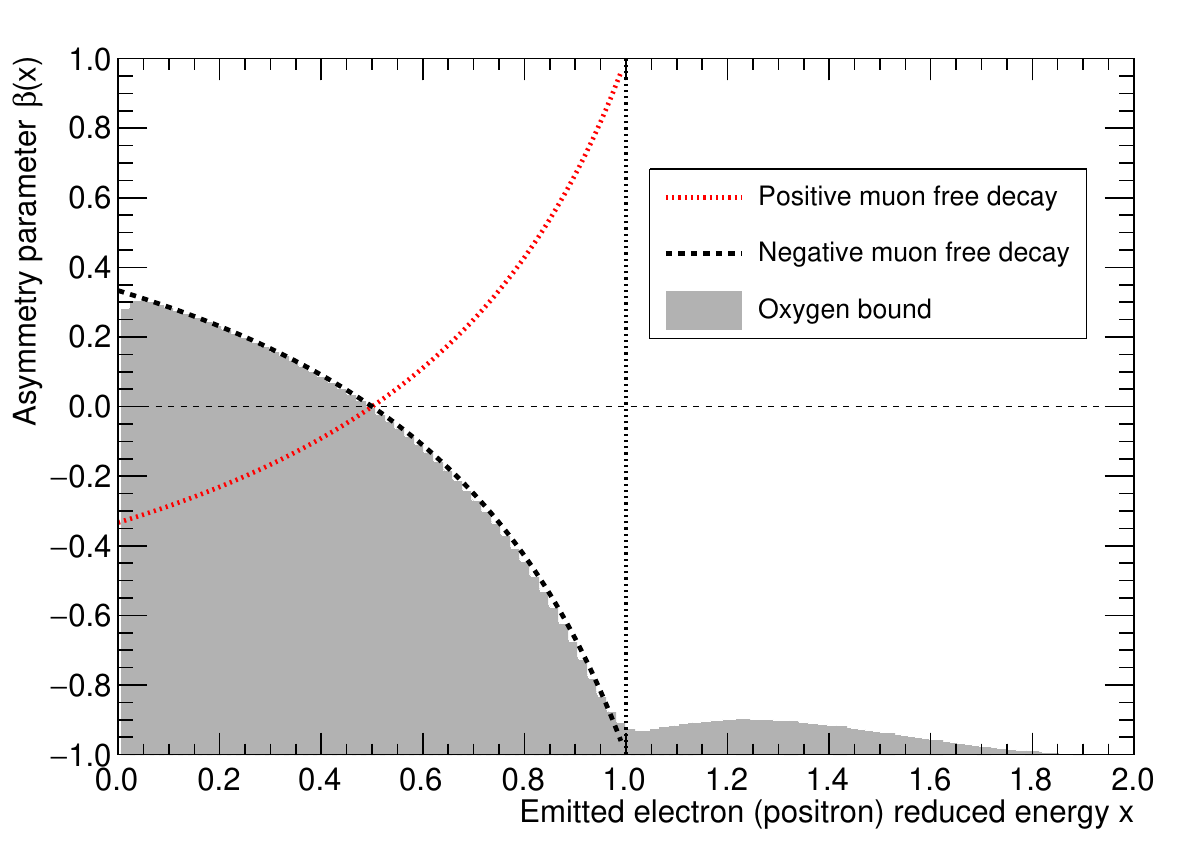}
    \end{center}
\caption{The degree of correlation between emitted electron momentum and parent muon spin direction~($\beta(x)$ in Eq.~(\ref{eq:free-rate})) for both free and bound muon decays, as a function of the reduced energy $x$~\cite{Watanabe:1993emp}. The red dotted line,  black dashed line, and gray filled histograms show the $\beta(x)$ parameter for free positive muon decays,  free negative muon decays, and decays of negative muons bound with oxygen, respectively. The maximum energy of an electron from free muon decay corresponds to $x=1.0~(E_{e}=52.8~\mathrm{MeV})$. For bound muon decays a reduced energy $x$ exceeding $x=1.0$ is permitted because of the recoiling nucleus. \label{fig:asym}}
\end{figure}

Figure~\ref{fig:asym} shows the impact of the modified asymmetry parameter~$\beta(x)$ on the decay electron energy; the energy distribution is distorted, most notably with an additional component from bound decays at $x>1.0$.

\begin{figure}[!h]
    \begin{center}
        \includegraphics[width=\linewidth]{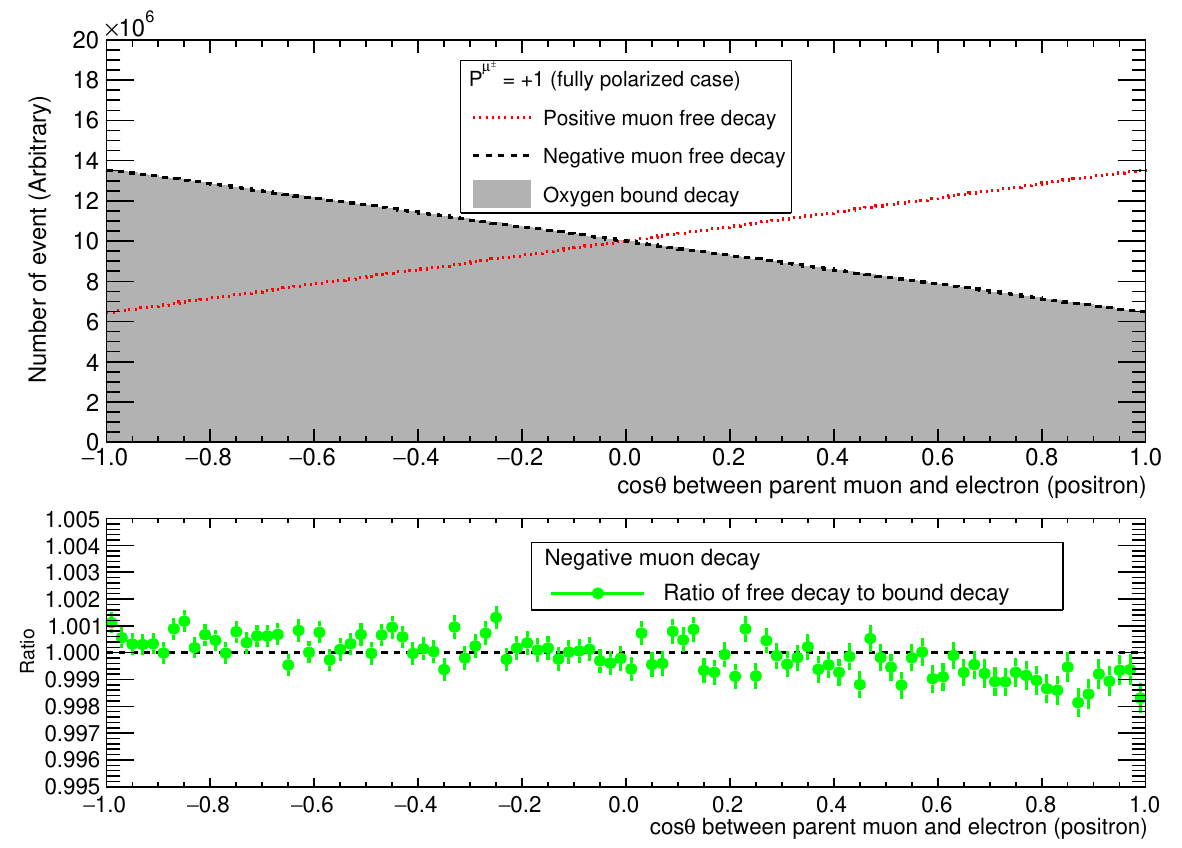}
    \end{center}
\caption{Distributions of $\cos \theta$ assuming fully polarized muon decays~($P^{\mu^{\pm}}_{0}=+1$), where $\theta$ describes the opening angle between the emitted electron direction and the direction of the parent muon spin. Because of their helicity, the slopes of the two distributions are inverted when the magnitude of polarization is the same. Top: The red dotted, black dotted, and gray filled histograms show $\cos \theta$ for electrons with energy greater than $15$~MeV emitted from positive muon free decay, negative muon free decay, and bound muon decay, respectively. Bottom: The ratio of negative muon free decay to bound decay as a function of $\cos \theta$; the distribution is slightly distorted by the $\beta(x)$ contribution from electrons with high energy~($x>1.0$) seen in Fig.~\ref{fig:asym}.\label{fig:fully-pol}}
\end{figure}

Figure~\ref{fig:fully-pol} shows the $\cos \theta$ distribution assuming fully polarized muon decays, again accounting for the distortion from bound muon decays. Because of the~$\beta(x)$ parameter decay electrons at lower energies are more likely to be emitted in the direction of the muon polarization, while those at higher energies are more likely to be emitted in the direction opposite to the muon polarization. The difference between free decays and bound decays is shown at the bottom of Fig.~\ref{fig:fully-pol}. 
The impact of the altered~$\beta(x)$ parameter from muon decay-in-orbit on both decay electron energy and angular distributions was included in the MC simulation, although the effect of the latter on the simulation results is only on the order of $\pm 0.2\%$.

\subsection{Muon de-polarization} \label{sec:de-pol}

While the polarization of cosmic ray muons at generation provides a handle on atmospheric $\nu_{e}$ production above $1$~TeV, cosmic-ray muons may lose their original polarization before stopping, through processes both in the rock overburden and in the detector itself~\cite{Percival:1976}. These de-polarization mechanisms affect positive and negative muons differently, so each must be considered separately.

\subsubsection{De-polarization during propagation}

The first stage of de-polarization originates from multiple Coulomb scattering (MCS) between the propagating muon and nuclei and electrons in matter. Table~\ref{tb:delta} summarizes the probability of de-polarization in the atmosphere and surrounding rock, estimated based on Ref.~\cite{Hayakawa:1957}. Here, we define the ratio of de-polarization during muon propagation as $\delta$. Owing to the lower density the amount of de-polarization in the atmosphere is negligible compared to that in rock, while de-polarization from MCS in the water is negligible due to the relatively short propagation distance inside the SK detector~\cite{Fermi:1947uv}. As the muon nears stopping other de-polarization processes, such as spin-flip~\cite{Demeur:1956ih}, spin precession~\cite{Akylas:1977}, and Auger effect~\cite{Ferrell:1960zz}, can occur, but the impact of these processes is similarly negligible.

\begin{table}[!h]
    \begin{center}
    \caption{Probability of muon de-polarization during propagation, based on Ref.~\cite{Hayakawa:1957}. These de-polarization effects during propagation affect both positive and negative muons equally. This parameter is defined as $\delta$.}
        \label{tb:delta}
            \begin{tabular}{ccc}
                \hline
                \hline
                Medium & Probability of losing \\
                          & polarization to decay~[$\%$]  \\ \hline
                Atmosphere & $1.8\times10^{-6}$ \\
                Rock & $0.3$ \\
                \hline
            \hline
        \end{tabular}
    \end{center}
\end{table}

\subsubsection{De-polarization of positive muons}
The depolarization of stopping positive muons in water has been previously studied experimentally. In Refs.~\cite{Swanson:1958zz, Walker:1978, Percival:1978}, a beam of polarized positive muons was used to evaluate the degree of residual polarization in muons that are captured in water. The experiment measured the asymmetry in the angular distribution of the emitted positron following positive muon decay.

The study found that $62\%$ of positive muons either decayed without capturing or after binding with a diamagnetic molecule, in both cases retaining polarization until decay. A further $20\%$ of muons were found to bind with electrons to form muonium. When forming muonium two spin-states~(a singlet or triplet) may be formed with equal probability~(since electrons in water are largely un-polarized). Those muons that form a singlet efficiently lose their original polarization in the process~\cite{Friedman:1957mz}, while those that form triplets retain their polarization in formation, but may lose it in subsequent chemical reactions of the muonium with the surrounding medium. The remaining $18\%$ of muons were observed to lose their polarization, but the mechanism of polarization loss was not determined~\cite{Percival:1978}. In total, the fraction of residual polarization of captured muons denoted $r^{+}_{\mathrm{H_{2}O}}$, was estimated to be $(71.8\pm0.7)\%$. Table~\ref{tb:de-pol-posi} summarizes the fraction of residual polarization in captured positive muons, measured by three different studies. In this study, we use the weighted mean of these measurements as $r^{+}_{\mathrm{H_{2}O}}$.

\begin{table}[!h]
    \begin{center}
    \caption{Summary of the parameter, $r^{+}_{\mathrm{H_{2}O}}$, describing the fraction of original polarization retained by polarized positive muons that stop in water. The combined value is calculated by taking the weighted mean of the three results.}
        \label{tb:de-pol-posi}
            \begin{tabular}{cc}
                \hline
                \hline
                Reference & Probability of retaining \\
                          & polarization to decay~[$\%$]  \\ \hline
                Ref.~\cite{Swanson:1958zz} & $68.2 \pm 5.3$ \\
                Ref.~\cite{Walker:1978} & $70.6 \pm 2.3$ \\
                Ref.~\cite{Percival:1978} & $72.0 \pm 0.6$ \\
                \hline
                Combined & $71.8 \pm 0.7$ \\ 
                \hline
            \hline
        \end{tabular}
    \end{center}
\end{table}

\subsubsection{De-polarization of negative muons} \label{sec:residual_negative}

Negative muons stopping in matter are initially captured by the Coulomb field of a nucleus into a highly excited bound state characterized by large orbital angular momentum. The resulting muonic atom then transitions through less excited states until it eventually reaches the ground state~\cite{Burbidge:1953, Eisenberg:1961, Vogel:1975dg}. In this de-excitation cascade the Auger process typically dominates at the beginning while radiative decay dominates in lower-energy orbitals. As the muonic atom transitions through several intermediate states the muon loses the majority of its original polarization~\cite{Mann:1961zz}. For nuclei with non-zero spin further de-polarization may occur following de-excitation through interactions between the magnetic moment of the muon and that of the nucleus, producing hyperfine level splitting in the energy levels of the muonic atom~\cite{Uberall:1959}. Since oxygen has zero spin, such de-polarization does not occur in water. The residual polarization upon reaching the K-orbit of the muonic atom is theoretically expected to be $1/6$ of the original polarization~\cite{Dzhrbashyan:1959, Shmushkevich:1959}. After the muon reaches the K-orbit of oxygen, a muonic atom with the electron shell of atomic nitrogen is produced. Hereafter, it is referred to as muonic nitrogen. This muonic nitrogen acquires electrons through collisions with the surrounding medium and compensates for the paramagnetism in its electron shells until the muon decays. This results in further depolarization through interactions between the muon and the magnetic moment of the electron shell. Table~\ref{tb:de-pol-nega} summarizes the resulting fraction of residual polarization for negative muons captured in water, denoted as $r^{-}_{\mathrm{H_{2}O}}$, as measured by four different studies. The earliest measurement in Ref.~\cite{Ignatenko:1961} observed a large fraction of residual polarization, while the other measurements cluster around a $5\%$ level. In this study, we used only the latter three measurements~\cite{Buckel:1968, Dzhuraev:1972, Dzhuraev:1974} and the combined value listed in Table~\ref{tb:de-pol-nega} is used for the analysis. We should note that additional captures by gadolinium and sulfur are expected after loading $\mathrm{Gd_{2}(SO_{4})_{3}}$ as mentioned in Sec.~\ref{sec:sk} while such captures can be ignored in this study. The detail is discussed later in Sec.~\ref{sec:gd}.

\begin{table}[!h]
    \begin{center}
    \caption{Summary of measurements of the parameter $r^{-}_{\mathrm{H_{2}O}}$, the fraction of original polarization retained by negative muons captured in water. The combined value is calculated by taking the weighted mean of the latter three measured results. The earliest study in Ref.~\cite{Ignatenko:1961} is not used in calculating the combined value.}
        \label{tb:de-pol-nega}
            \begin{tabular}{cc}
                \hline
                \hline
                Reference & Probability of retaining \\
                          & polarization to decay~[$\%$]  \\ \hline
                Ref.~\cite{Ignatenko:1961} & $12.9 \pm 1.5$ \\
                \hline
                Ref.~\cite{Buckel:1968} & $\phantom{0}5.1 \pm 1.5$ \\
                Ref.~\cite{Dzhuraev:1972} & $\phantom{0}4.8 \pm 0.6$ \\
                Ref.~\cite{Dzhuraev:1974} & $\phantom{0}5.3 \pm 0.4$ \\
                \hline
                Combined & $\phantom{0}5.1\pm 0.4$ \\
                \hline
                \hline
        \end{tabular}
    \end{center}
\end{table}

\subsection{Angular distribution in the detector} \label{sec:pol-detector}

The polarization observed in the detector, defined as $P_{\mathrm{obs}}$, needs to be corrected to account for these depolarization mechanisms to recover the polarization at production, defined as $P_{0}^{\mu^{\pm}}$. The angular distribution in water, which is defined as $I(\theta)$, can be represented as;

\begin{eqnarray*}
    I(\theta)  = && N^{\mu^{+}} I^{\mu^{+}}(\theta) + N^{\mu^{-}} I^{\mu^{-}}(\theta) \\
      = && N^{\mu^{+}}+N^{\mu^{-}} + \left(N^{\mu^{+}}P_{\mathrm{H_{2}O}}^{+}+N^{\mu^{-}}P_{\mathrm{H_{2}O}}^{-} \right)\cos \theta 
\end{eqnarray*}

\noindent where $N^{\mu^{}\pm}$ is the number of cosmic-ray muons, $I^{\mu^{\pm}}(\theta)$ is the angular distribution, which is the integration of Eq.~(\ref{eq:free-rate}) by $x$, and $P_{\mathrm{H_{2}O}}^{\pm}$ is the residual polarization in water. Hence, the observed polarization can be represented as,
\begin{eqnarray}
      P_{\mathrm{obs}} = && \frac{N^{\mu^{+}}P_{\mathrm{H_{2}O}}^{+}+N^{\mu^{-}}P_{\mathrm{H_{2}O}}^{-}}{N^{\mu^{+}}+N^{\mu^{-}}}, \nonumber \\
      = && \frac{R \, P_{\mathrm{H_{2}O}}^{+} + P_{\mathrm{H_{2}O}}^{-}}{1.0+R}, \nonumber
\end{eqnarray}
where the parameter $R=N^{\mu^{+}}/N^{\mu^{-}}$ is the charge ratio of cosmic-ray muons. After including de-polarization effects the parameters $P_{\mathrm{H_{2}O}}^{\pm}$ can be written as $P_{\mathrm{H_{2}O}}^{\pm} = (1.0 - \delta) \, r_{\mathrm{H_{2}O}}^{\pm} \, P^{\mu^{\pm}}_{0}$, where parameters $\delta$ and $ r_{\mathrm{H_{2}O}}^{\pm}$ are the ratio of de-polarization during the propagation listed in Table~\ref{tb:delta}, and those in water listed in Table~\ref{tb:de-pol-posi} and Table~\ref{tb:de-pol-nega}. Incorporating these into the equation above, the observed polarization in the SK detector can now be expressed as,

\begin{equation}
    P_{\mathrm{obs}} = (1.0 - \delta) \, \frac{R \, r_{\mathrm{H_{2}O}}^{+} \, P^{\mu^{+}}_{0}+ r_{\mathrm{H_{2}O}}^{-} \, P^{\mu^{-}}_{0}}{1.0+R}. \label{eq:p_obs}
\end{equation}

Although the absolute magnitude of polarization of positive muons is different from that of negative muons in the high momentum region above $1~\mathrm{TeV}/c$, as shown in Fig.~\ref{fig:mu_pol_expect}, the SK detector does not have the sensitivity to determine the two polarizations separately because of the small fraction of residual polarization of negative muons~($r_{\mathrm{H_{2}O}}^{-}$) as mentioned in Sec.~\ref{sec:residual_negative}. For this reason, we assumed in this analysis that the polarizations of positive and negative muons are equal while the sign is inverted. Hence, Eq.~(\ref{eq:p_obs}) can be expressed as

\begin{equation}
    P_{\mathrm{obs}} = (1.0 - \delta) \, \frac{P^{\mu}_{0} \, (R \, r_{\mathrm{H_{2}O}}^{+} + r_{\mathrm{H_{2}O}}^{-})}{1.0+R}, \label{eq:p-obs-equal}
\end{equation}

\noindent where $P^{\mu} = -P^{\mu^{+}} = P^{\mu^{-}}$.

In the analysis of data the number of tagged decay electrons, which is defined as $N^{e^{\pm}}$, is used instead of the number of muons~($N^{\mu^{\pm}}$) to calculate the opening angle. In the case of negative muons, nuclear capture by oxygen is expected through the interaction $\mu^{-}+p\to n + \nu_{\mu}$~\cite{Kaplan:1969bd}. 
The fraction of negative muons that undergo nuclear capture in water was experimentally measured as $\Lambda_{c}=0.184 \pm 0.001$~\cite{Suzuki:1987jf, Guichon:1979ga} and the charge ratio~$R$ can be expressed as,

\begin{equation}
R = \frac{N^{\mu^{+}}}{N^{\mu^{-}}} = \frac{N^{e^{+}}}{N^{^{e^{-}}}/\left(1-\Lambda_{c} \right)}. \label{eq:ratio}
\end{equation}

\begin{figure}[!h]
    \begin{center}
        \includegraphics[width=\linewidth]{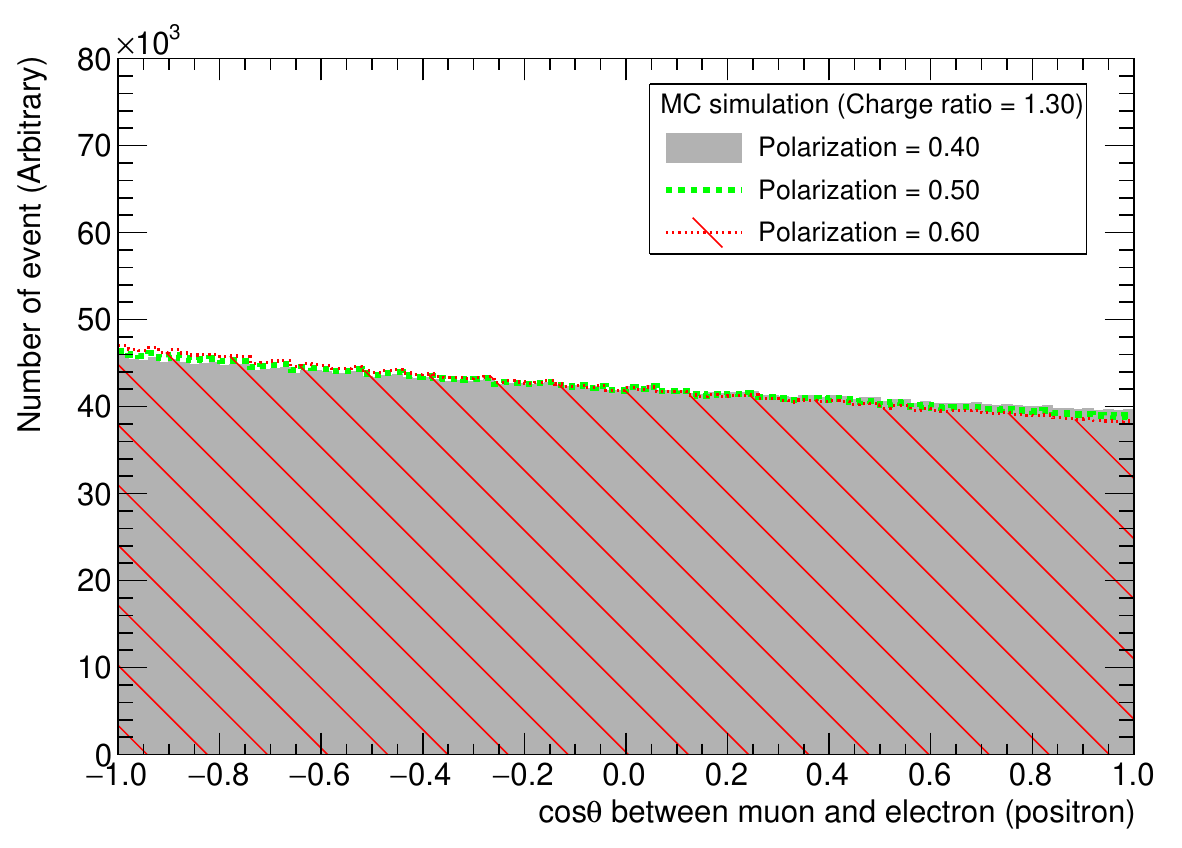}
    \end{center}
\caption{Simulated distributions of $\cos \theta$ for three different magnitudes of muon polarization in the SK detector~(ultra-pure water case). In these simulations, we assumed that the muon charge ratio is $R=1.30$, the reduced energy threshold is $x=15.5/52.8$, and the de-polarization mechanisms described in the previous subsections are considered. The filled gray histogram, green dashed line, and crosshatch red histogram show the expected $\cos \theta$ distributions for the parameters of $(R, P^{\mu}_{0})=(1.30,0.40)$, $(1.30,0.50)$, and $(1.30,0.60)$, respectively. The differences among them are small but visible when analyzing a decay electron sample with sufficiently high statistics. \label{fig:cos-mc}}
\end{figure}

Figure~\ref{fig:cos-mc} shows examples of $\cos \theta$ distributions obtained from the MC simulation with specific magnitudes of muon polarization at the SK detector. As the polarization at the production site increases, the slope of the distribution becomes steeper. Hence, the SK detector can determine the magnitude of polarization by measuring the opening angles between the direction of incoming muons and emitted decay electrons.

\section{Data analysis} \label{sec:opt}

In this section, we briefly describe the data selection procedure for identifying pairs of parent muons and decay electrons, and procedures for the rejection of backgrounds. The performance in SK-IV is described but we have confirmed similar performance in other phases~(SK-V and SK-VI). We also describe the $\chi^{2}$ method used in determining the charge ratio and polarization, and associated systematic uncertainties.

\subsection{Parent muon selection}

To select decay electrons from the observed data in the SK detector, the first step to analyze the data is to tag stopping muons.

\subsubsection{First reduction}

An event whose number of PMT hits exceeds $1000$ is enough to find the muon track and the muon reconstruction is applied to such events. Then, the events that MUBOY recognizes as stopping muons are selected. Since the cosmic-ray originated from the atmosphere, only down-going muons with a zenith angle~(angle with respect to the detector vertical axis) of $\cos \theta_{\mathrm{Zenith}} > 0.2$ are selected to reject muons from muon neutrino interactions in the rock around the detector~\cite{Super-Kamiokande:1998uiq}. We also rejected muon events whose $z$-position of the entering position is the bottom region of the detector.

\subsubsection{Second reduction}

After muon reconstruction, some reconstructed parameters are used to reject misfit events, those mainly originated from single through-going muons as well as the corner-clipping muons. Reconstructed muons whose track length ranges from $2.5$~m to $32.25$~m are selected. Furthermore, the reconstruction goodness parameter is also calculated by evaluating the Cherenkov light pattern and the track length. The events whose goodness parameter ranges from $0.51$ to $0.70$ are selected since the small~(large) value of this parameter mainly consists of corner-clipping muons~(single through-going muons). 

\begin{figure}[!h]
    \begin{center}
        \includegraphics[width=\linewidth]{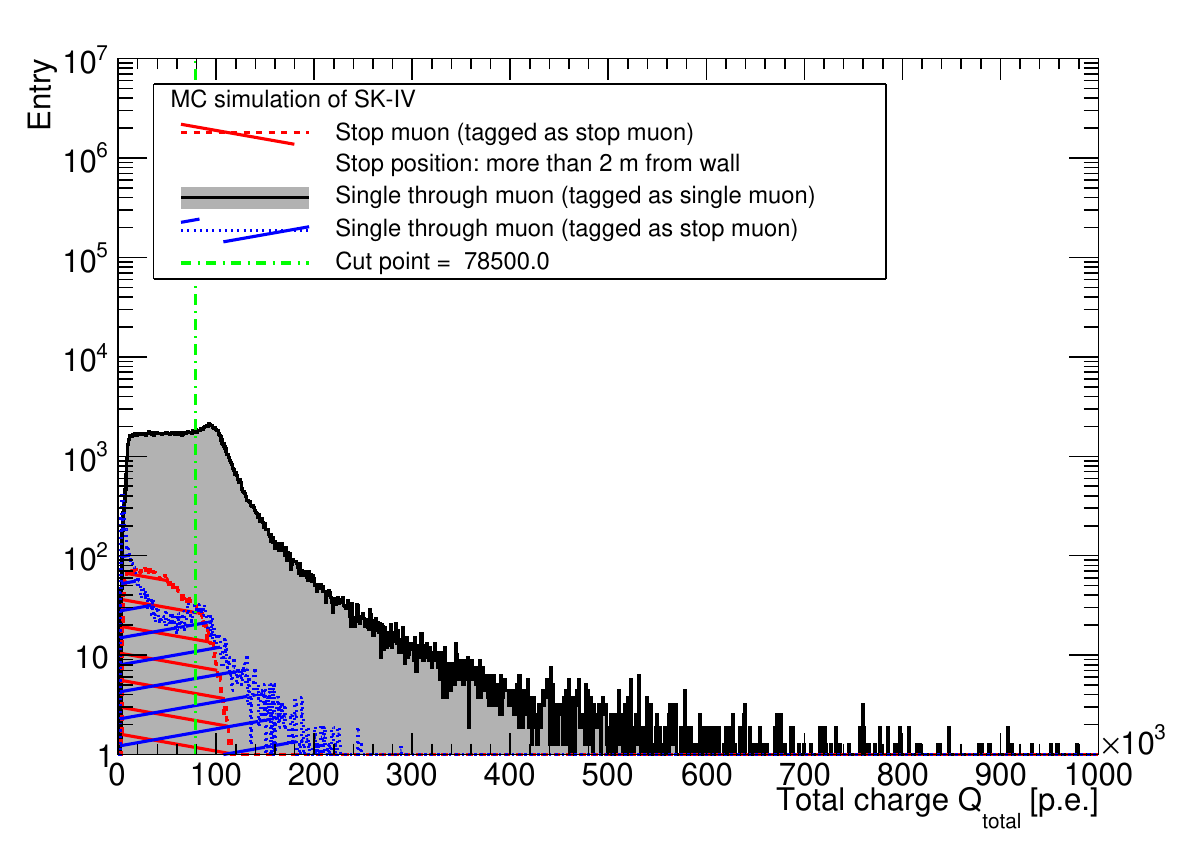}
    \end{center}
\caption{Typical distributions of total charge~($Q_{\mathrm{total}}$~[p.e.]) for stopping muons~(red
left-slanting line histogram), single through-going muons flagged as single muon~(gray filled histogram), and those flagged as stopping muon~(blue right slanting line histogram) from the MC simulation. The green vertical line shows the cut criterion on $Q_{\mathrm{total}}$.\label{fig:totq}}
\end{figure}

\begin{figure}[!h]
    \begin{center}
        \includegraphics[width=\linewidth]{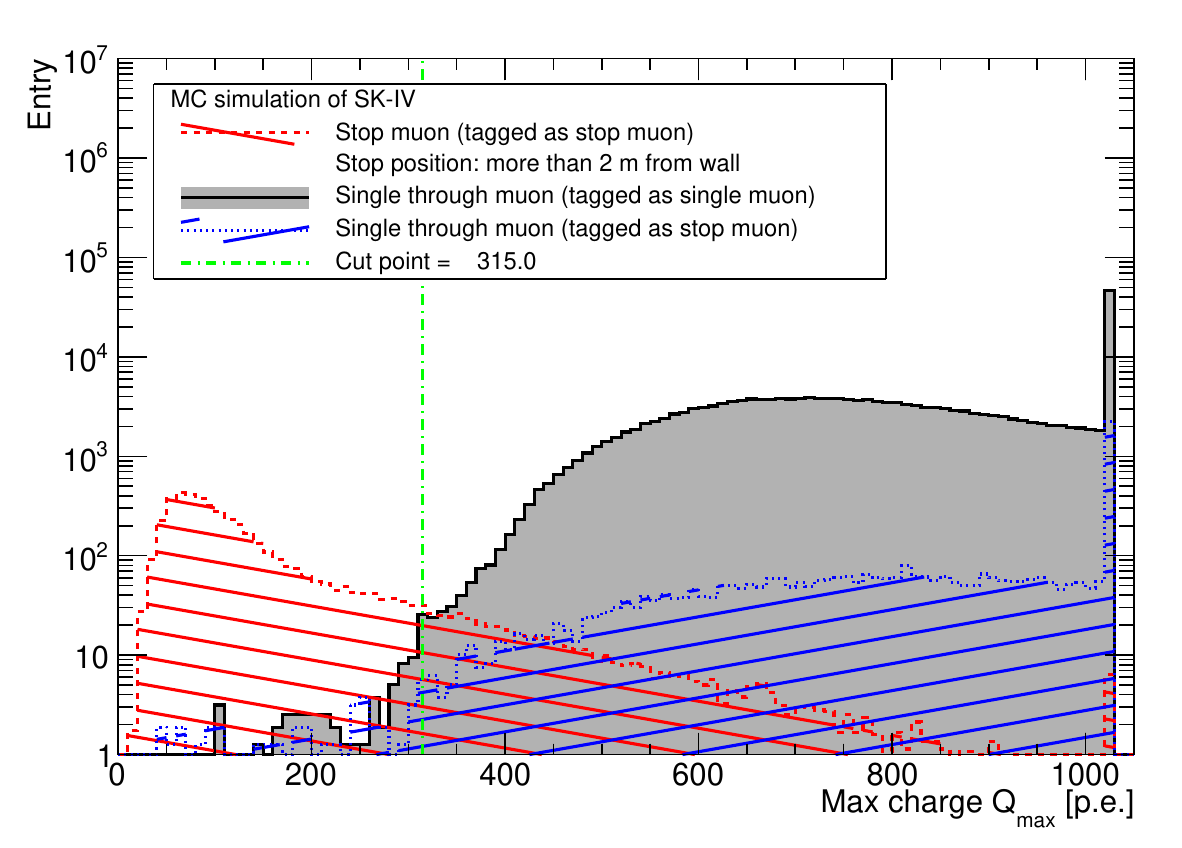}
    \end{center}
\caption{Typical distributions of maximum charge on a single PMT~($Q_{\mathrm{max}}$~[p.e.]) for stopping muons~(red left slanting line histogram), single through-going muons flagged as single muon~(gray filled histogram), and those flagged as stopping muon~(blue right slanting line histogram) from the MC simulation. The green vertical line shows the cut criterion of $Q_{\mathrm{max}}$. The large final bin count at $1020$~p.e. comes from the saturation of PMT electronics in the MC simulation. \label{fig:maxq}}
\end{figure}

In general stopping muons have a short track inside the detector and no exit point, resulting in a small total energy deposit in the detector. In the following sections charge is defined in units of the number of photoelectrons, which is obtained from the charge read out from a given PMT, converted by $2.658$~pC to $1$~p.e. for SK-IV~\cite{Abe:2013gga} and $2.460$~pc to $1$~p.e. for SK-V and SK-VI.  To quantitatively evaluate the deposited energy we define the total charge in an event, $Q_{\mathrm{total}}$, and the maximum charge on a single PMT in the event, $Q_{\mathrm{max}}$. For events in which the muon penetrates the SK detector the PMT nearest the exit point typically gives a large $Q_{\mathrm{max}}$, while stopping muons tend to have a small $Q_{\mathrm{max}}$. Figure~\ref{fig:totq} and Figure~\ref{fig:maxq} show typical distributions of $Q_{\mathrm{total}}$ and $Q_{\mathrm{max}}$ for stopping muons as well as penetrating muons using the MC simulation. 

Since some of the penetrating muons are incorrectly recognized as stopping muons, cut criteria for two parameters for stopping muons are optimized to maximize the selection efficiency for the stopping muon by evaluating the significance after the selection cuts. Due to the long operation of the SK detector, a gain shift of PMTs has been observed and this results in the gradual change of the parameters $Q_{\mathrm{total}}$ and $Q_{\mathrm{max}}$. To consider this gain shift effect, we optimized the cut criteria for each monthly data sample.   

After the stopping muon selection cuts, the total efficiency of finding stopping muon in the fiducial volume is $(71.86 \pm 0.01)\%$, where no difference can be seen between the negative and positive muons and the remaining fraction of other background events is less than $0.01\%$. The selection efficiencies of both the first and second reductions are summarized in Table~\ref{tb:eff-stop-mu}.

\begin{table*}[]
    \begin{center}
    \caption{Summary of stopping muon event selection efficiency, defined as the ratio of the number of events remaining after a given cut to the number of generated MC events assuming SK-IV. For the contamination of mis-reconstructed events, we applied the same cuts to the single through-going muon sample, and the remaining events after the cuts were evaluated. The mis-reconstructed events can be neglected in this analysis.}
        \label{tb:eff-stop-mu}
            \begin{tabular}{cccc}
                \hline
                \hline
                Selection cut for stopping muon & Stopping muons~[$\%$]& Stopping muons~[$\%$] & Other muons~[$\%$] \\ 
                Analysis volume & ID+OD & Fiducial volume~($22.5$~kton) & ID+OD \\ \hline
                First reduction & $57.54 \pm 0.01$ & $87.76\pm0.01$ & $ 1.85 \pm 0.01$ \\
                Second reduction & $44.32 \pm 0.01$ & $73.69\pm0.01$ & $ < 0.01$ \\ \hline
                Stopping muon selection efficiency  & $43.20 \pm 0.01$ & $71.86\pm0.01$ & $ <0.01$ \\
            \hline
            \hline
        \end{tabular}
    \end{center}
\end{table*}

\subsection{Decay electron selection} \label{sec:reduction}

After the selection of stopping muons, we then perform the search for decay electrons near the muon stopping position and reject background events.

\subsubsection{Delayed events search} \label{sec:delayed}

\begin{figure}[!h]
    \begin{center}
        \includegraphics[width=\linewidth]{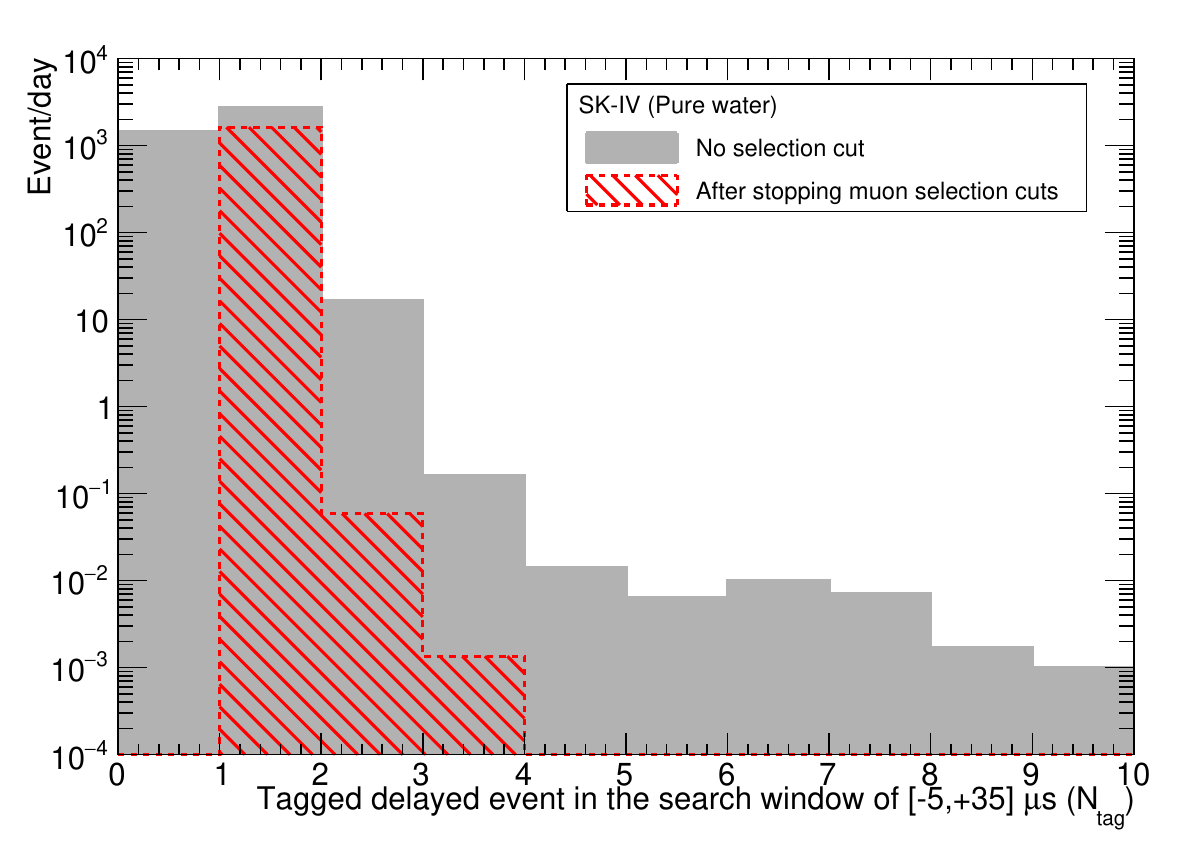}
    \end{center}
\caption{The typical distribution of $N_{\mathrm{tag}}$, which is defined as the number of tagged delayed events associated with single cosmic-ray muons, using the SK-IV data. The gray filled histogram~(red left slanting histogram) shows the number of events without~(with) the selection cuts described in Sec.~\ref{sec:reduction}. \label{fig:tag-sub-num}}
\end{figure}

In the process of delayed decay electron search, the number of tagged delayed events associated with a single cosmic-ray muon is expected to be $N_{\mathrm{tag}}=1$, where $N_{\mathrm{tag}}$ is defined as the number of tagged delayed events in the window. However, multiple accidental background events, whose number of hits PMTs within $200$~ns exceed the threshold, are also recorded within the search window of the single parent muon. In these cases, $N_{\mathrm{tag}}$ becomes more than one. Such accidental events originate from the decay of radon dissolved in the water~\cite{Nakano:2019bnr} or from spallation products induced by penetrating muons~\cite{Super-Kamiokande:2015xra}. Figure~\ref{fig:tag-sub-num} shows the typical distribution of $N_{\mathrm{tag}}$ using the SK-IV data set with and without the selection cuts described in Sec.~\ref{sec:reduction}.

Although selection cuts efficiently reject the accidental background events, some background events are still collected as data and result in $N_{\mathrm{tag}}\ge2$. To increase the purity of the decay electron data sample, we selected those events, where $N_{\mathrm{tag}}=1$.

Table~\ref{tb:num-decaye} categorizes the values of $N_{\mathrm{tag}}$ using the observed data. The fraction of tagging such accidental background events~($N_{\mathrm{tag}}\ge2$) is less than $0.01\%$ throughout three different phases after the selection cuts as shown in Fig.~\ref{fig:tag-sub-num}. This fraction is smaller than their statistical uncertainties. The relative increase of the number of delayed events with $N_{\mathrm{tag}} \ge 2$ is the result of $\gamma$-rays due to neutron capture with gadolinium-loaded water~\cite{Beacom:2003nk, Super-Kamiokande:2021the, Super-Kamiokande:2022cvw}. The details are discussed in Appendix~\ref{app:neutron-gd}.

\begin{table}[!h]
    \begin{center}
    \caption{Summary of the number of tagged events in the search window, which is defined as $N_{\mathrm{tag}}$. In this table, $N_{\mathrm{tag}}$ after the selection cuts with and without the energy cut described in Sec.~\ref{sec:gamma} is listed.}
        \label{tb:num-decaye}
            \begin{tabular}{cccc}
                \hline
                \hline
                SK phase & SK-IV & SK-V & SK-VI  \\ \hline 
                \multicolumn{4}{c}{No cut} \\
                $N_{\mathrm{tag}}=0$ & $4329334$ & $558240$ & $793876$ \\
                $N_{\mathrm{tag}}=1$ & $8062457$ & $1066656$ & $1550111$ \\
                $N_{\mathrm{tag}} \ge 2 $ & $\phantom{00}49983$ & $\phantom{00}26963$ & $\phantom{00}54210$ \\ \hline
                \multicolumn{4}{c}{Apply reduction cuts~(Sec.~\ref{sec:reduction})} \\
                $N_{\mathrm{tag}}=1$ & $5152331$ & $641761$ & $919853$ \\
                $N_{\mathrm{tag}} \ge 2 $ & $\phantom{000}191$ & $\phantom{000}25$ & $\phantom{000}319$ \\ \hline
                \multicolumn{4}{c}{Apply energy cut~(Sec.~\ref{sec:gamma})} \\
                 $N_{\mathrm{tag}}=1$ & $4823717$ & $600966$ & $860159$ \\
                $N_{\mathrm{tag}} \ge 2 $ & $\phantom{000}179$ & $\phantom{000}24$ & $\phantom{000}36$ \\
            \hline
            \hline
        \end{tabular}
    \end{center}
\end{table}

\subsubsection{Reduction cuts} \label{sec:reduction}

After the initial event reconstruction by BONSAI, several cuts are applied to select the decay electron events. Many radioactivity events from the PMTs and poorly reconstructed events are observed close to the ID wall. To reduce these backgrounds, events with reconstructed vertices within $2$~m horizontally from the ID wall are rejected. We also define a backward-projected distance~(the distance from the reconstructed vertex to the wall opposite from the direction of travel~\cite{Hosaka:2005um}) and reject events where this distance is less than $4$~m. When a decay electron is produced near the wall and is traveling toward the wall, some PMTs tend to observe multiple photons. This may lead to an underestimation of the number of PMTs that are hit, leading to an underestimation of the reconstructed energy of the decay electron. To mitigate this effect, we define a forward-projected distance~(the distance from the reconstructed vertex to the wall along the direction of travel) and reject events where this distance is less than $4$~m.

Next, a selection is applied to the distance between the stopping position of the parent muon and the vertex position of the decay electron candidate. For true muon-electron pairs, these vertices are expected to be close together, even if the vertex resolutions of the fitters mean the two are not the same. The distribution of separations, however, has a large tail arising from mis-tagged accidental backgrounds; events in which the decay of radon dissolved in the water~\cite{Nakano:2019bnr}, or of spallation products induced by penetrating muons~\cite{Super-Kamiokande:2015xra}, have been mis-identified as a decay electron. To remove these accidental coincidences a vertex separation cut of $<3$~m is applied. 

As described in Sec.~\ref{sec:tag_e}, when the time difference between the parent muon and the decay electron is shorter than $1.3~\mu$s it is usually a sign that light from the parent muon has been mis-identified as a decay electron. Such events pass the vertex separation cut, but the energy of decay electrons is generally not reconstructed correctly; we apply a time difference cut to avoid including such mis-reconstructed events. Since their lifetimes are different this cut results in a different selection efficiency for positive and negative muons in this analysis. We also reject events whose time difference is larger than $20~\mu$s because of their low statistics.

In addition to the vertex and timing cuts we also apply an event quality cut, where the quality of event reconstruction is quantified by two variables based on PMT hit timing~($g_{t}$) and hit pattern~($g_{p}$)~\cite{Abe:2016nxk}. Some radioactive background events, originating mainly from the PMT enclosures, PMT glass, and detector wall structure, are mis-reconstructed inside the fiducial volume even though the true vertex lies outside the fiducial volume. To reject such backgrounds we select events whose ${g_{t}}^{2}-{g_{p}}^{2}$ is larger than $0.22$. The selection efficiencies of each selection cut are summarized in Table~\ref{tb:eff-cut}.

\begin{table*}[]
    \begin{center}
    \caption{Summary of decay electron event selection efficiency in SK-IV. The first line gives the efficiency of finding the decay electron event in the fiducial volume~($22.5$~kton) in the window of $[-5,+35]$~$\mu$s, following the reduction cuts in Table~\ref{tb:eff-stop-mu}. Lines~$2$--$7$ describe selection efficiencies for the resulting subset of identified stopping muons, with line~$8$ being the resulting efficiency after the application of all of these selections. The final line gives the net efficiency, accounting for the reduction efficiency in Table~\ref{tb:eff-stop-mu} and the total selection efficiency in line~$8$.}
        \label{tb:eff-cut}
            \begin{tabular}{cccc}
                \hline
                \hline
                Selection cut & Positive muon & Negative muon & $\gamma$-rays  \\ 
                (In fiducial volume, $22.5$~kton) & [$\%$] & [$\%$] & [$\%$]\\ \hline
                Finding delayed event~($N_{\mathrm{tag}}=1$) & $76.31\pm0.02$ & $71.02\pm0.02$ & $16.05 \pm 0.07$ \\ \hline
                $\mu$-$e$ timing cut & $78.78\pm 0.02$ & $75.14 \pm 0.03$ & $92.98\pm0.12$\\
                Fiducial volume cut & $97.56 \pm 0.01$ & $97.60 \pm 0.01$ & $96.56\pm0.09$\\
                Effective wall cut & $91.50 \pm 0.02$ & $91.65 \pm 0.02$ & $91.57\pm0.13$\\
                $\mu$-$e$ distance cut & $95.23\pm0.01$ & $95.05 \pm 0.01$ & $95.31\pm0.13$\\
                Decay-e fit quality cut & $96.87 \pm 0.01$ & $96.43 \pm 0.01$ & $97.57\pm0.07$\\
                Energy cut & $87.17 \pm 0.02$ & $85.75\pm0.02$ & ${\phantom{0}}0.19\pm0.02$\\ \hline
                Decay electron selection efficiency & $64.66 \pm 0.03$ & $61.48 \pm 0.03$ & $0.0$\\ \hline
                Total efficiency (w/ stopping muon selection) & $35.53 \pm 0.02$ & $31.28 \pm 0.02$ & $0.0$ \\
            \hline
            \hline
        \end{tabular}
    \end{center}
\end{table*}

\subsubsection{Gamma-rays from oxygen capture} \label{sec:gamma}

As briefly explained in Sec.~\ref{sec:pol-detector}, negative muons sometimes are captured on oxygen in water. This reaction eventually produces $\mathrm{^{16}N}$, $\mathrm{^{15}N}$, or $\mathrm{^{14}N}$, depending on the number of neutrons simultaneously produced~\cite{Plett:1971qb, vanderSchaaf:1983ah, Measday:2001yr}. In the case of $\mathrm{^{15}N}$ and $\mathrm{^{14}N}$ de-excitation $\gamma$-rays are emitted soon after radioisotope production, which the trigger system can mis-identify as decay electron events. Since the charge ratio is determined by the numbers of decay electrons and positrons as described in Eq.~(\ref{eq:ratio}), the contamination from such background events decreases the sensitivity to the measurements of the charge ratio. 

\begin{figure}[!h]
    \begin{center}
        \includegraphics[width=\linewidth]{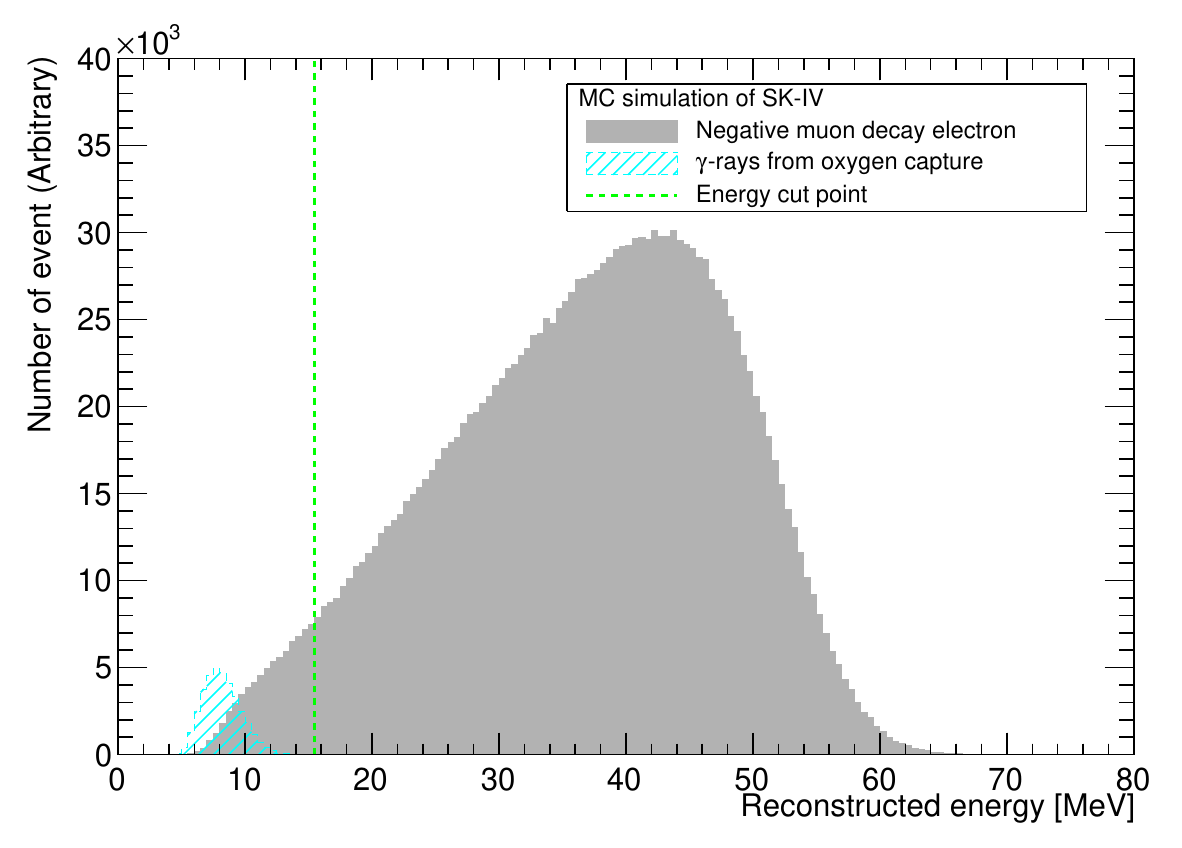}
    \end{center}
\caption{The reconstructed energy distribution for $\gamma$-rays from oxygen capture~(light-blue right slanting line histogram) as well as that of negative muon decay electron~(gray filled histogram), obtained from the MC simulation after applying the selection cuts listed in Table~\ref{tb:eff-cut}. The vertical green dashed line shows the optimized cut value used in the analysis. \label{fig:gamma-n50}}
\end{figure}

Figure~\ref{fig:gamma-n50} shows the reconstructed energy distribution for $\gamma$-rays events and negative muon decay events, from MC simulation. Since the energies of emitted $\gamma$-rays are less than about $15$~MeV the number of PMT hits is lower than those from true decay electron events. To eliminate such $\gamma$-ray events, we removed events whose reconstructed energy is less than $15.5$~MeV. Although this selection cut removes a small amount of decay electron events, $\gamma$-ray events are efficiently rejected from the analysis sample as listed in Table~\ref{tb:eff-cut}. In addition to this, accidental background events are also rejected by this cut as listed in Table~\ref{tb:num-decaye}.

\subsubsection{Summary of selection cuts}

After applying the cuts above the total selection efficiency is determined using MC simulated events.  The efficiency is evaluated for positive and negative muons separately, to account for the impact of nuclear capture incurred only by negative muons. Table~\ref{tb:eff-cut} summarises the resulting efficiencies including the stopping muon selection in Table~\ref{tb:eff-stop-mu}.

From MC simulation the total efficiency for selecting stopping positive~(negative) muon decay events in the fiducial volume~($22.5$~kton) is estimated to be $35.53 \pm 0.02\%$~($31.28 \pm 0.02\%$), where the difference originates from the difference in decay times, while the contamination due to $\gamma$-ray background events is completely rejected.

\subsection{Chi-square definition}

To determine the charge ratio and polarization from the observed data, the decay times of tagged decay electrons, the energy spectra of tagged decay electrons, and the $\cos \theta$ distribution between the direction of incoming cosmic-ray muons and that of emitted decay electrons are simultaneously fit to the distributions derived from the MC simulation. 
The definition of total chi-square~($\chi^{2}_{\mathrm{Total}}$) is

\begin{equation}
    \chi^{2}_{\mathrm{Total}}~(R, P^{\mu}_{0}) = \chi^{2}_{\mathrm{Time}}+\chi^{2}_{\mathrm{Energy}}+\chi^{2}_{\mathrm{\cos \theta}} \label{eq:chi2}
\end{equation}

\noindent where $\chi^{2}$ is chi-square for each distribution, $R$ is the given charge ratio, $P^{\mu}_{0}$ is the polarization of cosmic-ray muons at the production site. The $\chi^{2}$ for each distribution are defined as,

\begin{equation}
\left\{
\begin{array}{ll}
 \displaystyle \chi^{2}_{\mathrm{Time}} = \sum_{i}^{n_{\mathrm{Time}}} \frac{\left(N^{\mathrm{Data}}_{i}-N^{\mathrm{MC}}_{i} \right)^{2}}{(\sigma^{\mathrm{Data}}_{i})^{2}+(\sigma^{\mathrm{MC}}_{i})^{2}+(\sigma^{\mathrm{Syst.}}_{i})^{2}} \\
 \displaystyle \chi^{2}_{\cos\theta} = \sum_{i}^{n_{\cos \theta}} \frac{\left(N^{\mathrm{Data}}_{i}-N^{\mathrm{MC}}_{i} \right)^{2}}{(\sigma^{\mathrm{Data}}_{i})^{2}+(\sigma^{\mathrm{MC}}_{i})^{2}+(\sigma^{\mathrm{Syst.}}_{i})^{2}} \\
  \displaystyle \chi^{2}_{\mathrm{Energy}} = \sum_{i}^{n_{\mathrm{Energy}}} \frac{\left(N^{\mathrm{Data}}_{i}-N^{\mathrm{MC}}_{i} \right)^{2} }{(\sigma^{\mathrm{Data}}_{i})^{2}+(\sigma^{\mathrm{MC}}_{i})^{2}} + \left( \frac{1-p}{\sigma^{\mathrm{E\text{-}scale}}} \right)^{2}\\
\end{array}
\right.
\end{equation}

\noindent 
where $N_{i}^{\mathrm{Data}}$~($N_{i}^{\mathrm{MC}}$) is the number of selected events in $i$-th bin of the observed data~(MC) distribution, $n$ is the number of bins, $\sigma_{i}^{\mathrm{Data}}$~($\sigma_{i}^{\mathrm{MC}}$) is the statistical uncertainty on each bin of the observed data~(MC), and $\sigma_{i}^{\mathrm{Syst.}}$ is the systematic uncertainty on each bin described in the next subsection, respectively. Since the energy scale of decay electrons affects the value of $\chi^{2}_{\mathrm{Energy}}$, the pull term is introduced only for $\chi^{2}_{\mathrm{Energy}}$, where $\sigma^{\mathrm{E\text{-}scale}}$ is the systematic uncertainty of the energy scale determined from LINAC calibration~\cite{Nakahata:1998pz} and the details are described in the next subsection. 

In the presented analysis, we generated four kinds of MC simulations, where charged muons are fully polarized, e.g. $P_{0}^{\mu^{+}}=\pm 1.0$, and $P_{0}^{\mu^{-}} = \pm 1.0$ at the laboratory frame. These samples enable us to produce the expected distributions of decay time, energy, and $\cos \theta$ with any given combination of charge ratio~($R$) and polarization~($P_{0}^{\mu}$).

\subsection{Systematic uncertainties}

In this section, the systematic uncertainties associated with reconstruction methods are discussed. Since the selection cuts equally affect negative and positive muon decays, we have not included the systematic uncertainty due to the selection cuts.

\subsubsection{De-polarization during the propagation}

As we estimated the probability of de-polarization during the propagation in air and rock in Sec.~\ref{sec:de-pol}, such uncertainty propagates to the final result of the polarization measurement. Indeed, the parameter $\delta$ depends on the initial muon energy at the surface of the mountain, the propagation length, and the density profile of medium~\cite{Hayakawa:1957}. The dependencies originating from the energy and propagation length are canceled out because those two variables are anti-correlated. However, the density profile of the surrounding rock is difficult to fully understand. For conservatively considering such uncertainties, we assigned the systematic uncertainty of de-polarization during the propagation as $0.3\%$~(relatively $100\%$ of the parameter $\delta$).

\subsubsection{Accuracy of decay time} \label{sec:sys_time}

\begin{figure}[!h]
    \begin{center}
        \includegraphics[width=\linewidth]{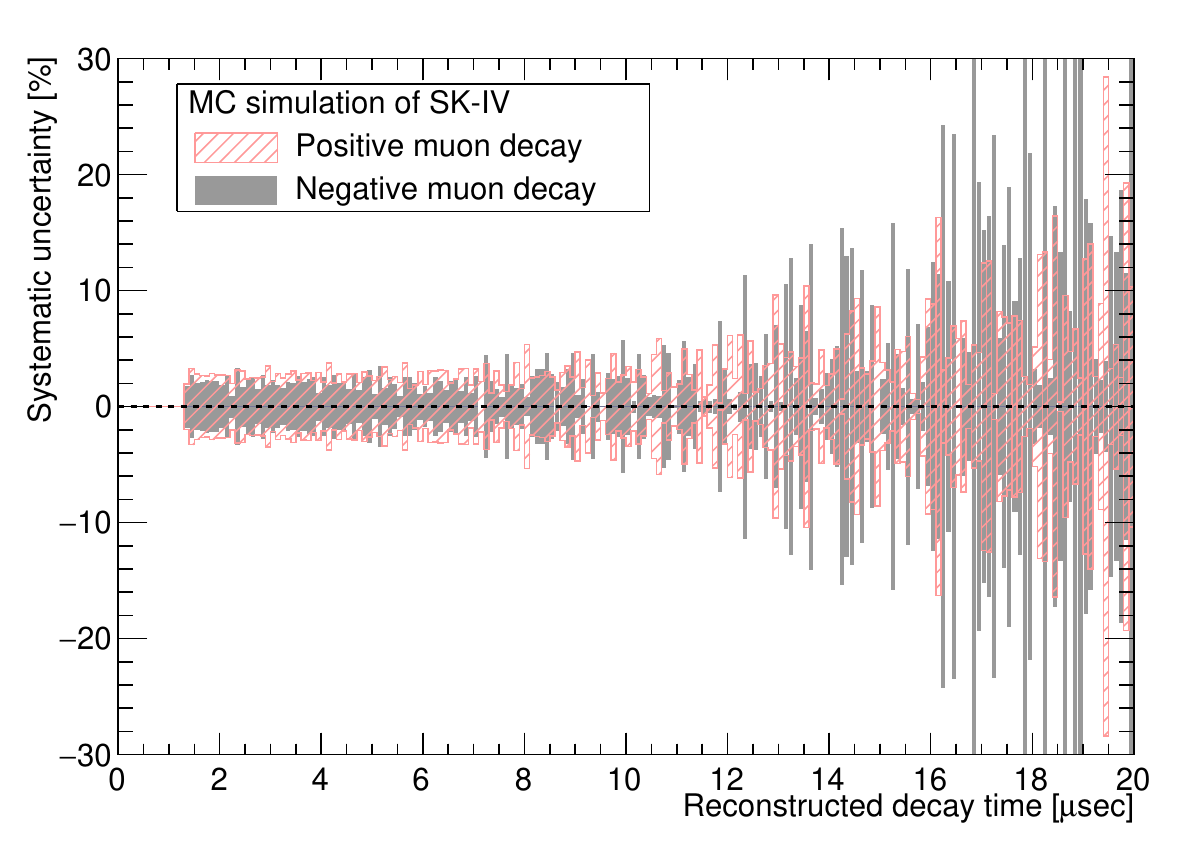}
    \end{center}
\caption{Systematic uncertainties, representing the difference in number of events between the true decay time sample and the reconstructed decay time sample, from decay time reconstruction in SK-IV. The gray filled~(red right slanting line) histogram shows the systematic uncertainty of the negative~(positive) muon decay sample. The relatively large uncertainty above $10$~$\mu$s is a result of low statistics on the selected events in the MC simulation. \label{fig:sys_time} }
\end{figure}

The time between the stopping muon event and the decay electron event is calculated by reconstructing the time of each event separately. To estimate the combined systematic uncertainty on the resulting decay time we prepared two samples. The first is produced by analyzing MC simulation data using the true decay time, while the second is produced using the reconstructed decay time. We then compared the number of events in each bin in Eq.~(\ref{eq:chi2}) between the data sets, and assigned their difference as the systematic uncertainty on that bin. Figure~\ref{fig:sys_time} shows the resulting systematic uncertainties from the timing reconstruction. Since the decay times are different between positive and negative muons, their systematic uncertainties are separately estimated.

\subsubsection{Track, direction, and vertex reconstructions}

As described in Sec.~\ref{sec:recon_mu}, the stopping position is estimated by MUBOY with an accuracy of $0.5$~m and this results in the uncertainty of the track length. In addition to the track reconstruction, the vertex reconstruction by BONSAI has about $0.3$~m of vertex resolution. To estimate their impact, we made an MC sample by applying the same selection cuts using the true vertex positions instead of the reconstructed track by MUBOY and the stopping position by BONSAI. By comparing the number of events after the selection cuts, their difference is estimated.

\begin{figure}[!h]
    \begin{center}
        \includegraphics[width=\linewidth]{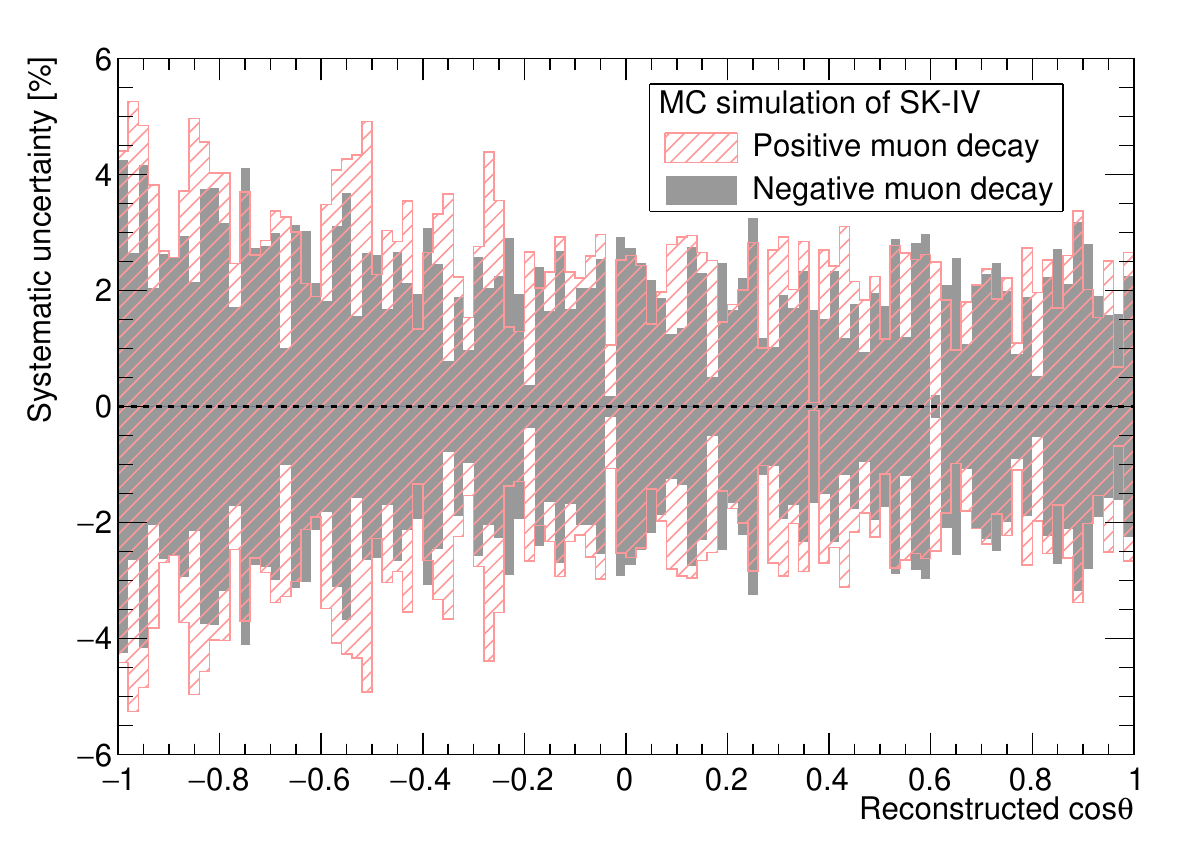}
    \end{center}
\caption{Systematic uncertainty on reconstructed $\cos \theta$ between the parent muon and the emitted electron in SK-IV. The gray filled~(red right slanting line) histogram shows the systematic uncertainty of the negative~(positive) muon decay sample. \label{fig:sys_cos} }
\end{figure}

In addition, the accuracy of directional reconstruction directly affects the measurement of muon polarization because the angle between the muon and the decay electron reflects the magnitude of the polarization. As mentioned in Sec.~\ref{sec:sk}, two separate algorithms are used for reconstructing the stopping muon track and the emitted decay electron. To estimate the systematic uncertainty on the angle between them we again prepared two samples according to the procedure described in Sec.~\ref{sec:sys_time}. Figure~\ref{fig:sys_cos} shows the systematic uncertainty caused by directional reconstruction. We evaluate the number of events in each bin of the $\cos \theta$ distribution, which is typically at the $\pm2\%$ level of systematic uncertainty due to the directional reconstruction, and consider those estimated values in Eq.~(\ref{eq:chi2}).

\subsubsection{Energy reconstruction}

This energy reconstruction is tuned by comparing calibration data against MC simulation with LINAC~\cite{Nakahata:1998pz} and deuterium-tritium neutron~(DT) generator~\cite{Blaufuss:2000tp} sources. The former determines the absolute energy scale by injecting mono-energy electron beams and the latter evaluates the directional dependence of the energy scale as well as the stability of the energy scale in time with high statistics radioactive $\beta$ decays.

Table~\ref{tb:escale} summarizes the systematic uncertainties on the energy scale determined by two calibration sources, for each of the SK phases analyzed~\cite{Super-Kamiokande:2023jbt}. The relatively large errors for SK-V and SK-VI are a result of the limited number of LINAC calibrations compared to SK-IV, due to their short running times.

\begin{table}[!h]
    \begin{center}
    \caption{Summary of systematic uncertainties on the energy scale, determined via LINAC and DT calibrations~\cite{Nakahata:1998pz, Blaufuss:2000tp, Super-Kamiokande:2023jbt}.}
        \label{tb:escale}
            \begin{tabular}{cc}
                \hline
                \hline
                SK phase & Systematic uncertainty~[$\%$] \\ \hline 
                SK-IV & $\pm0.48$ \\ 
                SK-V{\phantom{1}} & $\pm0.87$ \\ 
                SK-VI & $\pm1.32$ \\ 
            \hline
            \hline
        \end{tabular}
    \end{center}
\end{table}

\subsubsection{Gadolinium Addition after SK-VI} \label{sec:gd}

As briefly mentioned in Sec.~\ref{sec:dataset}, the SK-VI phase started after the first gadolinium loading in July 2020. During this loading work, $13$~tons of $\mathrm{Gd_{2}(SO_{4})_{3} \cdot 8 H_{2}O}$ was dissolved, resulting in $0.021\%$ of $\mathrm{Gd_{2}(SO_{4})_{3}}$ concentration in the SK tank~\cite{Super-Kamiokande:2021the}. These additional elements may produce additional muonic atoms instead of oxygen.

In a chemical compound, the ratio of muonic Coulomb capture on each element can be expressed by the number of nuclei and the relative capture probabilities~$P(Z)$, where $Z$ is the atomic number. The parameter $P(Z)$ for various elements has experimentally been measured~\cite{Daniel:1979ay}, where its values are normalized by $P(8)$, the value for oxygen. As stated in Sec.~\ref{sec:dcy_po}, the muons, which are captured on hydrogen, are immediately transferred to other nuclei with higher atomic numbers. Hence, the contribution from hydrogen is ignored in this estimation. By multiplying the capture probability and the element ratio, the relative ratio of muonic Coulomb capture in gadolinium-loaded water can be calculated. Table~\ref{tb:cap-ratio} summarizes the number ratio of elements in the SK tank, the muonic Coulomb capture probabilities relative to the oxygen capture~\cite{VonEgidy:1982pe}, the relative ratio of muonic Coulomb capture, and the lifetime of captured muons~\cite{Suzuki:1987jf}. 

\begin{table*}[]
    \begin{center}
    \caption{Summary of the element ratio in gadolinium-loaded water, the muonic Coulomb capture probability~$P(Z)$ normalized to~$P(8)$~\cite{Suzuki:1987jf, VonEgidy:1982pe}, the fraction of muonic Coulomb capture, and the lifetimes of the muonic atom~~\cite{ParticleDataGroup:2022pth}. Here the concentration of $\mathrm{Gd_{2}(SO_{4})_{3}}$ is $0.021\%$ as mentioned in the main text~\cite{Super-Kamiokande:2021the}.}
        \label{tb:cap-ratio}
            \begin{tabular}{ccccc}
                \hline
                \hline
                Elements & Element ratio & Relative capture  & Fraction of & Lifetime~[$\mu$s] \\ 
                 & in gadolinium-loaded water~[$\%$] & probability $P(Z)$ & muonic Coulomb capture~[$\%$] & \\ \hline
                Hydrogen & $66.6637$ & -- & -- & --\\
                Oxygen & $33.3352$ & $1.0$ & $99.990$ & $1.7954 \pm 0.0020$ \\
                Sulfur & $\phantom{0}0.0006$ & $1.23 \pm 0.05$ & $\phantom{0}0.002$ & $0.5447 \pm 0.0010$ \\
                Gadolinium & $\phantom{0}0.0004$ & $\phantom{0}5.8\pm0.5$ & $\phantom{0}0.007$ &  $0.0818 \pm 0.0015$ \\
                \hline
                Free muon & -- & -- & -- & $2.1969811 \pm 0.0000022$ \\ \hline
            \hline
        \end{tabular}
    \end{center}
\end{table*}

Because of low concentrations of sulfur and gadolinium, their relative ratio of forming muonic atoms is quite small. Furthermore, the $\mu$-$e$ timing cut listed in Table~\ref{tb:eff-cut} efficiently removes such events because of their shorter decay time. After the selection cuts, the number of pair of parent muon and decay electron in the SK-VI data sample is $860159$~events in $560.6$~days as listed in Table~\ref{tb:num-decaye}, where the stopping negative muon in the fiducial volume is about $1040667\pm34280$~events considering the selection efficiency and resulting charge ratio. Hence, the Coulomb capture by sulfur and gadolinium in the fiducial volume is expected to be $24.1\pm0.8$ and $75.9\pm2.5$, respectively. Based on the calculation above, their contamination after the timing cut of $1.3~\mu$s is about $7.7\pm0.3$~events for sulfur and $12.3\pm0.4$~events for gadolinium. We ignore their contamination in the analysis because the statistical uncertainties are larger than their contributions.

\section{Results and Discussion} \label{sec:result}

In this section, we present the measurement results for the muon charge ratio and the polarization. Then, we compare those results with the theoretical expectations and the results measured by other experiments.

\subsection{Chi-square map}

\begin{figure}[!h]
    \begin{center}
        \includegraphics[width=\linewidth]{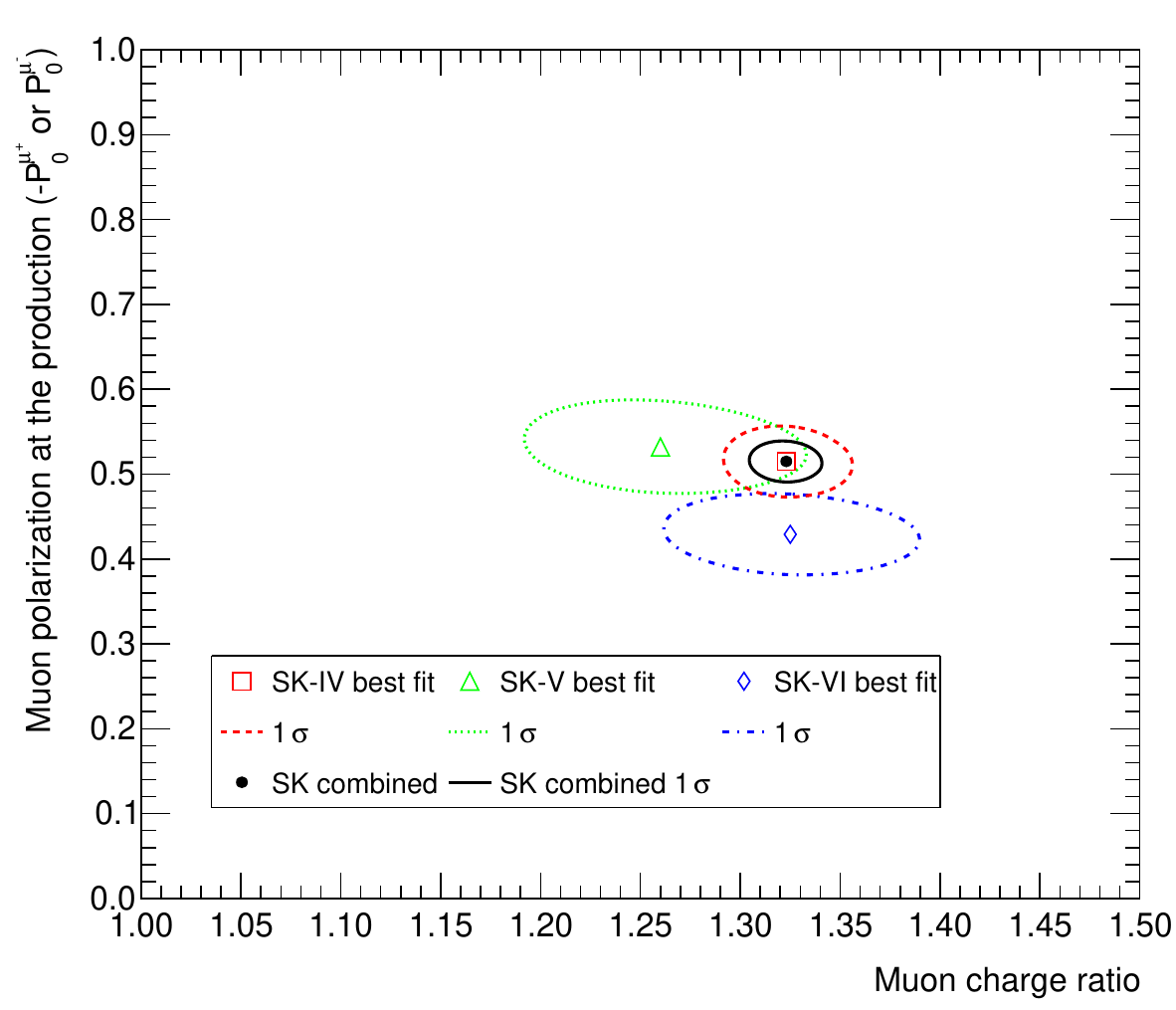}
    \end{center}
\caption{The allowed regions of the charge ratio and the polarization at the production site using SK-IV, SK-V, and SK-VI data sets. Red open square~(dashed line), green open upward-triangle~(dotted line), and blue open rhombus~(dashed-dotted line) show the best-fit values~(their $1\sigma$ allowed regions determined by $\Delta \chi^{2}(R, P^{\mu}_{0})=2.30$) of SK-IV, SK-V, and SK-VI, respectively. The black filled circle~(solid line) shows the combined value~($1\sigma$ allowed region). The measured values and the combined value are listed in Table~\ref{tb:chi2-results}. \label{fig:chi2-map}}
\end{figure}

To determine the charge ratio and the polarization at the production site, we calculated the $\chi^{2}$ defined in Eq.~(\ref{eq:chi2}) and then extracted the difference between each value and the minimum value of $\chi^{2}$, which is expressed as $\Delta \chi^{2}(R, P_{0}^{\mu}) = \chi^{2} (R, P_{0}^{\mu}) - \chi^{2}_{\mathrm{Min}}$, where $\chi^{2}_{\mathrm{Min}}$ is the minimum value of $\chi^{2}$. Figure~\ref{fig:chi2-map} shows the result of $\Delta \chi^{2}$ calculation using SK-IV, SK-V, and SK-VI data sets. The measured charge ratio and polarization among three different data sets are consistent with their estimated uncertainties. 

\begin{figure*}[]
    \begin{tabular}{cc}
        \begin{minipage}{0.33\textwidth}
            \centering
            \includegraphics[width=\textwidth]{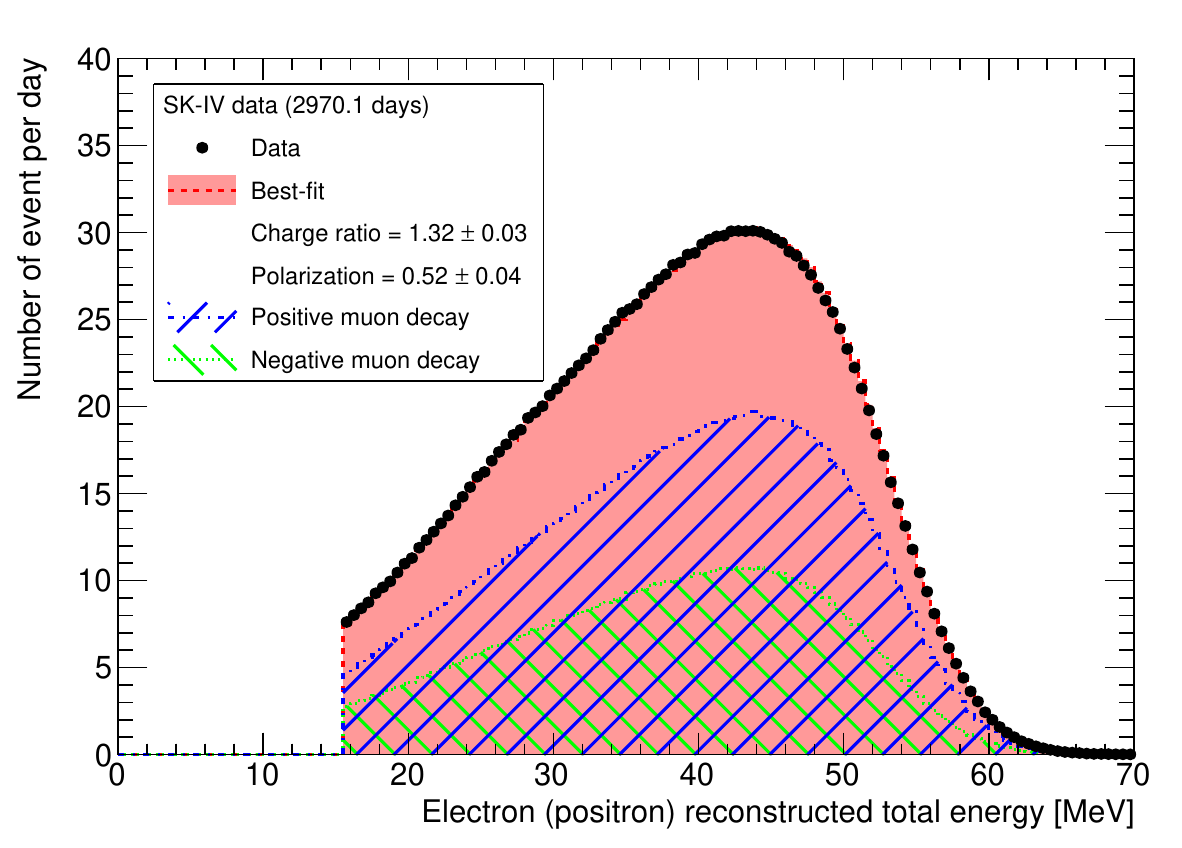}
        \end{minipage} 
        \begin{minipage}{0.33\textwidth}
            \centering
            \includegraphics[width=\textwidth]{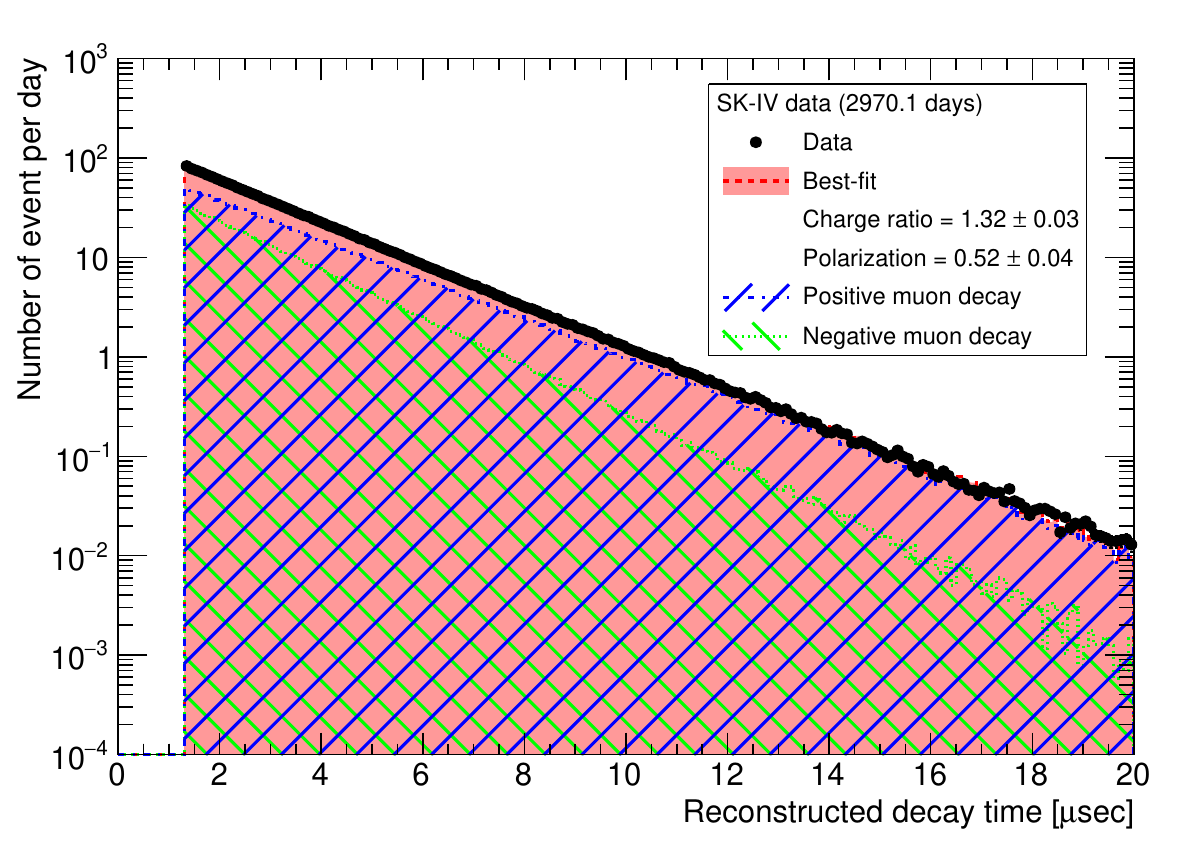}
	\end{minipage} 
         \begin{minipage}{0.33\textwidth}
            \centering
            \includegraphics[width=\textwidth]{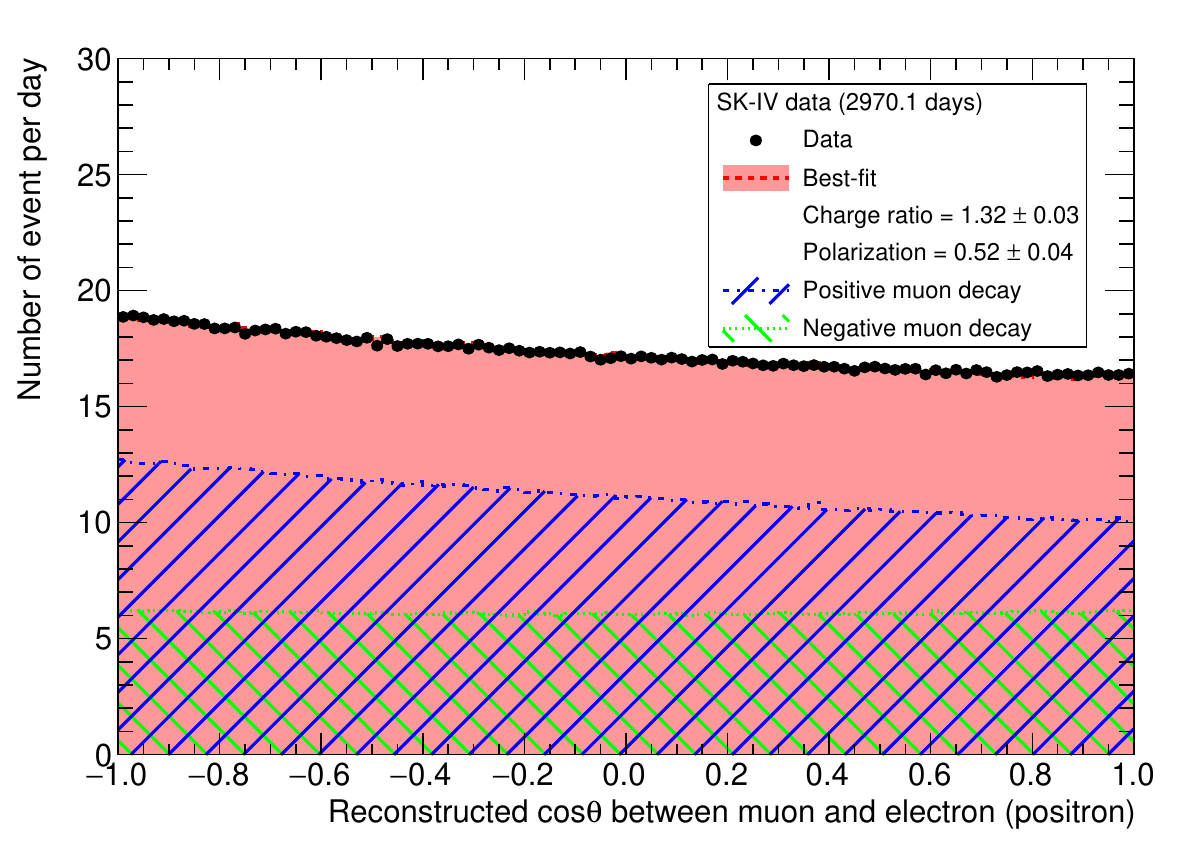}
	\end{minipage}
   \end{tabular}
\caption{The example of three distributions using the decay electron sample in SK-IV~(other phases are shown in appendix~\ref{app:best-fit}). The left, middle, and right panels show the distribution of the reconstructed energy, decay time, and $\cos \theta$ between the directions of the parent muon and emitted electron, respectively. Black points show the observed data, the blue dashed~(green dotted) histogram shows the contribution of positive~(negative) muons, and the red filled histogram shows the best-fit distribution by combining the positive and negative muon samples. \label{fig:dist-comp}}
\end{figure*}

Figure~\ref{fig:dist-comp} shows the example of three distributions of observed decay electron sample in SK-IV, i.e. the reconstructed energy distribution, time difference distribution, and $\cos \theta$ distribution together with the MC simulation. The same distributions using data taken in SK-V and SK-VI are shown in Appendix~\ref{app:best-fit}. Three distributions of the observed decay electron sample demonstrate good agreement with those of the best-fit MC simulation.

Table~\ref{tb:chi2-results} summarizes the charge ratio and the polarization at the production site among the three SK phases.

\begin{table}[!h]
    \begin{center}
    \caption{The summary of the charge ratio and the polarization at the production determined by the $\chi^{2}$ method. The statistical and systematic uncertainties are considered for the uncertainties of SK results.}
        \label{tb:chi2-results}
            \begin{tabular}{cc|cc}
                \hline \hline
                Phase & Charge ratio~($R_{\mu}$) & $P^{\mu}_{\mathrm{0}}$ & $P^{\mu}_{\mathrm{obs}}$ \\ \hline
                SK-IV & $1.32\pm0.03$ & $0.52\pm0.04$ & $0.22 \pm 0.02$   \\
                SK-V & $1.26\pm 0.07$ & $0.54^{+0.06}_{-0.05}$ & $0.22\pm 0.02$    \\
                SK-VI & $1.33\pm0.06$ & $0.43\pm0.05$ & $0.18 \pm 0.02$  \\
                \hline 
                SK combined & $1.32\pm0.02$ & $0.52\pm0.02$ & $0.22 \pm 0.01$  \\
            \hline
            \hline
        \end{tabular}
    \end{center}
\end{table}

By combining the results from three different SK phases, we determined the charge ratio and the polarization as $R=1.32\pm0.02~(\mathrm{stat.{+}syst.})$, and $P^{\mu}_{0}=0.52 \pm 0.02~(\mathrm{stat.{+}syst.})$, respectively.

\subsection{Periodicity search}

The production of atmospheric neutrinos depends on the intensity of primary cosmic-rays, the density of the atmosphere structure, geomagnetic field, etc. The density as well as the temperature in the stratosphere region, where cosmic-ray muons are produced, changes depending on the solar activity~\cite{solcli:2010}, and this results in the change of the production rate of secondary particles.

\begin{figure*}[]
    \begin{tabular}{cc}
	\begin{minipage}{0.50\textwidth}
		\centering
		\includegraphics[width=0.95\textwidth]{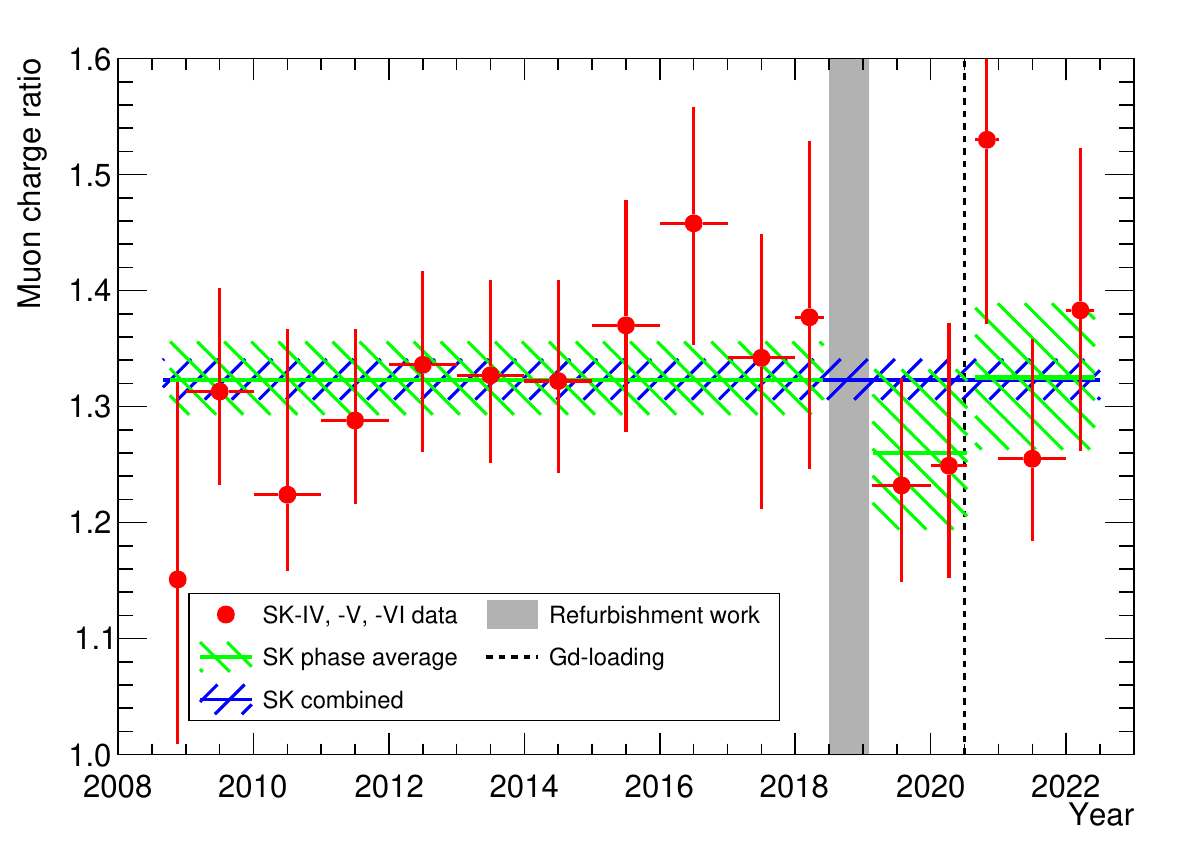}
    \end{minipage} 
        \begin{minipage}{0.50\textwidth}
       \centering
		\includegraphics[width=0.95\textwidth]{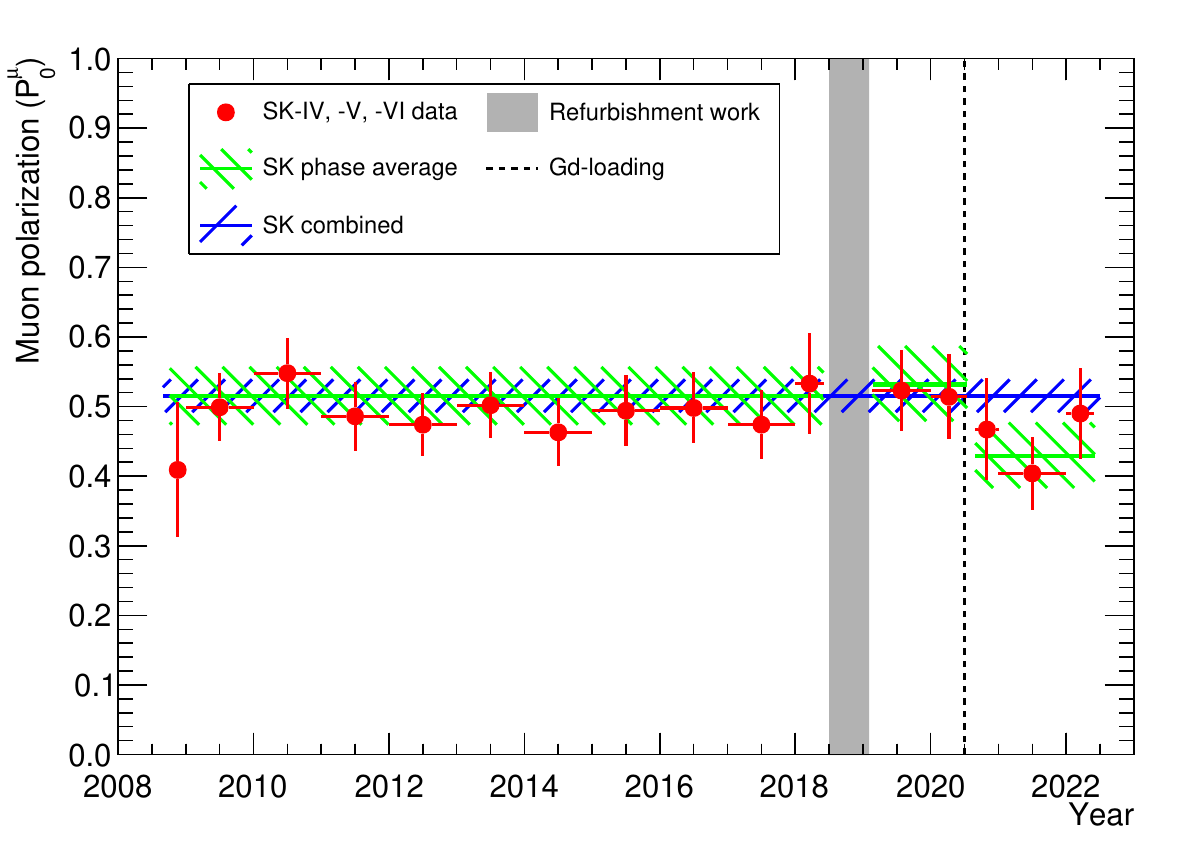}
	\end{minipage}
   \end{tabular}
\caption{Yearly data of the charge ratio~(left) and muon polarization~(right) from 2008 to 2022. The red filled circles show the yearly values obtained in this study and the green solid lines~(with a left slanting band) show the averages of each SK phase~(with their uncertainties). The blue solid line~(with a right slanting band) also shows the combined SK result~(with its uncertainty). The gray vertical band and the dashed vertical line show the period of the refurbishment work and the first gadolinium-loading~\cite{Super-Kamiokande:2021the}. \label{fig:year}}
\end{figure*}

The SK data presented in this article covers the period of the solar cycle~$24$, whose solar maximum was around April 2014. Such data can test possible correlations between the charge ratio~(polarization) of cosmic-ray muons and the solar activity whose periodicity is about $11$~years. Figure~\ref{fig:year} shows the yearly data of the charge ratio and the muon polarization at the production site from 2008 to 2018.

\begin{figure}[!h]
    \begin{center}
        \includegraphics[width=\linewidth]{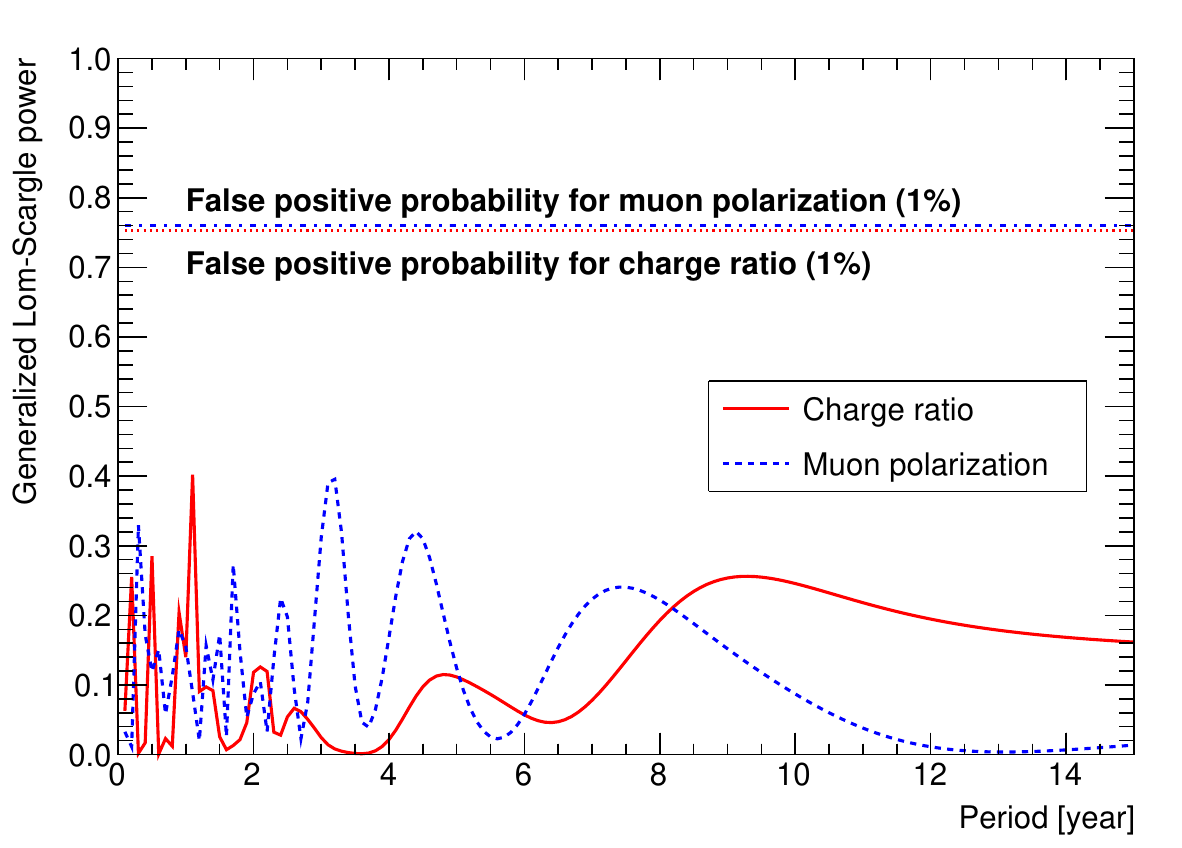}
    \end{center}
\caption{The power spectrum derived from the generalized Lomb-Scargle method~\cite{gls:2009}. The red solid~(blue dashed) line shows the power spectrum of charge ratio~(muon polarization). The maximum height of power spectra are $0.40$ at $1.1$~years for charge ratio and $0.40$ at $3.2$~years for muon polarization, where their probabilities are $6.9\%$, and $7.2\%$, respectively. We note that the false positive probability is defined as the probability of measuring a peak of a given height conditioned on the assumption that the data consists of Gaussian noise with no periodic component. The horizontal lines indicate the false positive probabilities for two measurements at $1\%$ according to Ref.~\cite{gls:2009}. Hence, no periodic change is observed in the SK data within the current measurement accuracy. \label{fig:lomb}}
\end{figure}

To test the periodicity of muon charge ratio and muon polarization, we analyzed yearly-binned data~(shown in Fig.~\ref{fig:year}) by using the generalized Lomb-Scargle method~\cite{gls:2009}, which is generally used to search for periodicity in frequency analysis of time series. Figure~\ref{fig:lomb} shows the power spectra by analyzing the yearly data of both charge ratio and muon polarization. The largest peak in the charge ratio power spectrum has an amplitude of $0.40$ and a periodicity of $1.1$~years, while that of the muon polarization power spectrum has an amplitude of $0.40$ and a periodicity of $3.2$~years. The probability of finding peaks of such magnitude by chance are $6.9\%$ and $7.2\%$, respectively. Therefore, the yearly data shown in Fig.~\ref{fig:year} can be explained by statistical fluctuation and no clear periodic change is found in both charge ratio and muon polarization measurements.

\subsection{Comparison with other experiments and simulations}
\subsubsection{Charge ratio} \label{sec:charge_result}

\begin{figure}[!h]
    \begin{center}
        \includegraphics[width=\linewidth]{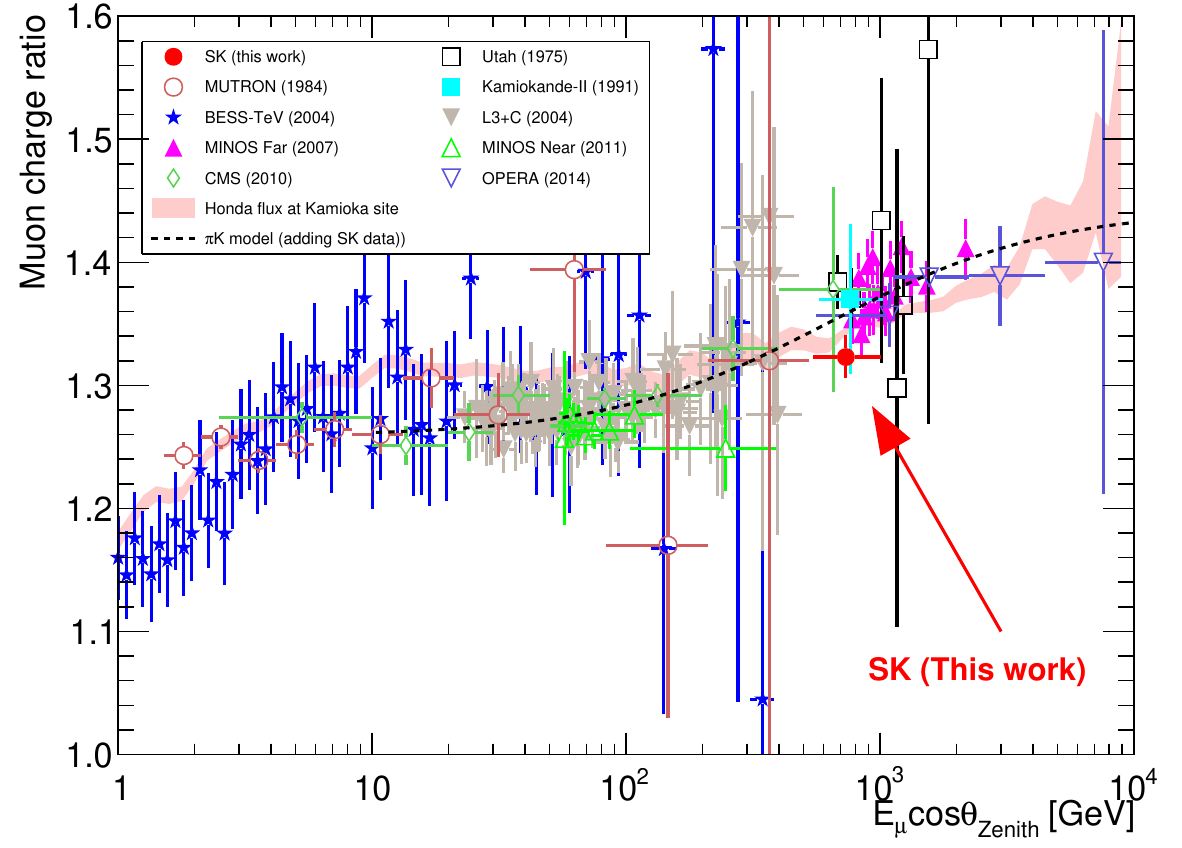}
    \end{center}
\caption{The muon charge ratio measured by the experiments together with the expectations from simulation models. The red filled circle shows the SK data and other symbols with different colors show the experimental data measured by other groups~\cite{Ashley:1975uj, Matsuno:1984kq, Yamada:1991aq, Haino:2004nq, Achard:2004ws, Adamson:2007ww, MINOS:2011amj, Khachatryan:2010mw, OPERA:2014blf}. The red band shows the expected muon charge ratio based on Honda flux simulation~\cite{Honda:2015fha} and the dashed curve shows the fitting result of Eq.~(\ref{eq:pika}) based on the $\pi K$ model~\cite{Schreiner:2009pf}. The experimental data below $10$~GeV are not used for the fitting with Eq.~(\ref{eq:pika}) according to Ref.~\cite{Gaisser:2011klf}.  \label{fig:pika-fit}}
\end{figure}

The muon charge ratio has been experimentally measured by several methods in the energy region of GeV to tens of TeV. Figure~\ref{fig:pika-fit} shows the comparison within the experimental results from other detectors~\cite{Ashley:1975uj, Matsuno:1984kq, Yamada:1991aq, Haino:2004nq, Achard:2004ws, Adamson:2007ww, MINOS:2011amj, Khachatryan:2010mw, OPERA:2014blf}. Around the energy range of the SK detector, the charge ratio has been measured by the experiment of Utah~\cite{Ashley:1975uj}, MINOS~(far detector)~\cite{Adamson:2007ww}, CMS~\cite{Khachatryan:2010mw}, and OPERA~\cite{OPERA:2014blf}. Those experimental data are consistent among their estimated uncertainties. 

The charge ratio is interpreted in terms of the primary cosmic-ray spectrum and composition. As mentioned in Sec.~\ref{sec:intro}, the contribution from kaon decays increases relative to that from pion decays when $E_{\mu}$ is larger than $\varepsilon_{\pi}/\cos \theta_{\mathrm{Zenith}}$. In the $\pi K$ simulation model~\cite{Schreiner:2009pf}, the charge ratio of cosmic-ray muons is described as,

\begin{equation} 
        R_{\mu} = \frac{\frac{f_{\pi}}{1+1.1E_{\mu}\cos\theta/\varepsilon_{\pi}}+\frac{\eta f_{K}}{1+1.1E_{\mu}\cos\theta/\varepsilon_{K}}}{\frac{1-f_{\pi}}{1+1.1E_{\mu}\cos\theta/\varepsilon_{\pi}}+\frac{\eta (1-f_{K})}{1+1.1E_{\mu}\cos\theta/\varepsilon_{K}}}, \label{eq:pika}
\end{equation}

\noindent where $f_{\pi}/(1-f_{\pi})$ is the charge ratio of muon from pion decays, $f_{K}/(1-f_{K})$ is that from kaon decays, $\eta$~($=0.054$) is the Gaisser constant, and $\varepsilon_{\pi}$~($\varepsilon_{K}$) is the critical energy defined in the first section. To determine the parameters~($f_{\pi}$ and $f_{K}$), Eq.~(\ref{eq:pika}) is fitted with the experimental results of charge ratio shown in Fig.~\ref{fig:pika-fit}. By adding the SK's result, the experimental results are fitted with Eq.~(\ref{eq:pika}), and the parameters are determined as $f_{\pi}=0.550\pm0.001$ and $f_{K}=0.693\pm0.006$ with $\chi^{2} = 159.4/160 = 1.0$. We should note that we did not include the experimental data below $10$~GeV for the fitting with Eq.~(\ref{eq:pika})~\cite{Gaisser:2011klf}.

The muon charge ratio measured by the SK detector is consistent with that by the Kamiokande-II detector, which was located at almost the same depth in the same mountain, within their uncertainties. The result from the SK detector is consistent with the prediction from the Honda flux model while it deviates by $1.9\sigma$ from the $\pi K$ model at $E_{\mu} \cos\theta_{\mathrm{Zenith}}=0.7^{+0.3}_{-0.2}~\mathrm{TeV}$. This tension between the measured charge ratio and the $\pi K$ model should lead to further improvement of atmospheric neutrino simulations.

\subsubsection{Muon polarization} \label{sec:pol_result}

\begin{figure}[!h]
    \begin{center}
        \includegraphics[width=\linewidth]{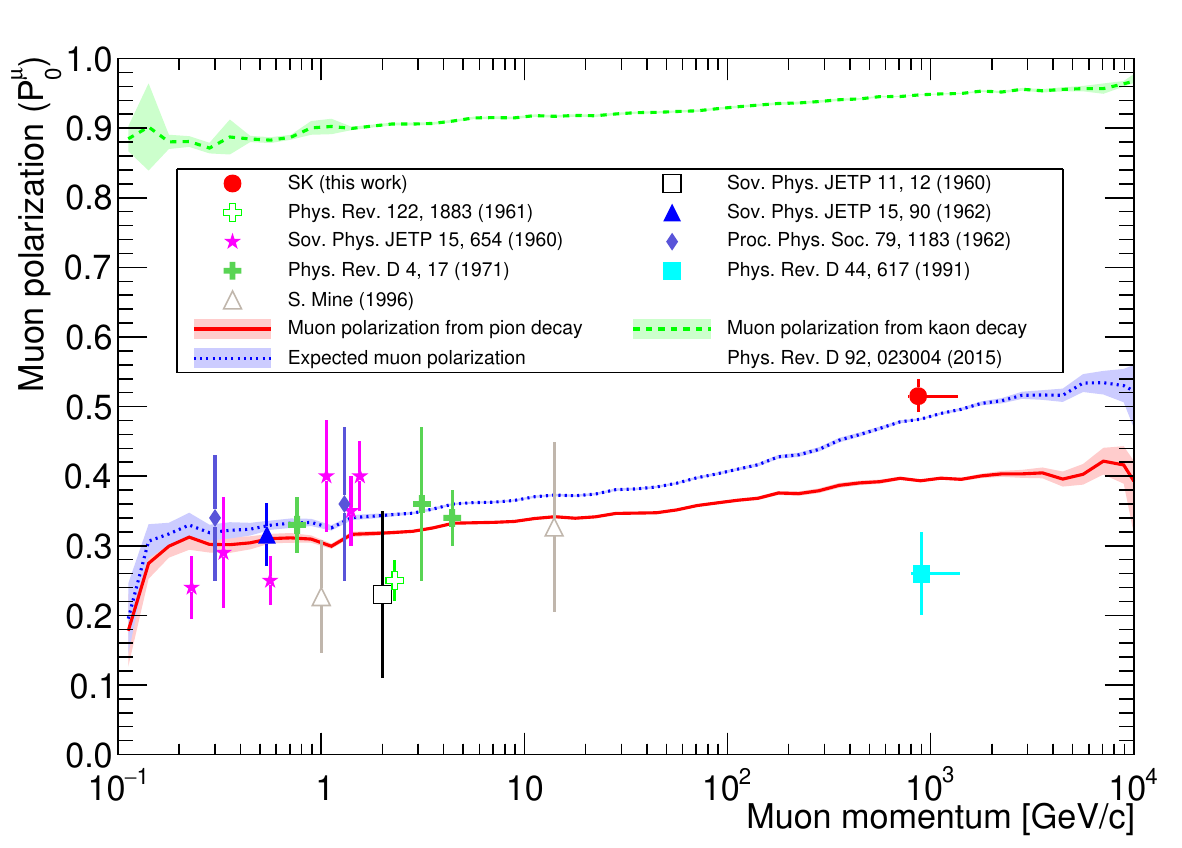}
    \end{center}
\caption{The experimental data of the polarization of the cosmic-ray muon at the production site as a function of the muon momentum. The red filled circle shows the SK data and other symbols with different colors show the experimental data measured by other groups~\cite{Kocharyan:1960, Johnson:1961, Alikhanyan:1962, Dolgoshein:1962, Gupta:1962, Turner:1971, Yamada:1991aq, Mine:1996kb}. Red solid and green dashed lines with band show the expected muon polarization from pion decay only and kaon decay only, where the details are described in Appendix~\ref{sec:mc-pol}. The blue line shows the expected muon polarization based on the Honda flux simulation~\cite{Honda:2015fha}. \label{fig:pol-comp}}
\end{figure}

The muon polarization has been experimentally measured at various locations. It has been mostly measured below $15~\mathrm{GeV}/c$ by experiments that measured the decay asymmetry of muons in a magnetic metal absorber on the ground~\cite{Kocharyan:1960, Johnson:1961, Alikhanyan:1962, Dolgoshein:1962, Gupta:1962, Turner:1971} as well as underground~\cite{Mine:1996kb}. Those experiments measured the muon polarization in the energy region where pion decay is dominant. On the other hand, the SK detector, as well as the Kamiokande-II detector~\cite{Yamada:1991aq}, uniquely measured the polarization of cosmic-ray muons with momentum around $1.0~\mathrm{TeV}/c$ at sea level because of their shared underground location. Such higher energy measurement of muon polarization can evaluate the contribution from kaon decays. Figure~\ref{fig:pol-comp} shows the comparison of the polarization of cosmic-ray muons with different momenta. 

As summarized in Table~\ref{tb:chi2-results}, the polarization measured by the SK detector is $0.52\pm0.02~(\mathrm{stat. {+}syst.})$, which is the most precise measurement of muon polarization ever because of the large statistics and well-controlled analysis method. However, this measured polarization is largely different from the result from Kamiokande-II~\cite{Yamada:1991aq} despite the consideration of estimated total uncertainties. A possible explanation of this discrepancy is discussed in Appendix~\ref{app-kam}. 

The muon polarization measured by the SK detector deviates by $1.5\sigma$ from the expectation simulated by the Honda flux model~\cite{Honda:2015fha} as shown in Fig.~\ref{fig:pol-comp}. This measurement completely excludes a ``pure-pion scenario" based on the Honda flux model with a significance of $5.1$--$5.2\sigma$ level at the energy near $1~\mathrm{TeV}/c$. 

\subsubsection{Directional dependence of charge ratio and muon polarization} \label{sec:zenith}

\begin{figure*}[]
    \begin{tabular}{cc}
        \begin{minipage}{0.50\textwidth}
            \centering
            \includegraphics[width=0.95\textwidth]{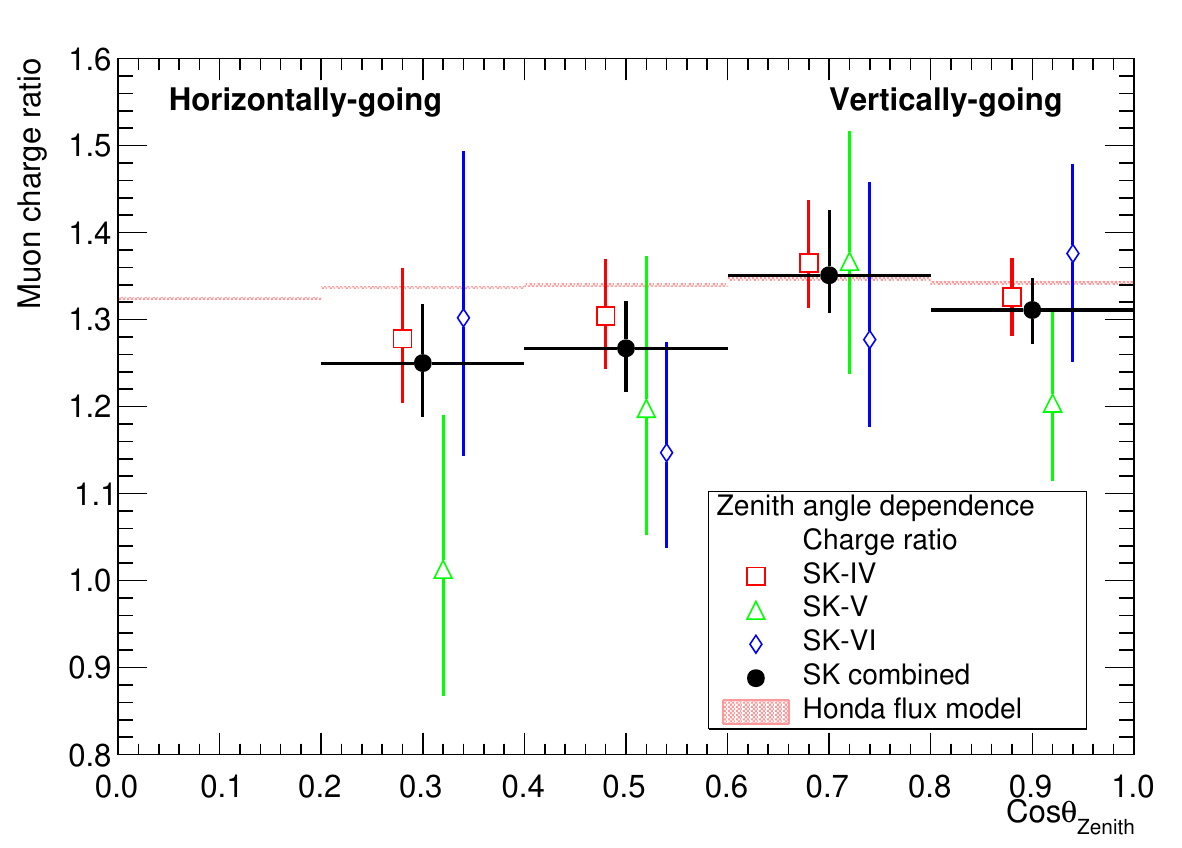}
        \end{minipage} 
        \begin{minipage}{0.50\textwidth}
            \centering
            \includegraphics[width=0.95\textwidth]{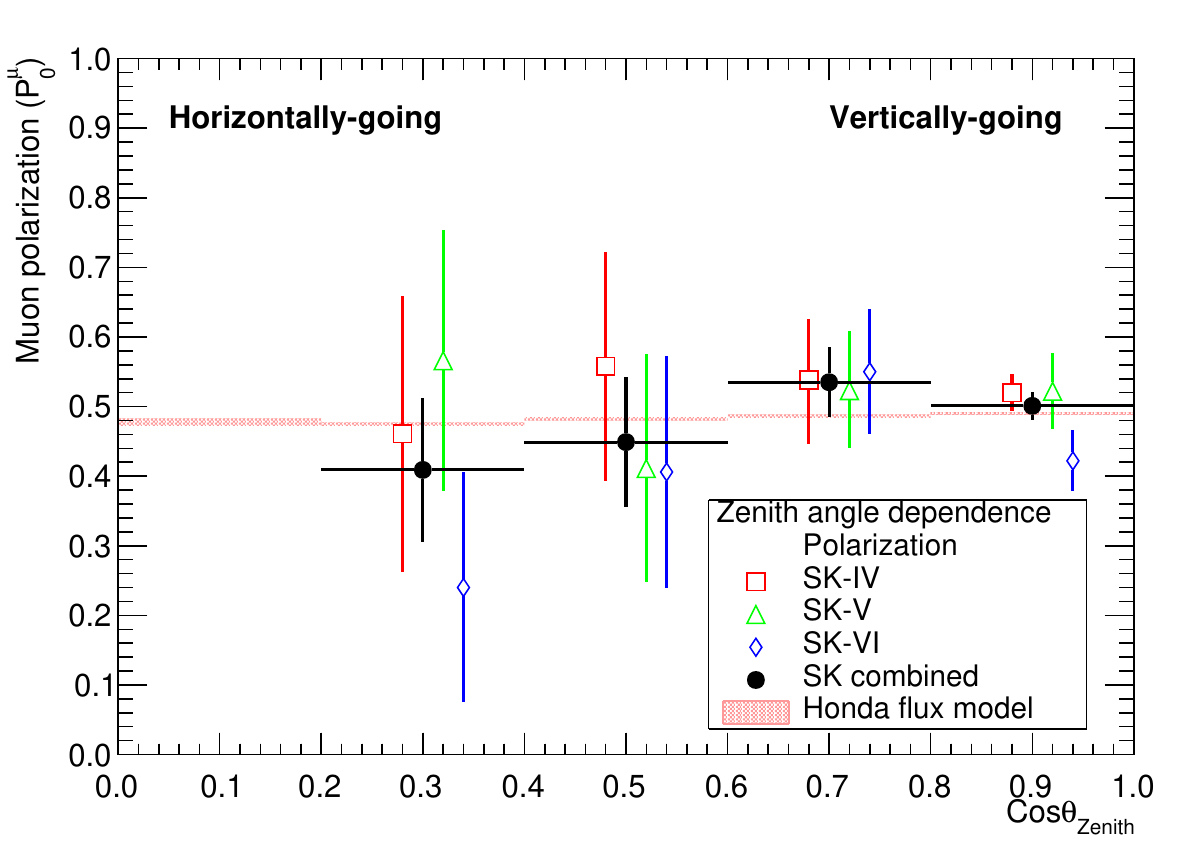}
	\end{minipage} \\
 	\begin{minipage}{0.50\textwidth}
		  \centering
		  \includegraphics[width=0.95\textwidth]{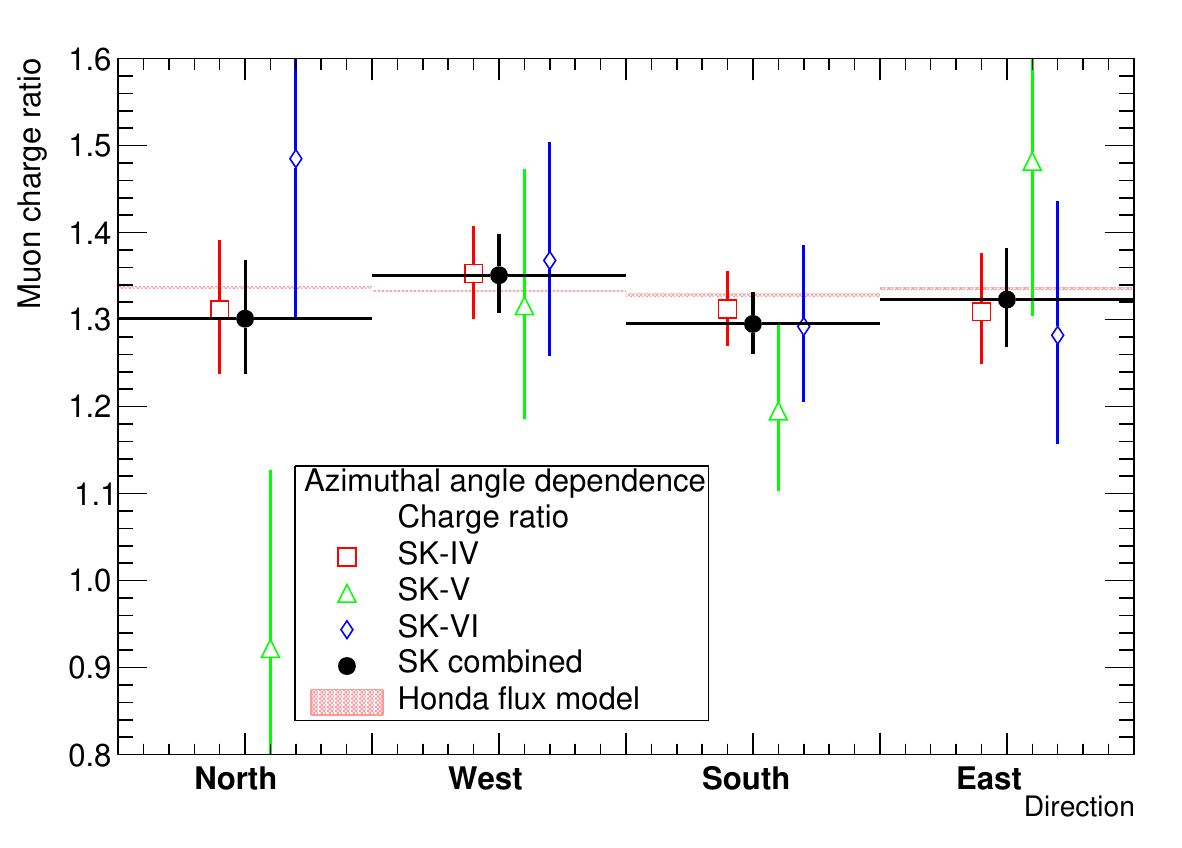}
        \end{minipage} 
        \begin{minipage}{0.50\textwidth}
            \centering
		  \includegraphics[width=0.95\textwidth]{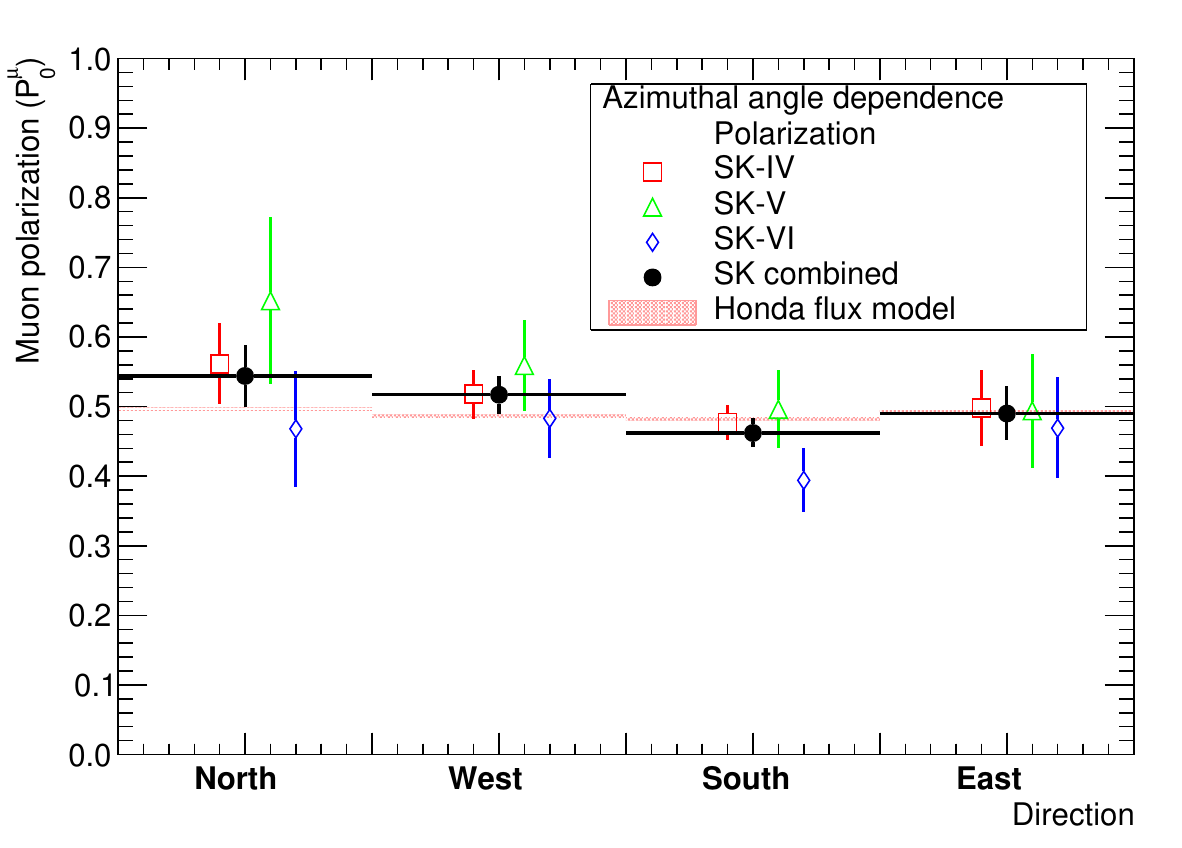}
	\end{minipage}
   \end{tabular}
\caption{Zenith and azimuthal angular dependences of the charge ratio and the muon polarization. The top-left~(right) panel shows the measured charge ratio~(muon polarization) as a function of $\cos\theta_{\mathrm{Zenith}}$ and the bottom-left~(right) panel shows their directional dependence. The red open square, green open upward-triangle, and blue open rhombus show the measured charge ratio using SK-IV, SK-V, and SK-VI, respectively. The black filled circle shows the combined value and the filled band shows the expectation derived from the Honda flux model~\cite{Honda:2015fha}. For ease of visualization, data points of each SK phase shift horizontally from their original positions. \label{fig:zenith}}
\end{figure*}

The propagation length in the mountain of cosmic-ray muons depends on its direction as shown in Fig.~\ref{fig:e-threshold}. That directional dependence selects different energies of muons as summarized in Table~\ref{tb:ranges} and this means the measured charge ratio and muon polarization will change depending on the muon direction. To test such directional dependence, we created sub-groups depending on the muon direction and analyzed the sub-groups to determine the charge ratio and muon polarization. Figure~\ref{fig:zenith} shows the zenith angle and azimuthal angle dependences of the measured charge ratio and the muon polarization. For the zenith angle dependence, a tension between the observed charge ratio and its expectation exists in the range of $0.2 < \cos \theta_{\mathrm{Zenith}} \le 0.6$ while the muon polarization measurement shows no clear discrepancy between the observed data and the expectation. For the azimuthal angle dependence, both observed data show no clear difference between them.

The contribution of muons from kaon decays changes depending on the zenith angle because of the difference of the relative propagation length of parent mesons in the atmosphere~\cite{Sanuki:2006yd}. However, this dependence is canceled out by the longer propagation length in the rock of the mountain, which efficiently selects cosmic-ray muons with high energies, as shown in Fig.~\ref{fig:mu-initial}. 

\begin{figure*}[]
    \begin{tabular}{cc}
        \begin{minipage}{0.50\textwidth}
            \centering
            \includegraphics[width=0.95\textwidth]{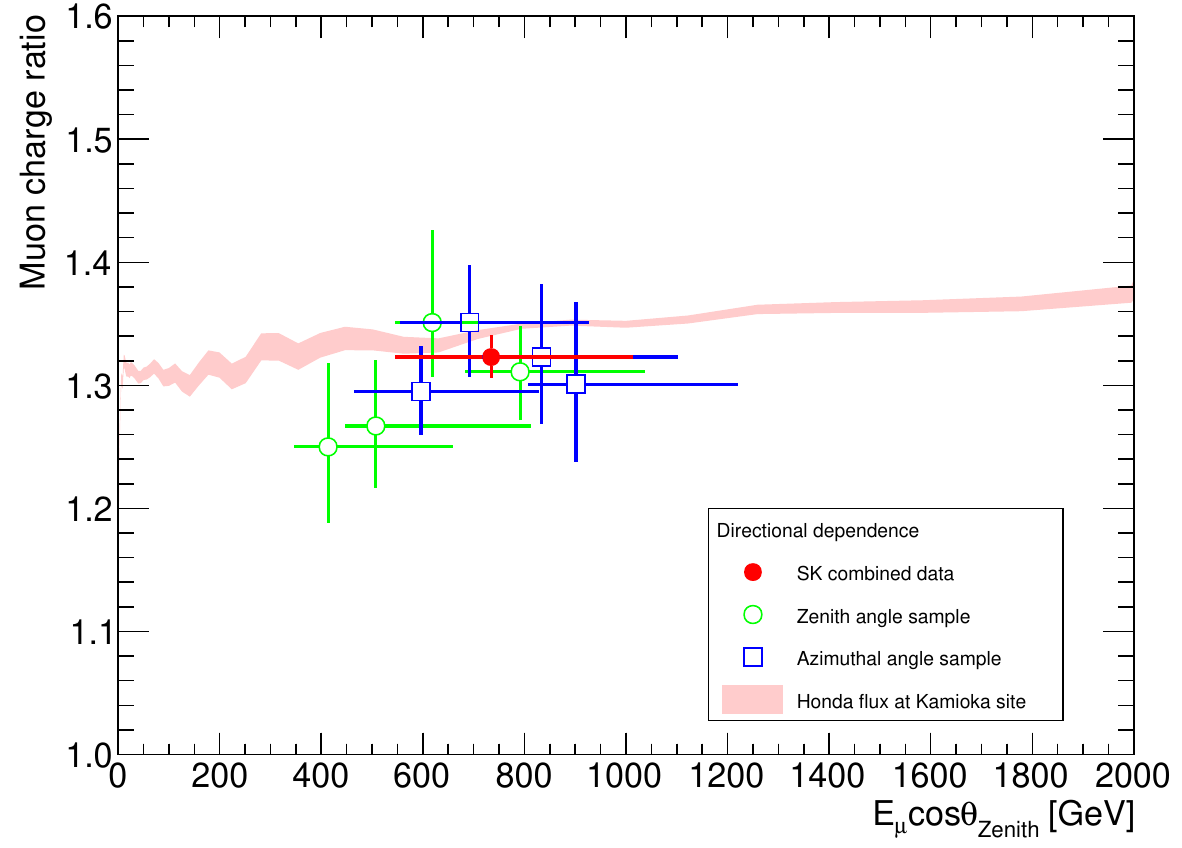}
        \end{minipage} 
        \begin{minipage}{0.50\textwidth}
            \centering
            \includegraphics[width=0.95\textwidth]{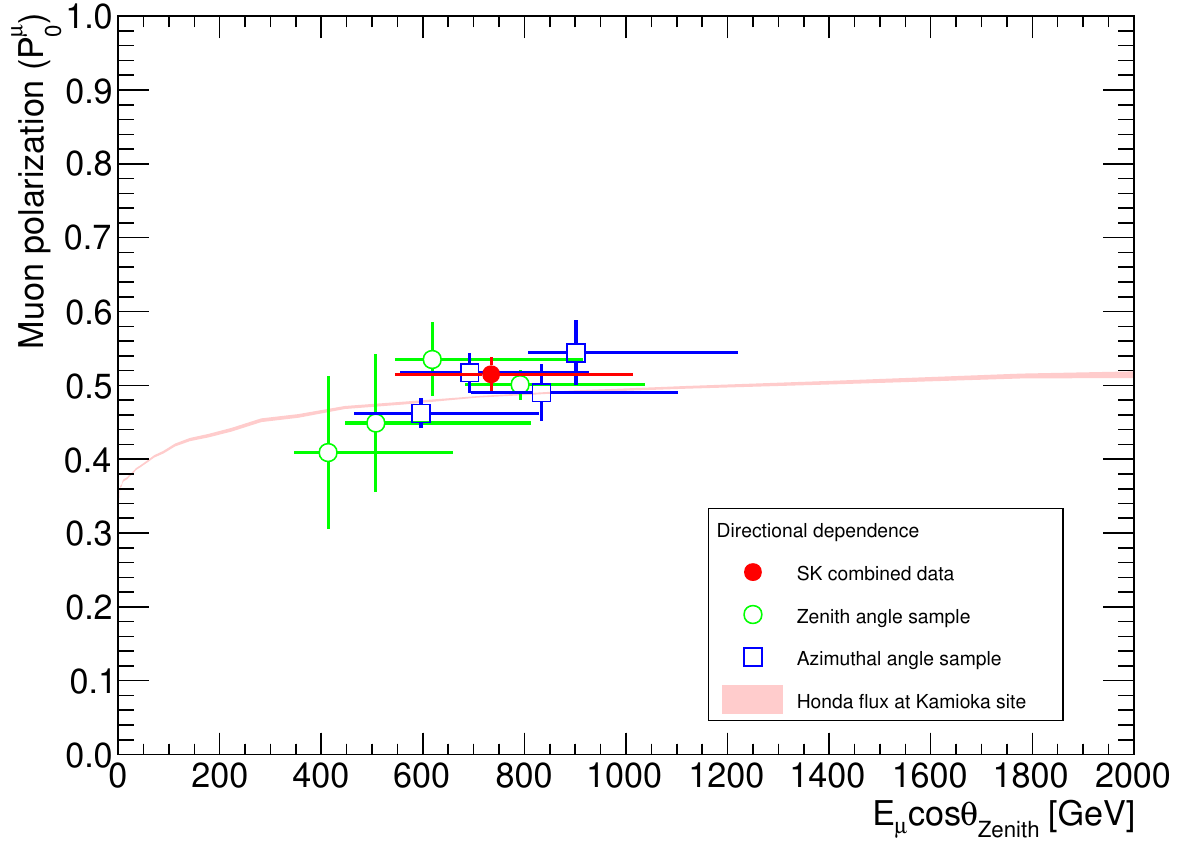}
	\end{minipage} 
   \end{tabular}
\caption{Directional dependence of the charge ratio and the muon polarization. The left~(right) panel shows the measured charge ratio~(muon polarization) as a function of $E_{\mu}\cos\theta_{\mathrm{Zenith}}$ using the sub-groups of different muon directions summarized in Table~\ref{tb:ranges}. The green open circle shows the combined SK data shown in Fig.~\ref{fig:zenith} of zenith angle sample, the blue open square shows that of azimuthal angle sample, the red filled circle shows the combined SK data from all directions listed in Table~\ref{tb:chi2-results}, and the red band shows the expectation simulated by the Honda flux model~\cite{Honda:2015fha}, respectively. \label{fig:scatter}}
\end{figure*}

Figure~\ref{fig:scatter} shows the directional dependence of the charge ratio and muon polarization to test the consistency between the observed data and the expectation with different ranges of $E_{\mu} \cos \theta_{\mathrm{Zenith}}$. Comparing the measured values with the expectation from the Honda flux simulation model~\cite{Honda:2015fha}, the accuracy of measurements is not enough to test the consistency between them. Hence further statistics are required to precisely evaluate the directional dependence of the charge ratio and the muon polarization by analyzing the decay electron sample. However, we should mention that the measured values are useful as the inputs for future atmospheric neutrino flux simulation models.

\section{Summary and prospects}

For improving the sensitivity to the neutrino oscillation parameters, precise modeling of atmospheric neutrino flux is highly required. The energy spectrum of atmospheric neutrinos reflects the charge ratio and polarization of cosmic-ray muons because of their origin. For providing new inputs for atmospheric neutrino simulations, the charge ratio of positive to negative cosmic-ray muons and the muon polarization are experimentally measured in the SK detector using data collected from 2008 September to 2022 June. Because of its long operation, these precision measurements were performed with large statistics of accumulated decay electron events.

The muon charge ratio is measured through the analysis of the decay time of stopping muons and found to be $R=1.32 \pm 0.02\,(\mathrm{stat. {+}syst.})$ at $E_{\mu} \cos \theta_{\mathrm{Zenith}}=0.7^{+0.3}_{-0.2}~\mathrm{TeV}$. This result is in agreement with past measurements at the energy range around $1$~TeV and the prediction from the Honda flux model~\cite{Honda:2015fha} within their uncertainties while it deviates by $1.9\sigma$ from the $\pi K$ model~\cite{Schreiner:2009pf}.

The polarization of the cosmic-ray muons is also measured by evaluating the angle between the parent muon and the decay electron. By assuming the magnitude of polarization for the negative and positive muons is equal while the sign is inverted~($P^{\mu}=-P^{\mu^{+}}_{0}=P^{\mu^{-}}_{0}$), the magnitude of the polarization at the production site is obtained to be $P^{\mu}_{0} = 0.52\pm0.02\,(\mathrm{stat. {+}syst.})$ at the muon momentum of $0.9^{+0.5}_{-0.1}~\mathrm{TeV}/c$. This is the most precise measurement ever to experimentally determine the cosmic-ray muon polarization near $1~\mathrm{TeV}/c$ because of large statistics. The measured polarization also deviates by $1.5\sigma$ from the expectation simulated by the Honda flux model~\cite{Honda:2015fha}. 

We also searched for a possible periodicity in charge ratio and muon polarization by analyzing the yearly-binned observed data with the generalized Lomb-Scargle method. No clear periodic change is found in the period from $2008$ to $2018$, which covers the solar cycle~$24$.

As for prospects, the measurement result of charge ratio constrains the ratio of atmospheric neutrinos to anti-neutrinos~($\nu/\bar{\nu}$), and the measurement of muon polarization also constrains the energy spectrum of atmospheric neutrinos in neutrino flux simulation models. These measurements can therefore contribute to more precise determination of neutrino oscillation parameters. In addition, the precise measurement of muon polarization in the underground environment could prove helpful in understanding the asymmetric changes in helical biopolymers, which may account for the emergence of biological homochirality~\cite{Globus:2020gud, Globus:2021kau}, due to spin-polarized cosmic radiation.

\begin{acknowledgments}

We would like to thank P.~Lipari from INFN Roma for answering our questions about muon polarization. We would also like to thank H.~Kurashige from Kobe University for suggestions on technical issues when developing the simulation of muon decays with spin. We gratefully acknowledge the cooperation of the Kamioka Mining and Smelting Company. The Super-Kamiokande experiment has been built and operated from funding by the Japanese Ministry of Education, Culture, Sports, Science and Technology; the U.S. Department of Energy; and the U.S. National Science Foundation. Some of us have been supported by funds from the National Research Foundation of Korea~(NRF-2009-0083526 and NRF 2022R1A5A1030700) funded by the Ministry of Science, Information and Communication Technology~(ICT); the Institute for Basic Science~(IBS-R016-Y2); and the Ministry of Education~(2018R1D1A1B07049158, 2021R1I1A1A01042256, 2021R1I1A1A01059559); the Japan Society for the Promotion of Science; the National Natural Science Foundation of China under Grants No.11620101004; the Spanish Ministry of Science, Universities and Innovation~(grant PID2021-124050NB-C31); the Natural Sciences and Engineering Research Council~(NSERC) of Canada; the Scinet and Westgrid consortia of Compute Canada; the National Science Centre~(UMO-2018/30/E/ST2/00441 and UMO-2022/46/E/ST2/00336) and the Ministry of Education and Science~(2023/WK/04), Poland; the Science and Technology Facilities Council~(STFC) and Grid for Particle Physics~(GridPP), UK; the European Union's Horizon 2020 Research and Innovation Programme under the Marie Sklodowska-Curie grant agreement no.754496; H2020-MSCA-RISE-2018 JENNIFER2 grant agreement no.822070, H2020-MSCA-RISE-2019 SK2HK grant agreement no.872549; and European Union's Next Generation EU/PRTR grant CA3/RSUE2021-00559.

\end{acknowledgments}

\bibliography{main}

\appendix

\section{Muon polarization prediction with Honda flux calculation} \label{sec:mc-pol}

For considering the polarization of cosmic-ray muons in the MC simulation, relativistic transformations of the polarization vector should be formalized.  In this section, we briefly describe the formulation of the muon polarization after its production by assuming the natural unit of $c= \hbar = 1$. 

In the Honda flux model calculation~\cite{Honda:2015fha}, the polarization of each muon is calculated based on the kinematics of its parent particle.
Cosmic-ray muons in air showers are dominantly produced from the four decay modes, summarized in Table~\ref{tab:decay2mu}. 

\begin{table} [h]
  \caption{Decay modes producing muons in air showers. Branching ratios used in the Honda calculation are also shown.}
\label{tab:decay2mu}
  \begin{center}
   \begin{tabular}{c c c }
   \hline \hline 
   Parent particle & Decay mode & Branching ratio~[$\%$] \\
    \hline 
    $\pi^{\pm}$ & $\pi^\pm \to \mu^\pm + \nu_{\mu} (\bar{\nu}_{\mu})$  & $100.0$ \\
    \hline 
    $K^{\pm}$ & $K^\pm \to \mu^\pm + \nu_{\mu}(\bar{\nu}_{\mu})$  & $\phantom{0}63.5$ \\
            & $K^{\pm} \to \pi^{0} + \mu^{\pm} + \nu_{\mu}(\bar{\nu}_{\mu})$ & $\phantom{00}3.2$ \\
    \hline 
    $K_{L}$     & $K_{L}\to \pi^{\mp} + \mu^{\pm} + \nu_{\mu}(\bar{\nu}_{\mu})$ & $\phantom{0}27.1$ \\
    \hline \hline
       \end{tabular}
  \end{center}
\end{table}
 
If the decay is a two-body decay ({\it i.e.}, $\pi \to \mu + \nu$ or $K \to \mu + \nu$), the muon polarization $P^{*}_{\mu}$ in the parent meson's rest frame~(or center-of-mass~(hereafter CM) frame) is fully-polarized, {\it i.e.} $+1$ or $-1$. Here, an asterisk~($^*$) distinguishes quantities referenced in the CM frame. In the case of three-body decays ({\it i.e.}, $K^{\pm} \to \pi^{0} + \mu + \nu$ or $K_{L} \to \pi + \mu + \nu$), $P^{*}_{\mu}$ is given as a function of muon energy according to Ref.~\cite{BRENE1961553}.

At first, we define three vectors according to Ref.~\cite{Hayakawa:1957}; (i)~$p_{\sigma} = (E_{\mu}, \vb{p})$ is the muon energy-momentum vector, where $\sigma$ is a dummy index, which runs $0$ to $3$, $E_{\mu}$ is the muon energy, and $\vb{p}$ is the spatial component of $p_{\sigma}$, respectively. (ii)~$s_{\sigma} = (s_{0},\vb{s})$ is the spin four-vector, which forms a pseudo-vector. (iii)~$\boldsymbol{\zeta}$ is a unit vector to express the polarization of muon. If the muon is at rest, $\vb{s}$ corresponds to $\boldsymbol{\zeta}$ while $\vb{s}$ has the direction of the momentum if the muon is in motion.

The muon polarization in the laboratory frame, which is defined as $P_{\mu}$, is obtained by introducing $\boldsymbol{\zeta}$. We randomly sample a unit vector $\boldsymbol{\zeta}^{*}$ which represents the polarization direction in the CM frame, satisfying 

\begin{equation}
 \label{eq:pol_in_CM}
 P^{*}_{\mu} = \boldsymbol{\zeta}^{*} \cdot \frac{\vb{p}^{*}}{|\vb{p}^{*}|}.
\end{equation} 

\noindent Then, the spin vector in the laboratory frame, $s_{\sigma}$, is calculated to be

\begin{eqnarray}
\label{eq:spin4vec}
    \vb{s} &=& \boldsymbol{\zeta}^{*} + \frac{\boldsymbol{\zeta}^{*} \cdot \vb{p}^{*}}{E^{*}_{\mu}(m_{\mu}+E^{*}_{\mu})}\vb{p}^{*},\\ \label{eq:spin4vec_t} 
    s_{0} &=&  \boldsymbol{\zeta}^*\cdot\vb{p}^{*} / m_{\mu},
\end{eqnarray}

\noindent where $m_{\mu}$ represents the muon rest mass. After applying the Lorentz transformation to the $s_{\sigma}$, the polarization in the laboratory frame $P_{\mu}$ is obtained from Eqs.~(\ref{eq:pol_in_CM}) and~(\ref{eq:spin4vec_t}), as

\begin{equation}
 P_{\mu} = s_{0}\frac{ m_{\mu} }{|\vb{p}|}, \label{eq:mc-pol}
\end{equation}

\noindent where the $s_{0}$ and $\vb{p}$ are quantities in the laboratory frame. The relation of Eq.~(\ref{eq:mc-pol}) is satisfied in both two and three-body decays. In the case of the two-body decays, the result can be simplified to be

\begin{equation}
 P_{\mu} = \frac{ E^{*}_{\mu} E_{\mu} }{ p^{*}_{\sigma} p_{\sigma} } - \gamma_{\mathrm{prt}} \frac{ m_{\mu}^{2} }{ p^{*}_{\sigma} p_{\sigma} }, \label{eq:two-body}
\end{equation}

\noindent where $\gamma_{\mathrm{prt}}$ is the gamma factor of the parent meson in the laboratory frame. 

\begin{figure}[!h]
    \begin{tabular}{cc}
        \begin{minipage}{0.5\textwidth}
            \centering
		  \includegraphics[width=0.95\textwidth]{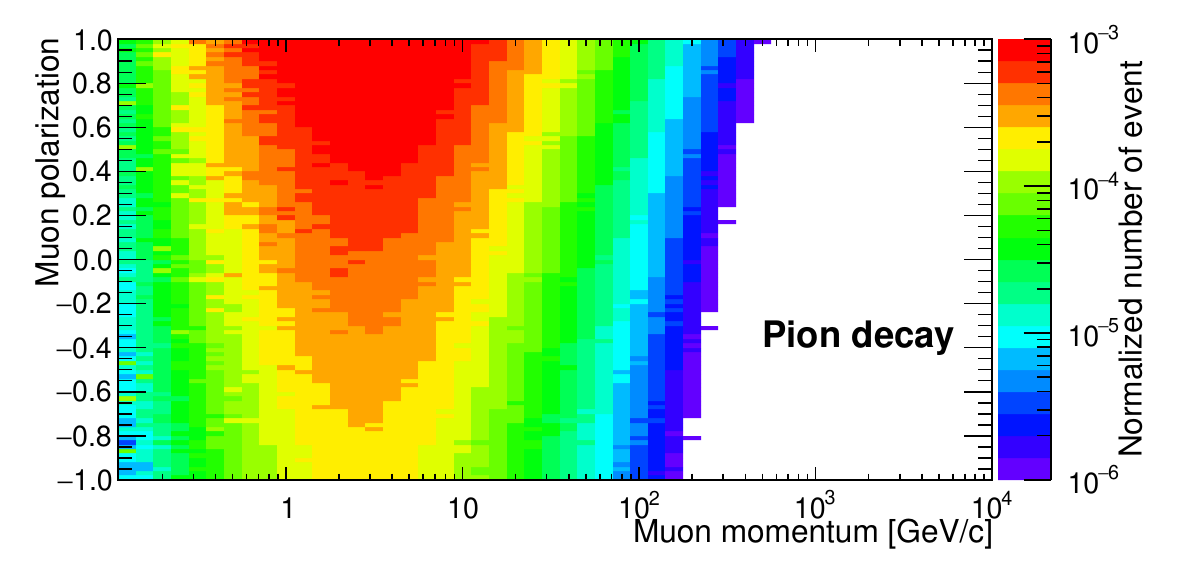}
	\end{minipage} \\
	\begin{minipage}{0.5\textwidth}
		  \centering
		  \includegraphics[width=0.95\textwidth]{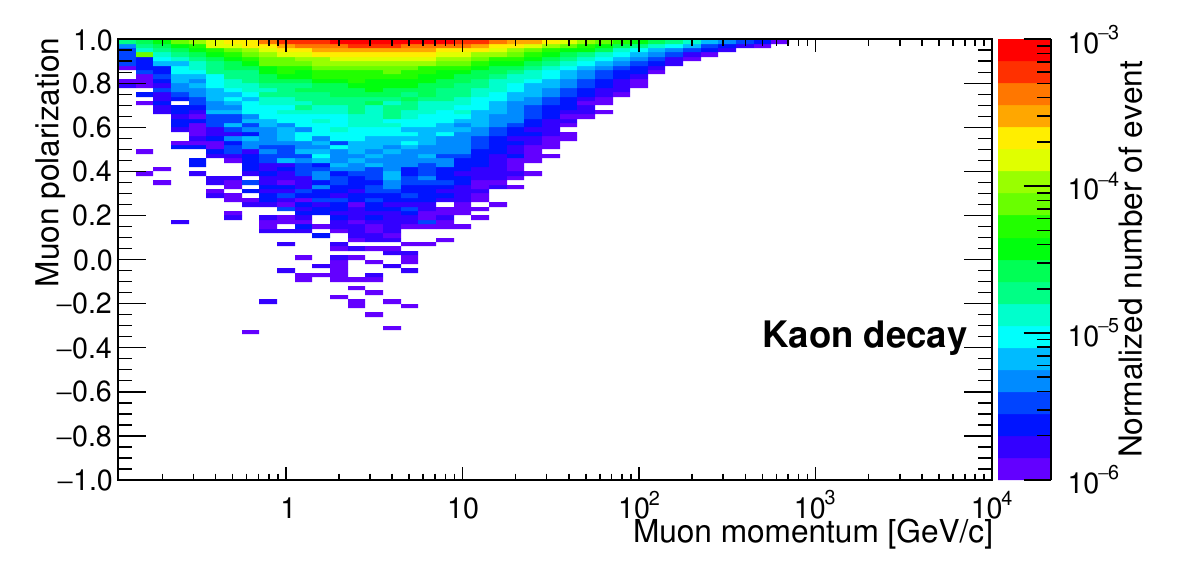}
	\end{minipage}
    \end{tabular}
\caption{Expected muon polarization distribution from pion decay~(top) and kaon decay~(bottom) predicted by the Honda flux model calculation~\cite{Honda:2015fha}
\label{fig:pol-honda-dist}}
\end{figure}

Figure~\ref{fig:pol-honda-dist} shows the produced muon energy dependence of the expected muon polarization. The muon polarization from pion decay is not only in the forward direction but also in a wide variety of directions. On the other hand, the muon polarization from kaon decay tends to be in the forward direction in the direction of momentum. Although both decays are described in the two-body decay, the different distributions originate from the mass of mesons in Eq.~(\ref{eq:two-body}).

\section{Delayed signal after the gadolinium-loading in SK-VI} \label{app:neutron-gd}

\begin{figure}[!h]
    \begin{tabular}{cc}
        \begin{minipage}{0.5\textwidth}
            \centering
		  \includegraphics[width=0.95\textwidth]{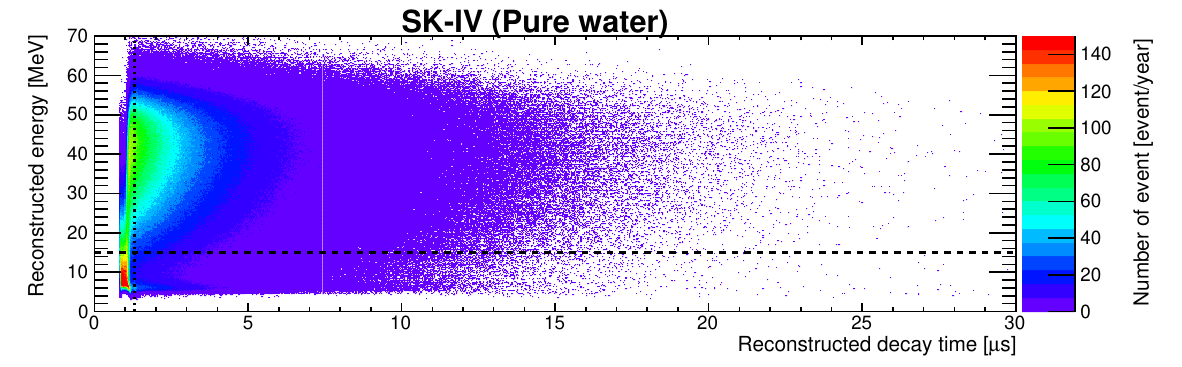}
	\end{minipage} \\
	\begin{minipage}{0.5\textwidth}
		  \centering
		  \includegraphics[width=0.95\textwidth]{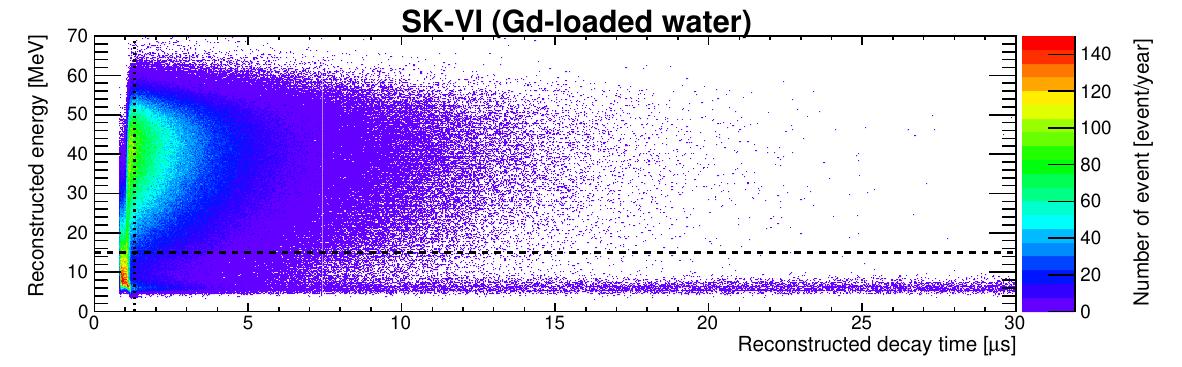}
	\end{minipage}
    \end{tabular}
\caption{The time difference between the stopping muon and the reconstructed events in the SK-IV~(top) and -VI~(bottom) data sample. In the SK-VI data sample, $8$~MeV $\gamma$-rays originating from the neutron capture by gadolinium in water are observed after tagging the decay electron. \label{fig:tdiff_sk6}}
\end{figure}

In general, neutrons are produced along the muon track as spallation products~\cite{Super-Kamiokande:2022cvw}. In addition to this, neutrons are also emitted from nitrogen~($\mathrm{^{14}N}$ and $\mathrm{^{15}N}$) via the muon capture on oxygen~\cite{Measday:2001yr} as explained in Sec.~\ref{sec:pol-detector}. Because of high efficiency for neutron detection after the gadolinium-loading~\cite{Beacom:2003nk, Super-Kamiokande:2021the}, such neutrons are tagged within the decay electron search window of [$-5,+35$]~$\mu$s. Figure~\ref{fig:tdiff_sk6} shows the relationship between the muon timing and the reconstructed energies of tagged events, where the delayed signals consist of decay electron events and $\gamma$-rays from the neutron capture by gadolinium. Gadolinium de-excitation $\gamma$-rays are observed with a longer time difference compared with the decay electrons. Those additional events are rejected by the energy cut as described in Sec.~\ref{sec:gamma}.

Although the additional delayed events are found in the window of [$-5,+35$]~$\mu$s, the occurrence rate of $N_{\mathrm{trg}}$ is smaller than the statistical fluctuation as summarized in Table~\ref{tb:num-decaye}. Hence, the contamination due to  $\gamma$-rays from gadolinium capture in decay electron sample presented in Sec.~\ref{sec:reduction} is quite small and such contamination does not affect the measurements for charge ratio and muon polarization.

\section{The energy, time difference, and $\cos \theta$ distributions in SK-V and SK-VI} \label{app:best-fit}

\begin{figure*}[]
    \begin{tabular}{cc}
        \begin{minipage}{0.33\textwidth}
            \centering
            \includegraphics[width=\textwidth]{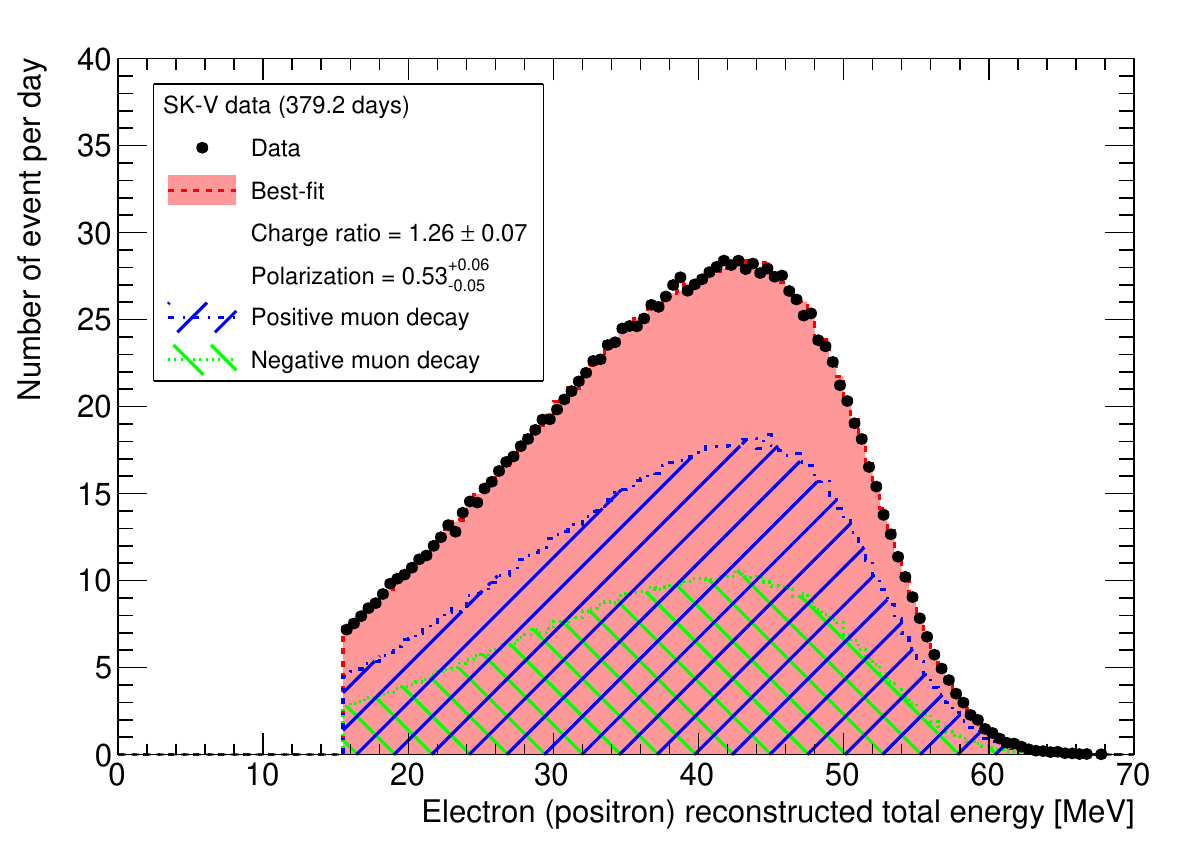}
        \end{minipage} 
        \begin{minipage}{0.33\textwidth}
            \centering
            \includegraphics[width=\textwidth]{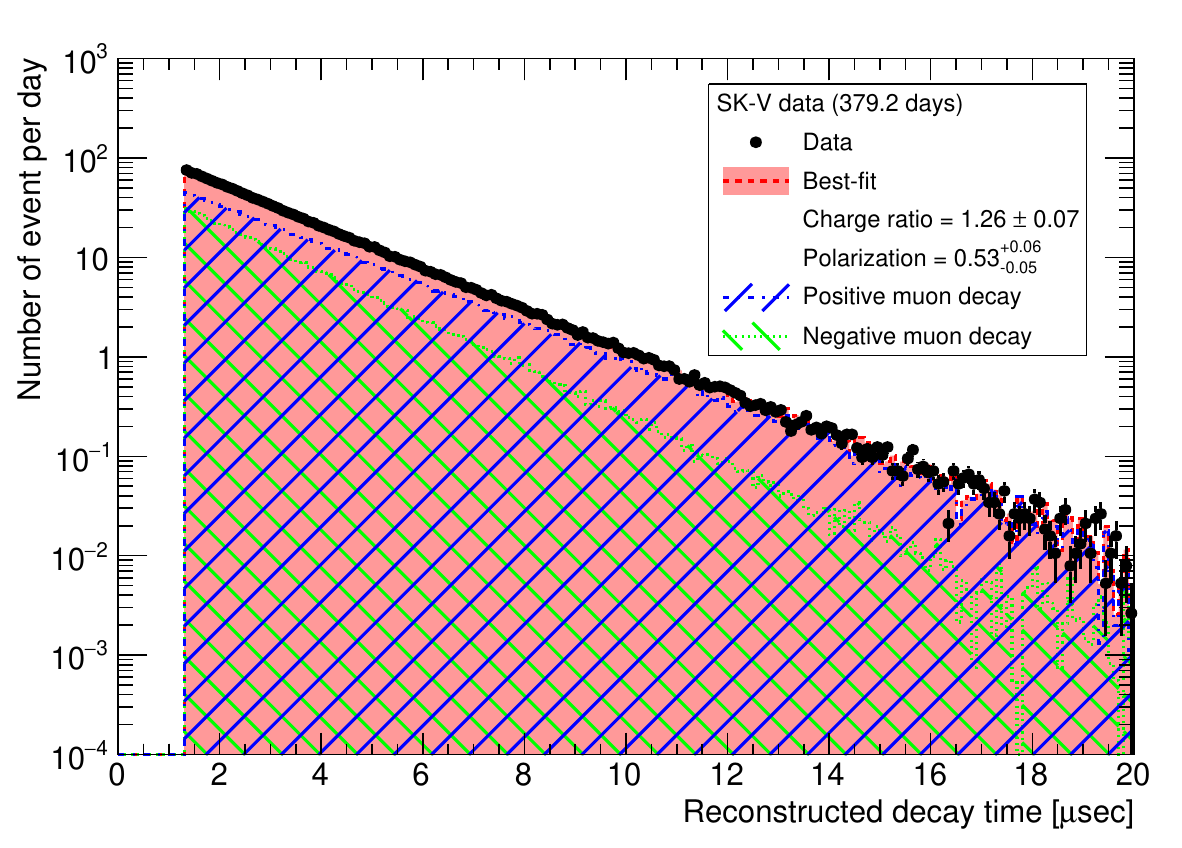}
	\end{minipage} 
         \begin{minipage}{0.33\textwidth}
            \centering
            \includegraphics[width=\textwidth]{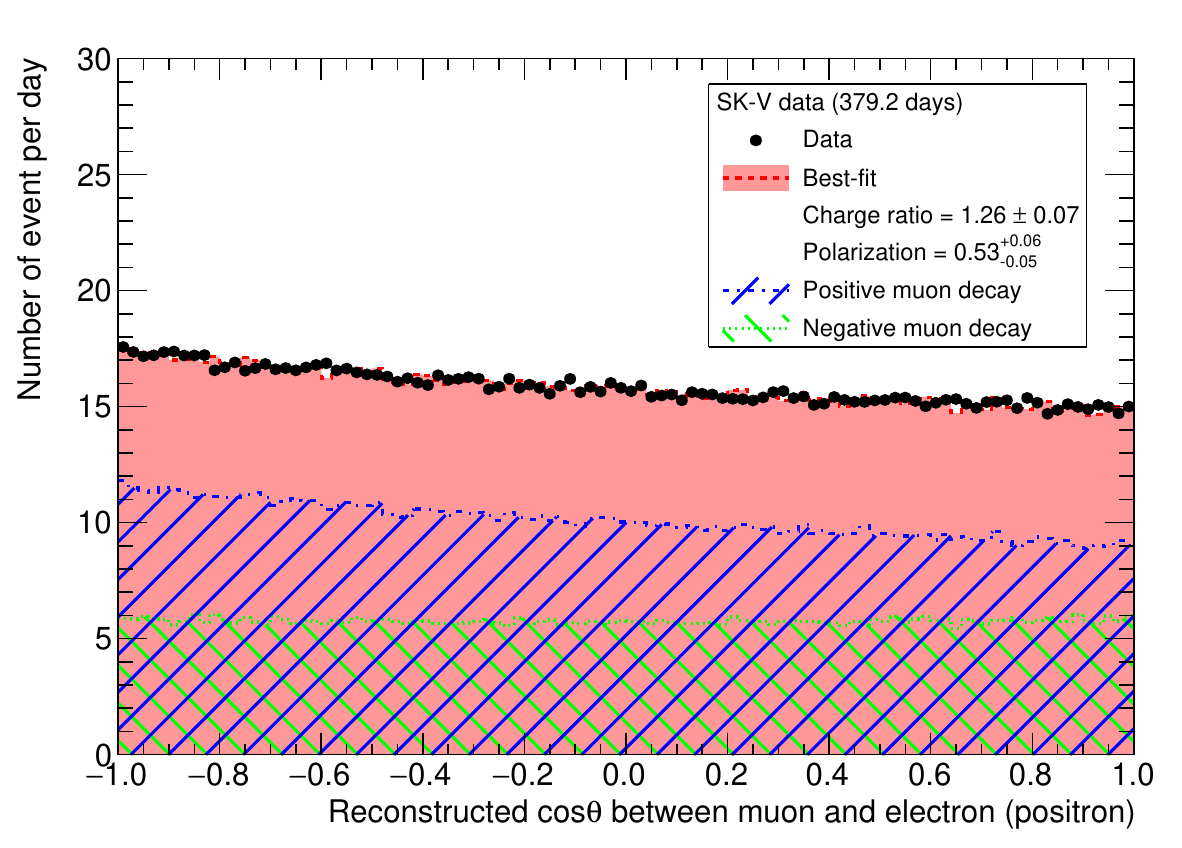}
	\end{minipage}
   \end{tabular}
\caption{The example of three distributions using the decay electron sample in SK-V~(purified water and $379.2$~days). The definition of colors is the same as Fig.~\ref{fig:dist-comp}. \label{fig:dist-comp-sk5}}
\end{figure*}

\begin{figure*}[]
    \begin{tabular}{cc}
        \begin{minipage}{0.33\textwidth}
            \centering
            \includegraphics[width=\textwidth]{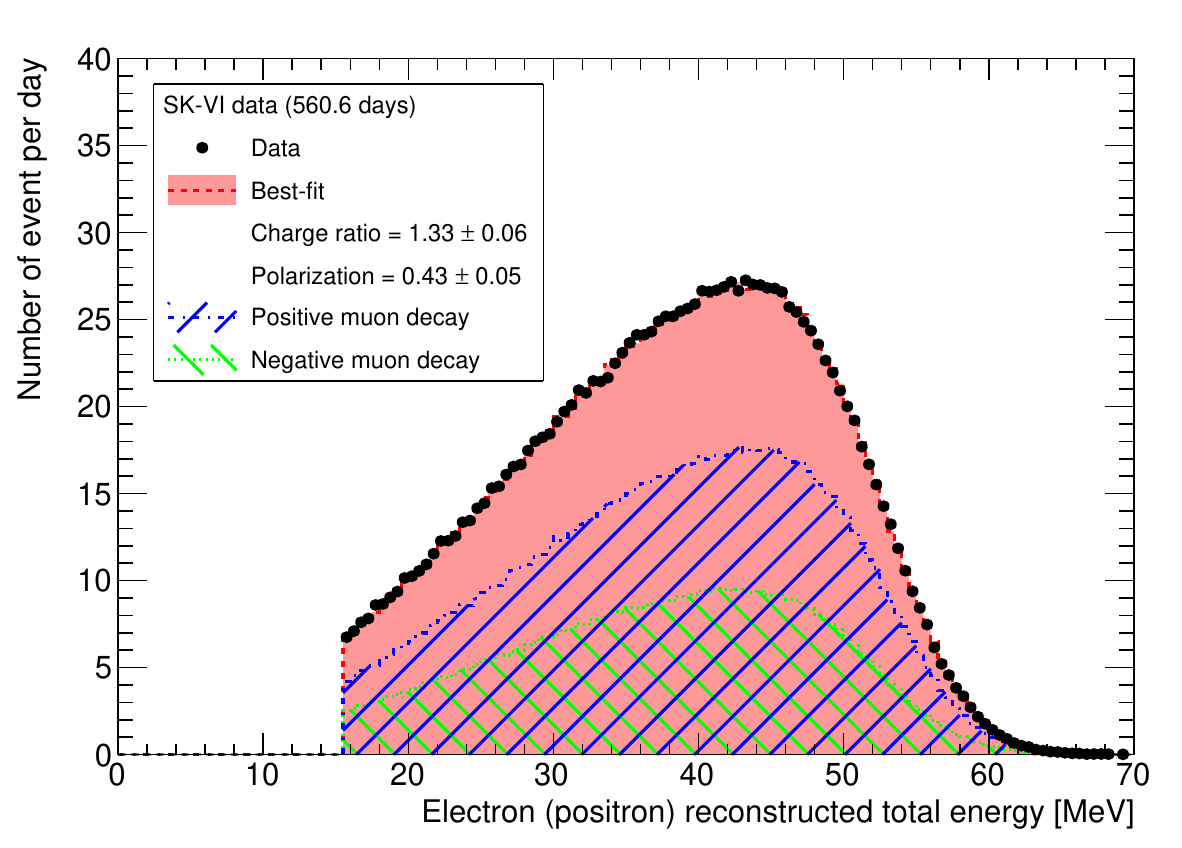}
        \end{minipage} 
        \begin{minipage}{0.33\textwidth}
            \centering
            \includegraphics[width=\textwidth]{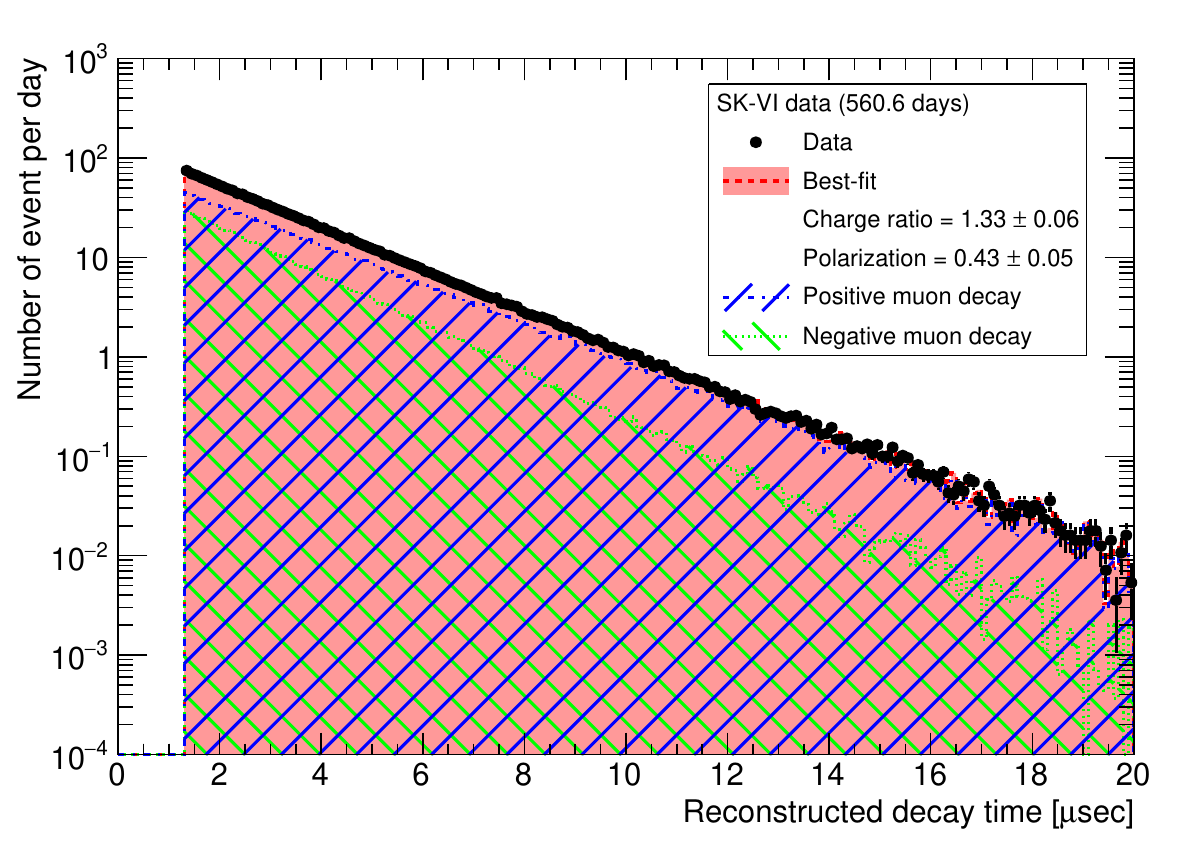}
	\end{minipage} 
         \begin{minipage}{0.33\textwidth}
            \centering
            \includegraphics[width=\textwidth]{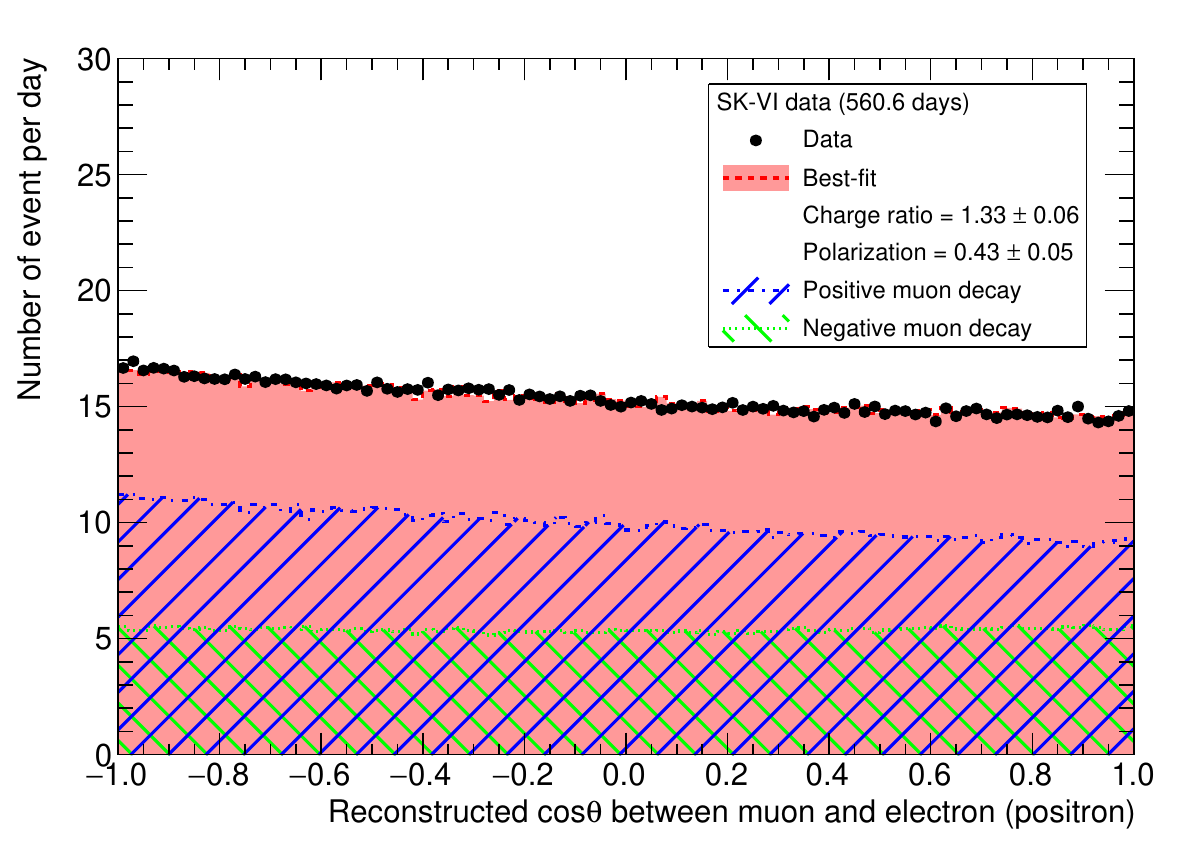}
	\end{minipage}
   \end{tabular}
\caption{The example of three distributions using the decay electron sample in SK-VI~(gadolinium loaded water and $560.6$~days). The definition of colors is the same as Fig.~\ref{fig:dist-comp}. \label{fig:dist-comp-sk6}}
\end{figure*}

Figures~\ref{fig:dist-comp-sk5} and~\ref{fig:dist-comp-sk6} show three distributions of observed decay electron sample in SK-V and SK-VI, i.e. the reconstructed energy distribution, time difference distribution, and $\cos \theta$ distribution together with the MC simulation. 

\section{Comments on the polarization measurement by the Kamiokande-II detector} \label{app-kam}

As presented in Sect.~\ref{sec:pol_result}, the muon polarization measured by the SK detector is higher than that measured by the Kamiokande-II detector~\cite{Yamada:1991aq} and the measured values deviate by more than their estimated uncertainties despite their being at almost the same location~(depth). According to MUSIC simulations, the expected muon momentum at the surface of the mountain for the Kamiokande-II detector is $0.9^{+0.5}_{-0.1}~\mathrm{TeV}/c$, which is consistent with that of the SK detector as summarized in Table~\ref{tb:ranges}.

The method used in Ref.~\cite{Yamada:1991aq} fits the distribution of $\cos \theta$ between the incoming muon and the emitted electrons assuming Michel parameters without MC simulation. The following function was used:

\begin{eqnarray}
    \frac{dN}{d(\cos \theta)} & \propto & (1-2x^{3}_{0}+x^{4}_{0}) \nonumber \\ 
     & & - \left(\frac{1}{3}+\frac{2x^{3}_{0}}{3}-x^{4}_{0}\right) P_{\mathrm{obs}}^{\mu} \cos \theta, \label{eq:fit}
\end{eqnarray}

\noindent where the parameters $P_{\mathrm{obs}}^{\mu}$ and $\cos \theta$ have the same definition as in Sect.~\ref{sec:dcy_po} and $x_{0}$ is the relative threshold of energy, which we selected as $x_{0}=15.5/52.8$ for the SK analysis to avoid contamination by $\gamma$-rays from oxygen capture. This fitting method can determine the muon polarization without MC simulation.

To validate the method above, we also performed this fitting procedure with Eq.~(\ref{eq:fit}) using the SK data sets. Table~\ref{tb:pol-comp} summarizes the measurement results of the muon polarization at the detector using the two different methods.  When analyzing SK data we found a discrepancy of at most $2\sigma$ between the two different methods, with the method without MC simulation yielding a lower polarization value. Hence, we suspect that the analysis procedure used in Ref.~\cite{Yamada:1991aq} may cause a systematic shift of polarization measurement. This systematic shift may be explained by the resolution of the opening angle between the incoming muon and the decay electron, the resolution of energy reconstruction of decay electrons, and the choice of $x_{0}$ in Eq.~(\ref{eq:fit}). For these reasons, we conclude that the analysis procedure to measure cosmic-ray muon polarization by the Kamiokande-II collaboration is not appropriate.

\begin{table}[!h]
    \begin{center}
    \caption{The summary of the polarization at the detector measured by the $\chi^{2}$ method and the fitting method. $x_{0}$ in Eq.~(\ref{eq:fit}) is $x_{0}=15.5/52.8$ for the SK analysis while $x_{0}=14.0/52.8$ in Ref.~\cite{Yamada:1991aq}. The uncertainty in the $\chi^{2}$ method includes the statistical and systematic uncertainties while that in the fitting method is statistical error only.}
        \label{tb:pol-comp}
            \begin{tabular}{c|cc}
                \hline \hline
                 & $P^{\mu}_{\mathrm{obs}}$  & $P^{\mu}_{\mathrm{obs}}$ \\
                 & $\chi^{2}$ method~(Eq.~(\ref{eq:p-obs-equal})) & fitting method \\
                 & (stat. and syst.) & (stat. only)\\ \hline
                SK-IV &  $0.22\pm0.02$ & $0.191 \pm 0.002$   \\
                SK-V &  $0.22\pm0.02$ & $0.198\pm 0.007$    \\
                SK-VI & $0.18\pm0.02$ & $0.184 \pm 0.005$  \\
                \hline 
                SK combined & $0.22\pm0.01$ & $0.191 \pm 0.002$  \\
                Kamiokande-II & -- & $0.12\pm0.02$   \\
            \hline
            \hline
        \end{tabular}
    \end{center}
\end{table}

\end{document}